\documentclass{article}

\usepackage{arxiv}

\usepackage[utf8]{inputenc} 
\usepackage[T1]{fontenc}    
\usepackage{hyperref}       
\usepackage{url}            
\usepackage{booktabs}       
\usepackage{amsfonts}       
\usepackage{nicefrac}       
\usepackage{microtype}      
\usepackage{lipsum}
\usepackage{graphicx}
\graphicspath{ {./images/} }
\usepackage{subcaption}
\usepackage{mathtools}
\usepackage{algorithm} 
\usepackage{algpseudocode}

\title{Pi-ViMo: Physiology-inspired Robust Vital Sign Monitoring using mmWave Radars}

\author{
 Bo Zhang \\
  Dept. of Computing and Software\\
  McMaster University\\
  Hamilton, Ontario, Canada \\
  \texttt{zhanb59@mcmaster.ca} \\
   \And
 Boyu Jiang \\
  Dept. of Computing and Software\\
  McMaster University\\
  Hamilton, Ontario, Canada \\
  \texttt{jiangb11@mcmaster.ca} \\
  \And
 Rong Zheng \\
  Dept. of Computing and Software\\
  McMaster University\\
  Hamilton, Ontario, Canada \\
  \texttt{rzheng@mcmaster.ca} \\
  \And
 Xiaoping Zhang \\
  Department of Electrical, Computer, and Biomedical Engineering\\
  Toronto Metropolitan University\\
  Toronto, Ontario, Canada \\
  \texttt{xzhang@ee.ryerson.ca} \\
  \And
 Jun Li \\
  Huawei Human Machine Interaction Lab\\
  Markham, Ontario, Canada \\
  \texttt{jun.li5@huawei.com} \\
  \And
 Qiang Xu \\
  Huawei Human Machine Interaction Lab\\
  Markham, Ontario, Canada \\
  \texttt{qiang.xu1@huawei.com} \\
}

\begin{document}
\maketitle
\begin{abstract}
Continuous monitoring of human vital signs using non-contact mmWave radars is attractive due to their ability to penetrate garments and operate under different lighting conditions. Unfortunately, most prior research requires subjects to stay at a fixed distance from radar sensors and to remain still during monitoring. These restrictions limit the applications of radar vital sign monitoring in real life scenarios. In this paper, we address these limitations and present Pi-ViMo, a non-contact \underline{P}hysiology-\underline{i}nspired Robust \underline{Vi}tal Sign \underline{Mo}nitoring system, using mmWave radars. We first derive a multi-scattering point model for the human body, and introduce a coherent combining of multiple scatterings to enhance the quality of estimated chest-wall movements. It enables vital sign estimations of subjects at any location in a radar's field of view (FoV). We then propose a template matching method to extract human vital signs by adopting physical models of respiration and cardiac activities. The proposed method is capable to separate respiration and heartbeat in the presence of micro-level random body movements (RBM) when a subject is at any location within the field of view of a radar. Experiments in a radar testbed show average respiration rate errors of $6\%$ and heart rate errors of $11.9\%$ for the stationary subjects, and average errors of $13.5\%$ for respiration rate and $13.6\%$ for heart rate for subjects under different RBMs. 
\end{abstract}


\section{Introduction}
Non-contact sensing for vital signs and human activities has gained a lot of attention in  recent years. Compared to vision-based solutions, RF sensing is attractive in its ability to penetrate garments or walls, operate under different lighting and weather conditions, and better preserve people’s privacy. mmWave radars transmitting frequency modulated continuous wave (FMCW) signals have some unique advantages over narrow-band technologies such as WiFi and RFID for vital sign monitoring. The carrier frequency of mmWave radar signals lies in a wide frequency range from 30GHz to up to 300GHz, with wavelength on the order of millimeters, and bandwidth on the order of GHz. Such a wide bandwidth implies robustness against noise, interference and high range resolutions. Additionally, short pulse duration and multiple pulses in a short period time allow for the detection of subtle movements such as hand gestures.  

Since both respiration and heartbeats cause movements of human chest wall, a typical pipeline for mmWave radar-based vital sign monitoring consists of two stages: estimating chest-wall displacements and extracting vital signs. To estimate chest-wall displacements, one needs to identify one or a few range bins that contain the most informative signals, also called {\it range bin selection}. With few exceptions, many existing work treats a human chest as a point target and selects {\it a single} range bin containing vital signs based on criteria such as maximum magnitude~\cite{JM:2019} or phase changes~\cite{MostafaA:2019}, coherency between magnitude and phase fluctuation~\cite{Ho-IkC:2020}. These methods suffer from several drawbacks. First, a range bin selected by these criteria may not correspond a fixed chest position over time. Second, a fixed position on a chest wall may fall into different range bins at different time. Third, magnitude or phase fluctuations of the reflected FMCW signals may be dominated by one's body movements. 

\begin{figure}[th]
    \centering
    \begin{subfigure}[b]{0.49\columnwidth}
        \centering
        \includegraphics[width=\linewidth]{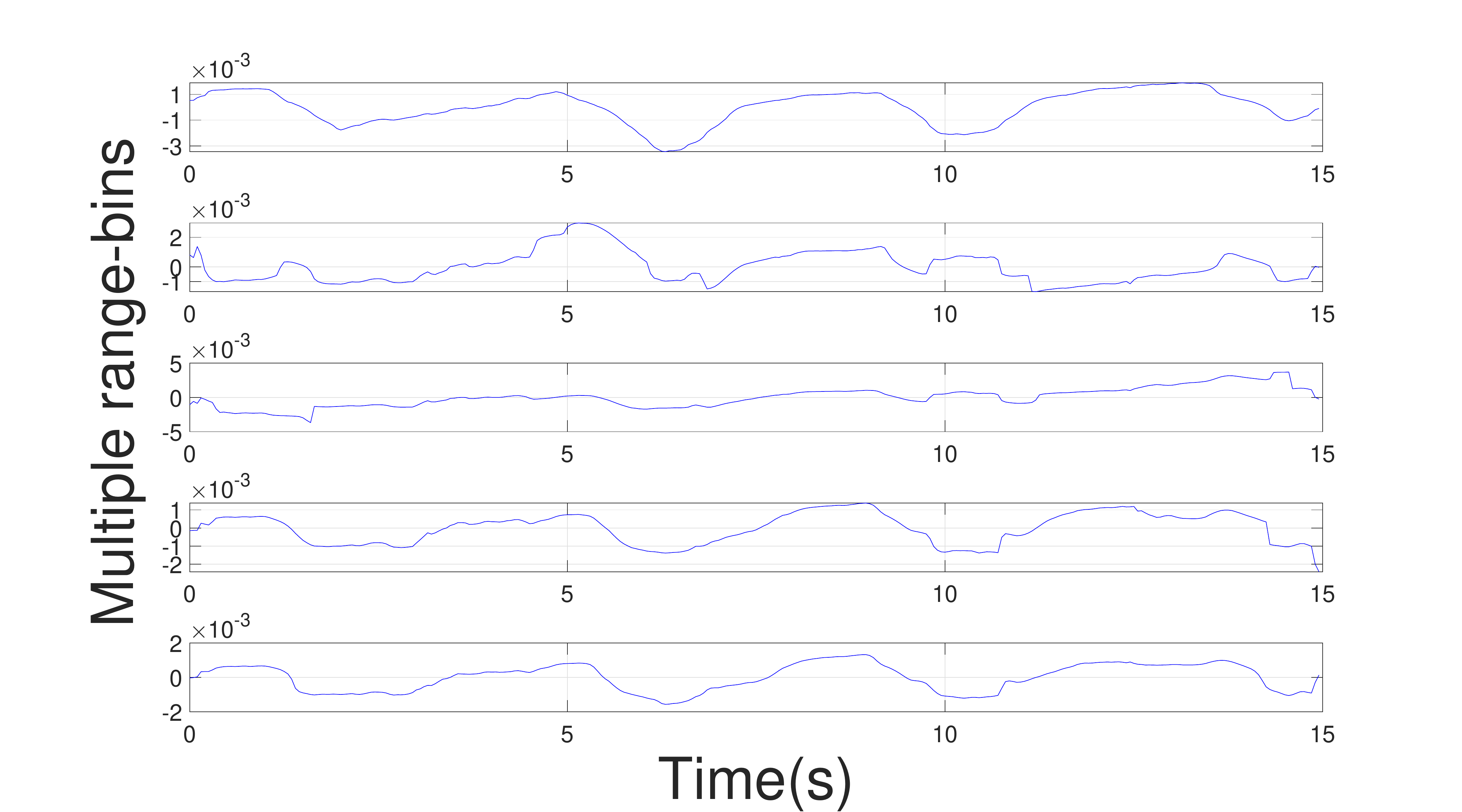}
        \caption{Sitting still}
        \label{fig:example_a}
    \end{subfigure}
    \begin{subfigure}[b]{0.49\columnwidth}
        \centering
        \includegraphics[width=\linewidth]{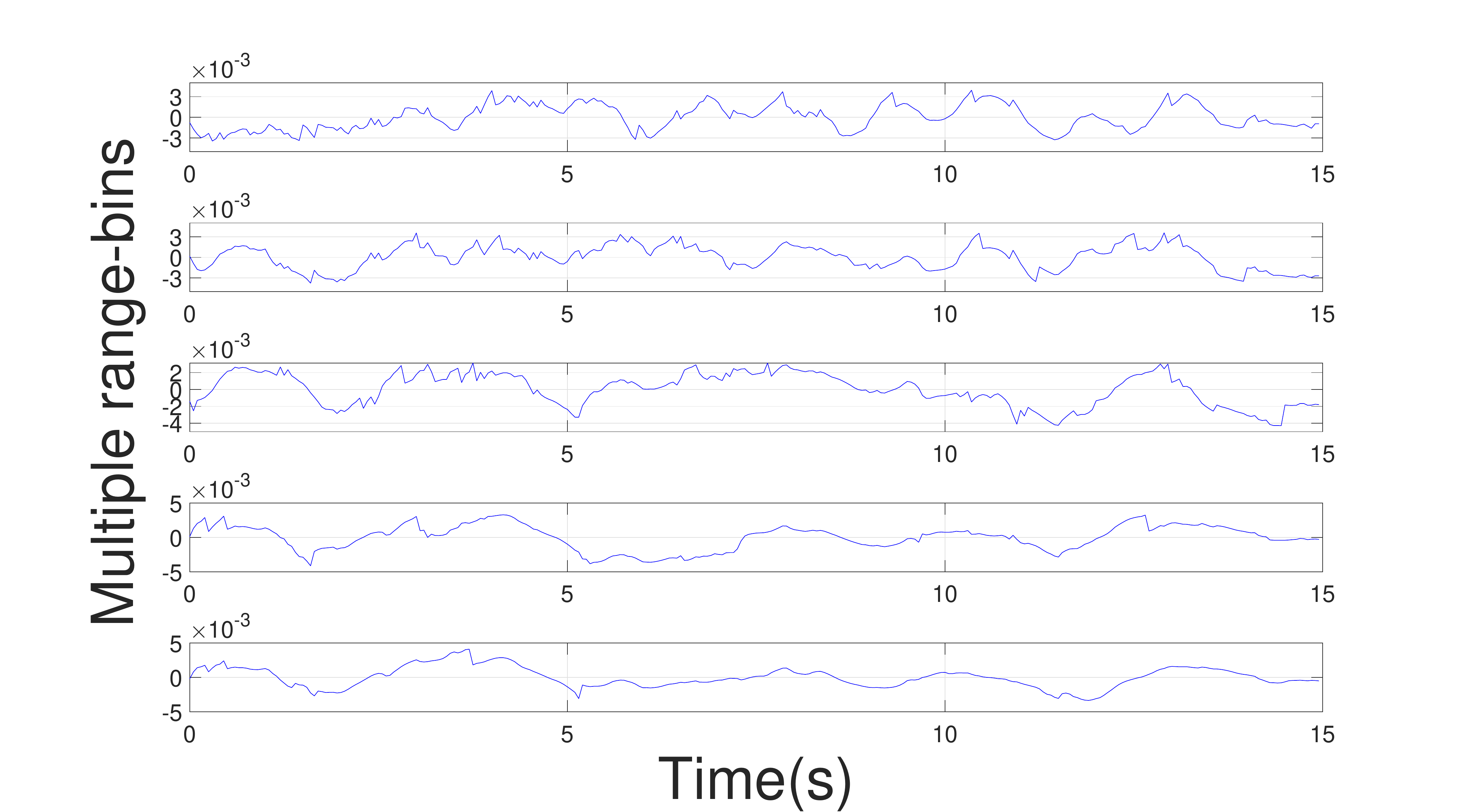}
        \caption{Standing with body sway}
  \label{fig:example_b}

    \end{subfigure}
    \caption{Chest-wall displacement of multiple range bins at a short sensing distance.}
\label{fig:examples1}

\end{figure}

Figure~\ref{fig:examples1} provides an illustrative example of the displacement measurements extracted from FMCW signals from different range bins. 
In Figure~\ref{fig:example_a}, a subject sits still with her back against the back of chair at 0.3 meters distance from a radar. In Figure~\ref{fig:example_b}, the same subject stands with slight body sways. $5$ neighboring range bins are shown in figures. All range bins contain similar amounts of signal variations. Single range bin selection may wrongly select the second or the third range bin with severe distortion in Figure~\ref{fig:example_a}. In contrast, the first, fourth, fifth range bins contain cleaner waveforms, all being distorted replicas of the  same vital sign signals, which can be used for better chest-wall estimations.
Similar observations can be made from  Figure~\ref{fig:example_b}. In this case, even subtle movements such as unconsciously body sways can lead to non-negligible distortions to the displacement measurements. Therefore, utilizing a single range bin to estimate chest wall displacement represents both a source of error and a missed opportunity -- combining noisy signals from multiple range bins can lead to more robust estimations. In fact, the point target assumption holds only when subjects are in the far field of a radar. 

To extract vital signs, one needs to separate displacement caused by heartbeats from respiration, which is an order of magnitude larger, as well as from physical movements of the body. Though typically at different frequencies, heartbeat signals can be easily buried in the harmonics of respiration signals in frequency domain. Large movements such as walking and running pose significant challenges to non-contact vital sign monitoring. Even conscious or unconscious small movements such as body sway, head turning, leg shaking during sitting or standing (called {\it micro-level random body movements (micro-RBM)}) can introduce large noise in displacement measurements as evident from Figure~\ref{fig:example_b}.  Methods based on STFT and Wavelet transform fail to work robustly since they assume clear separation of heart beat, respiration and mRBM signals in frequency and/or time domains~\cite{Liang:2018,He:2017}. Signal decomposition methods such as EMD, VMD and their variations~\cite{NordenEH:1998,KonstantinD:2013,MariaET:2011,DmytroI:2015} decompose a composite signal into a collection of intrinsic mode functions (IMF), and have been adopted in vital sign extraction in \cite{ArindamR:2021,ZhenyuL:2020,FYWang:2022,TianyueZ:2020}. However, they assume the IMFs are narrow-band signals and automatic selecting proper modes that correspond to vital signals is in itself a non-trivial problem. 
 
\begin{figure}[tb]
  \centering
  \includegraphics[width=\linewidth]{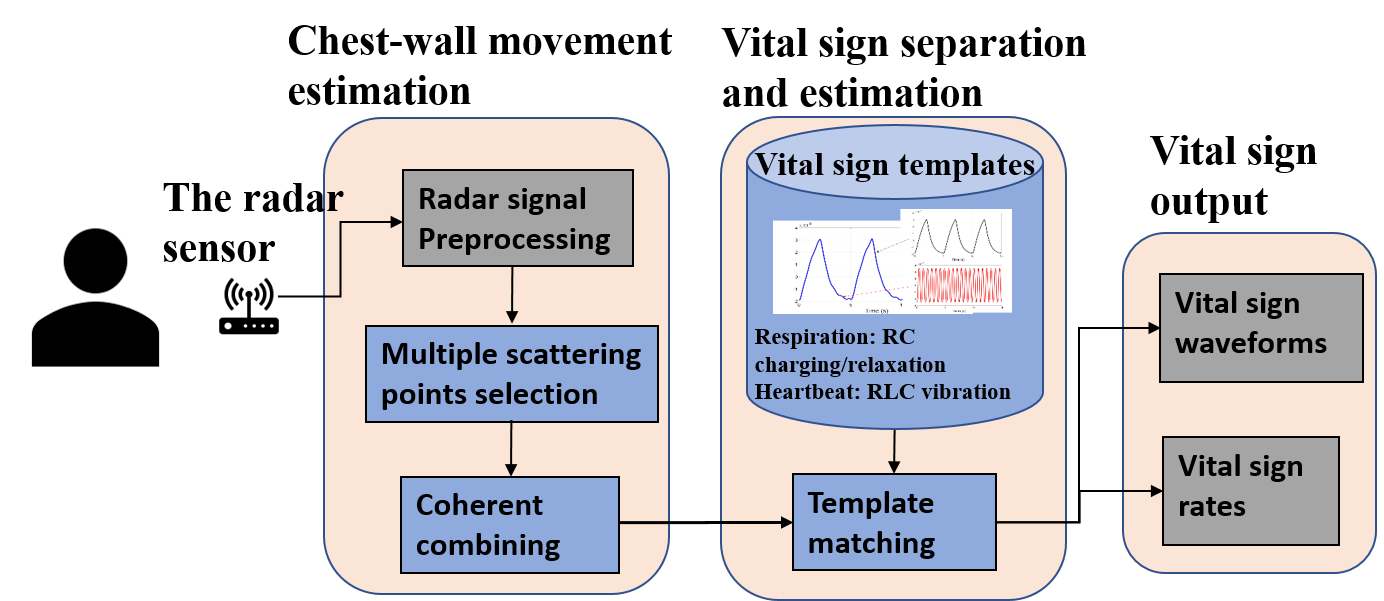}
  \caption{The system diagram of Pi-ViMo. New processing blocks introduced in this paper are highlighted in blue boxes.}
  \label{fig:sys1}
\end{figure}

In this paper, we present Pi-ViMo, a novel \underline{P}hysiology-\underline{i}nspired Robust \underline{Vi}tal Sign \underline{Mo}nitoring system, using mmWave radars (Figure~\ref{fig:sys1}). Pi-ViMo is driven by the understanding of the interplay between the geometric shape of human chests and radar signals, and characteristics of the bio-mechanical processes behind respiration and cardiac activities. To estimate chest-wall movements, we first introduce a multi-scattering point (MSP) model of back-scattering RF signals from human body. A coherent combining (CC) scheme of signals from multiple range bins is proposed to enhance the accuracy and robustness of displacement estimations regardless of the physical distance between monitored subjects and the radar. To combat micro-RBM, we adopt physical models of respiration and heartbeat from prior physiology research, and propose a template matching (TM) optimization method to extract vital signs. Compared to data-driven methods for signal composition in \cite{ZheC:2021,TianyueZ:2021}, Pi-ViMo is a white-box approach incorporating domain-specific knowledge, and thus has better interpretability and generalizability. 

We have implemented a prototype of Pi-ViMo using a Texas Instrument IWR6843ISK mmWave radar. 11 subjects with different genders, ages and body mass indices (BMI) have been recruited to collect data from two environments, a large lab space and a lounge room with many furniture. During the data collection, subjects are asked to perform different micro-RBMs while sitting and standing. Pi-ViMo achieves average errors of  6\% for respiration rate and 11\% for heart rate estimates over all locations among stationary subjects; and average errors of 12\% for respiration rate and 13\% for heart rate estimates for subjects under different micro-RBMs.
It consistently outperforms two state-of-the-art (SOTA) baseline methods at all distances and under micro-RBMs.

In summary, we make the following new contributions in this work and bring mmWave-based vital sign monitoring one step closer to real-world adoption and deployment.  
\begin{itemize}
\item Pi-ViMo is the first work, to the best of our knowledge, that models a human chest as multi-scattering points for chest-wall displacement estimations.
\item Coherent combining of reflective signals from multi-scattering points enables Pi-ViMo to monitor subjects at arbitrary locations in a radar's FOV. 
\item Pi-ViMo incorporates physical models of respiration and cardiac activities and introduces a novel template matching based signal decomposition method. 
\end{itemize}

The rest of the paper is organized as follows. In Section \ref{sec:SystemModel}, we introduce the MSP model and propose the coherent combining algorithm. In Section \ref{sec:Modelsandtemplates}, we introduce the physical models of chest-wall movements due to both respiration and heartbeat and provide detailed description of the template matching optimization method. Section \ref{sec:Experiments} provides experiments setups and detailed results. A review of recent development of vital sign monitoring using modern radar sensors is presented in Section \ref{sec:Relatedwork}. Section \ref{sec:Discussion} discusses the scope and limitations in this work. Finally, we conclude this paper and point out future directions in Section \ref{sec:Conclusions}.

\section{Chest-wall motion estimation with Multi-scattering Point Model}
\label{sec:SystemModel}
In this section, we first introduce radar signal fundamentals and the preprocessing steps of radar data. Next, we formally define the MSP model. Finally, inspired by the measurements from a motion capture system, we present the proposed coherent combining algorithm.

\subsection{RF Fundamentals and Preprocessing}
In FMCW radars, a transmitted chirp signal at time $t$ is given by
\begin{eqnarray}
s(t)=A_{t}e^{j(2 \pi {f}_{min}t+\pi St^{2})}, \quad 0\leq t\leq T_s, \label{eq:chirpsig}
\end{eqnarray}
where $A_{t}$ is the complex amplitude, $f_{min}$ is the start frequency, $T_s$ is the chirp signal duration, and $S = (f_{max}-f_{min})/T_s$ is the slope of linear chirps. 

From a single scattering point target in far field, the reflected signal at the receiver with a delay $t_d(t)$ is given by,
\begin{align}
r(t) &= A^{'}_{r}s(t-t_d(t)) \nonumber \\
&=A_{t}A_{r}(t)e^{j(2 \pi {f}_{min}(t-t_d(t))+\pi S(t-t_d(t))^{2})}, \quad t_d(t)\leq t\leq T_s, 
\label{eq:receivedsig}
\end{align}
where $A_{r}(t)$ is the received complex amplitude, $t_d(t) = 2R(t)/c$ is the reflected time delay, $R(t)$ is the range of the point target that reflects transmitted signals, and $c = 3\times 10^8 m/s$ is the speed of light.

After a signal mixer (matched filter), we have an output IF signal as
\begin{align}
y(t) = A_{t}A_{r}(t)e^{j( 4 \pi \frac{R(t)}{\lambda_{max}} + 4 \pi S\frac{R(t)}{c} t - 4\pi S\frac{R^2(t)}{c^2})}, \quad t_d(t)\leq t\leq T_s,
\label{eq:IFsig}
\end{align}
where $\lambda_{max} = c/{f}_{min}$.
There are three terms in the phase of the IF signal. For a TI IWR6843ISK radar,  $B=f_{max}-f_{min}\approx 4GHz$, $T_s \approx 60 \mu s$, $S \approx 2/3 \times 10^{14}$, $\lambda_{max} \approx 5\times 10^{-3}$, and $R(t)\approx 2m$ . In this setting, the first term of phase is around $5\times 10^3$, the second term is around $0.8 rad$ per ADC sample duration, and the last term is around $4\times 10^{-2}$, which is negligible. Furthermore, $R(t)=R_0+x(t)$, where $R_0$ stands for the time-invariant distance between human chest-wall area and the radar sensor, and $x(t)$ corresponds to time-variant vital sign movements. Since $x(t)$ barely changes in one chirp duration, it is safe to treat the first term as a constant within $T_s$.

A typical respiration cycle lasts 4 to 5 seconds. Thus, an observation window lasting multiple cycles is needed to estimate respiration rates. In the window, $M$ chirps are transmitted. With $K$ samples per chirp, the discrete received signals can be expressed as
\begin{align}
y(t_k,t_m) = A_{t}A_{r}(t_m)e^{ j4 \pi \frac{R(t_m)}{\lambda_{max}}}e^{ j4 \pi S\frac{R(t_m)}{c} t_k}, 
\quad 1\leq k\leq K, \quad 1\leq m\leq M,
\end{align}
where, index $k$ stands for samples within a chirp signal (fast time dimension) and index $m$ stands for chirps sent over time (slow time dimension). The fast time samples are used to map the target and background environment into a range map (or a range-angle map with multiple antennas) while the slow time samples are used to track the dynamic movements of the target and background environment. Actually, $A_{r}(t_m)$ is also time varying due to target movements $R(t_m)$. For vital sign monitoring, it is clear that the received IF signal behaves as an amplitude-modulated-phase-modulated (AM-PM) signal modulated by chest-wall movements. Since amplitudes generally suffer more distortion (noise) than  phases, phase information is commonly used in displacement estimation. The second phase terms $e^{ j4 \pi S\frac{R(t_m)}{c} t_k}$ translates frequencies into a range map; while the first phase term is used to estimate the change in $R(t_m)$. Specifically, the estimation of $R(t_m)$ depends on the ratio $\frac{R(t_m)}{\lambda_{max}}$, where $\lambda_{max}$ is around 5 millimeters. Therefore, the sensitivity of the estimation of $R(t_m)$ could achieve sub-millimeter level. 

It is straight-forward to extend the system model to reflections from $N$ scattering points,
\begin{align}
 y(t_k,t_m) \approx  \sum_{n=1}^{N} A_{t,n}A_{r,n}(t_m)e^{ j4 \pi \frac{R_n(t_m)}{\lambda_{max}}}e^{ j4 \pi S\frac{R_n(t_m)}{c} t_k}, 
 \quad 1\leq k\leq K, 1\leq m\leq M,
\label{eq:multiReflection}
\end{align}
where we assume the scattering points are at different ranges($>R_{res} = c/2B$). Reflections for static environment (e.g. furniture, wall, desk, etc.) are time invariant. Techniques such as DC compensation and clutter removal can be applied to remove the static background before applying range-FFT to fast-time samples. Range-FFT behaves like matched filters to output spectral peaks at $f_n$, which have a one-to-one mapping to the range $R_n$. At the end of the preprocessing, a ``range map'' (range-FFT along slow-time) is calculated for further processing.

\subsection{Multi-scattering Point Model for Human Chests}
\subsubsection{Geometric Interpretations}
\begin{figure}[th]
    \centering
    \begin{subfigure}[b]{0.49\columnwidth}
        \centering
        \includegraphics[width=\textwidth]{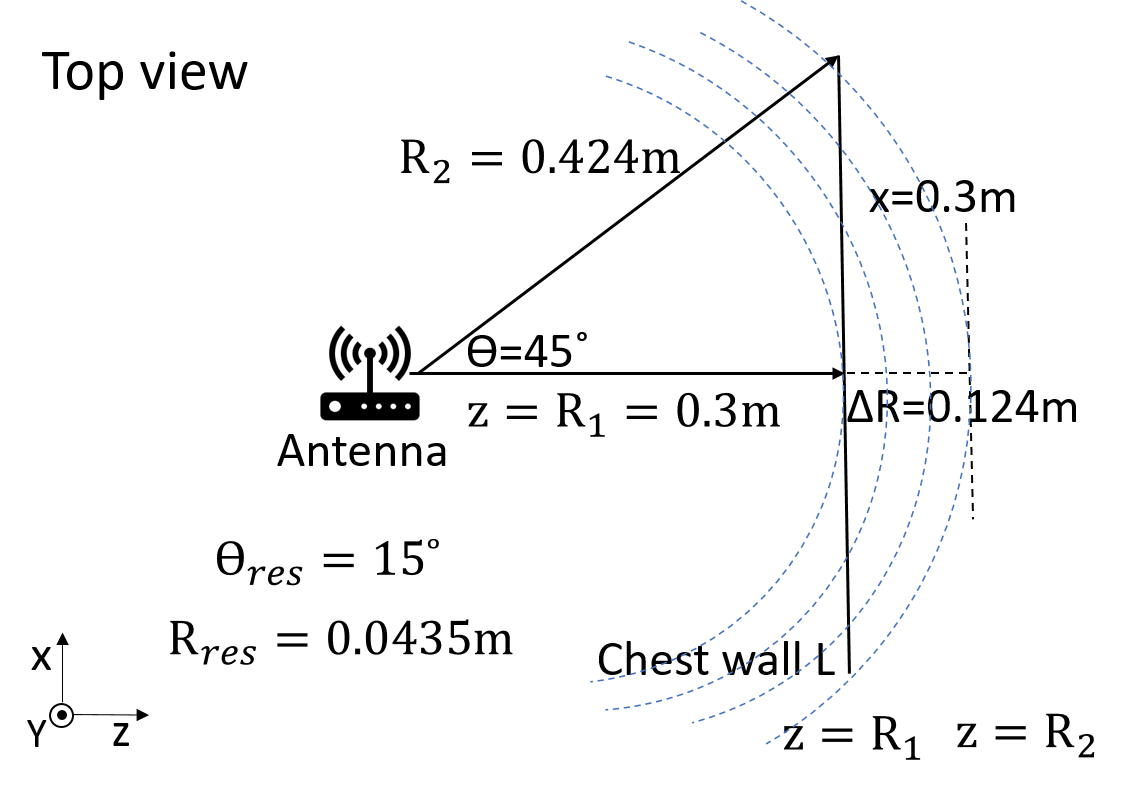}
        \caption{Sensing at range $0.3m$.}
    \end{subfigure}
    \begin{subfigure}[b]{0.49\columnwidth}
        \centering
        \includegraphics[width=\textwidth]{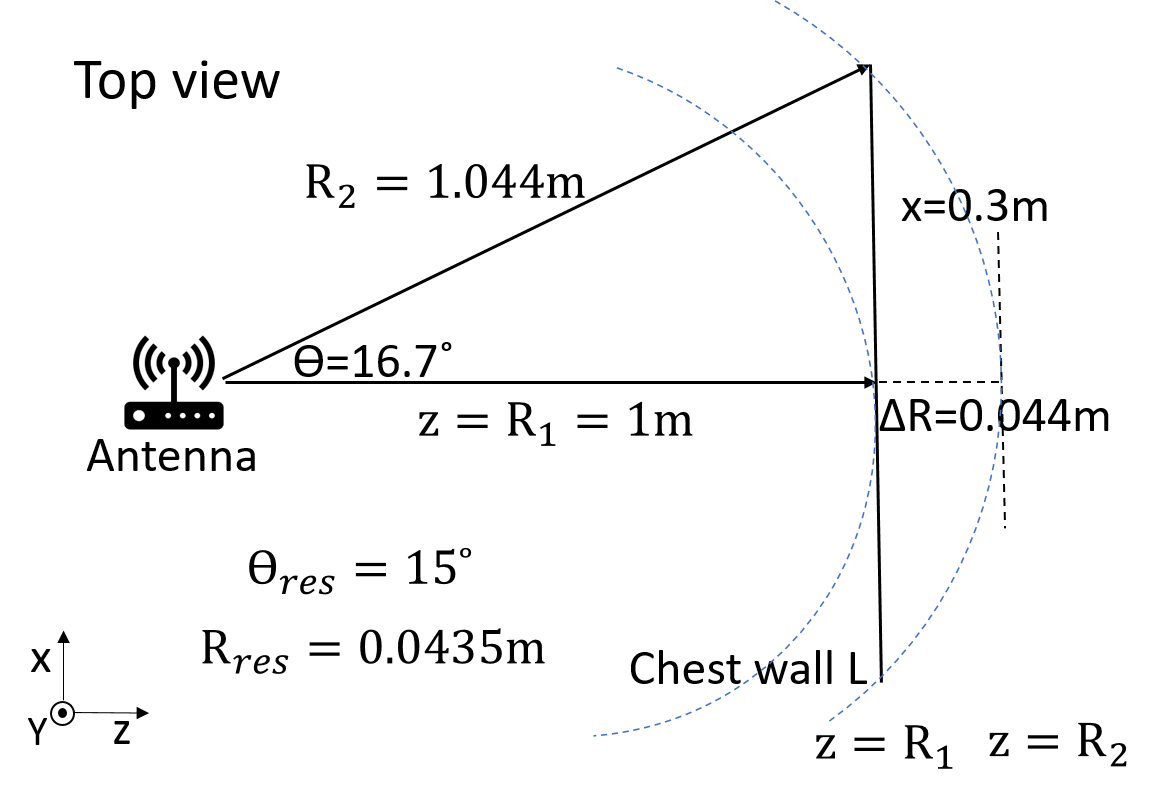}
        \caption{Sensing at range $1m$.}
    \end{subfigure}
\caption{Geometric interpretation: sensing in a close range.}
\label{fig:geo1}
\end{figure}
The average width of adult torsos is 60cm. When radars are placed at a close range, human chests cannot be approximated as point targets. 
Figure~\ref{fig:geo1}a illustrates the top-down view of a scenario where a radar is  at $0.3$ distance from a subject facing the radar. The center of her chest-wall lies along the middle line from the radar sensor. Blue dot lines represent spherical radiation wave fronts. It is easy to show that the boundary of the chest wall is at distance $R_2 = 0.42m$ away from the radar. Therefore, the human chest-wall spans $\Delta R = 0.124m$ in range and $90^{\circ}$ in azimuth angle. Considering a typical hardware resolution setting ($\theta_{res} = 15^{\circ}$, and $R_{res} = 0.0435m$), the human chest-wall spans roughly three range bins and six angle-bins in a range-angle radar map. In fact, the wave front lines partition the chest-wall area into multiple sections, each falling into one range bin.
 
In Figure~\ref{fig:geo1}b, a radar is at a distance of $1$ meters from a subject's chest wall. By the same argument, we find that her human chest-wall occupies $0.044m$ in range and $33^{\circ}$ in angle, or equivalently, one range bin and two angle-bins in a range-angle radar map. Similarly, $\Delta R = 0.022m$, $\theta = 6.3^{\circ}$ at a sensing distance of $2m$. The chest-wall is contained in a single range bin and a single angle-bin. Beyond $2m$, received signals contain reflections from surrounding stationary or moving objects. This may explain most existing work reports results in the range of $1 - 2$ meters. 
  
\subsubsection{MSP Model}
With the insights from the geometric interpretation, we now derive the multi-scattering point Model. 
The received signal from a single point at distance $R(t)$ can be re-writte  as:
\begin{align*}
y(R,t) = \frac{\rho(R(t))}{R(t)^4} e^{ j4 \pi \frac{R(t)}{\lambda_{max}}}e^{ j4 \pi S\frac{R(t)}{c} t}, 
\end{align*} 
where $\rho(R)$, is the reflectivity of a human chest-wall surface; $R$ stands for the range from the chest-wall to a radar sensor, which is a function of $t$. Here we do not differentiate fast time and slow time. 
 The typical value of $\rho(R)$ of human skins is around $0.6 \sim 0.8$~\cite{ZhichengY:2016,AmaniYousefO:2020}. We assume that it is a constant for the whole chest-wall area and drop its dependency on $R$.
To further simplify the notations, we use wave number $k$ in the signal expression. Let $k=\frac{2\pi f}{c}$, where $f=f_{min} + St$. Thus, $k=2\pi \frac{1}{\lambda_{max}} + 2\pi S\frac{t}{c}$. Using wave number $k$, and drop the time indexes for convenience, we have
\begin{align*}
y(R,t) = \frac{\rho}{R^4} e^{j2 \pi kR}.
\end{align*} 
  
Consider a 3-D top view geometric as shown in Figure~\ref{fig:geo1}a. Denote the  chest-wall surface $\mathcal{S}$ by $z(x,y)$, where the range $R^2 = x^2+y^2+z^2$. Note that $z(x,y)$ can be any continuous and differentiable function over $x, y$.  The received signal reflected from the chest wall surface is given by, 
\begin{align}
y(k) &= \iint_{\mathcal{S}} f(x,y,z) d_\mathcal{S} = \iint_{\mathcal{T}} f(x,y,z) \sqrt{1+(\frac{\partial z}{\partial x})^2+(\frac{\partial z}{\partial y})^2} d_xd_y, \nonumber 
\end{align}
where $f(x,y,z) = \rho \frac{exp(j2kR)}{R^4} = \rho \frac{exp(j2k\sqrt{x^2+y^2+z^2})}{(x^2+y^2+z^2)^2}$, and $\mathcal{T}$ stands for the region of a projection from $\mathcal{S}$ onto the $X-Y$ plane. 

If we divide $\mathcal{T}$ into $I$ areas of width and height roughly equal to the range resolution, the corresponding patches can be approximated as a flat surface with $z=z_i$'s, a constant independent of $x$ and $y$ within each patch.
By Mean Value Theorem (MVT), for patch $i$, there exists a point $(x_i,y_i) \in \mathcal{T}$ and $(x_i,y_i,z_i) \in \mathcal{S}$, and with its range $R_i^2 = x_i^2+y_i^2+z_i^2$, such that
\begin{align}
y_i(k) &\approx \rho \frac{exp(j2k\sqrt{x_i^2+y_i^2+z_i^2})}{(x_i^2+y_i^2+z_i^2)^2}\Delta \mathcal{S}_i, \nonumber \\
&= \alpha'_i exp(j2k\sqrt{R_i^2}) = \alpha'_i exp(j2kR_i). \nonumber
\end{align} 

Plugging in wave-number $k$ and slow-time/fast-time sampling indices, we have the corresponding MSP model as
\begin{align}
y(t_k,t_m) = \sum_{i=1}^{I} \alpha'_i e^{j(4 \pi \frac{R_i(t_m)}{\lambda_{max}} + 4 \pi S\frac{R_i(t_m)}{c} t_k )},
\quad 1\leq k\leq K,  1\leq m\leq M.
\label{eq:multiScatter}
\end{align}

For the derivation, we approximate measurements in the same range bin as from the same point target, and $I$ is the number of range bins that a human body occupies. Intuitively, the MSP model is a more accurate approximation of real geometry than the single point model, especially when sensing distances are short and thus targets can not be treated as a single point. In the case that the sensing distance is large and one's chest area can be approximated as a single-point target, MSP also has some advantages as will be discussed in Section~\ref{sec:discussion}. 

\subsubsection{ MSP selection}
\label{sec:MSPselection}
MSP selection aims to select multiple range bins on the radar range map that contain relevant vital sign information. The selection criteria is associated with the signal quality in each range bin, instead of energy level alone. We propose a two-step process to choose range bins containing the chest-wall area with non-negligible vital sign signals.   

The first step is to select range bins that a human body occupies. The conventional constant false alarm rate (CFAR) algorithm works on a range map and outputs local maximums with energy higher than an adaptive threshold. Each local maximum corresponds to one single range bin. Two problems arise in CFAR when a subject occupies multiple contiguous range bins. First, it fails to include useful signals from neighboring range bins. Second, it may misclassify a single subject as multiple ones if the energy responses in the range map contain saddle points. To deal with these issues, we modify CFAR by taking into account of the energy in surrounding range bins of the local maximums with energy large than an adaptive threshold. The threshold is set to be the sum of the average energy of the range map and $\alpha$ ratio of the maximum value. The ratio is a tunable parameter that trades off useful signals captured and the amount of interference, and is set to $20\%$ in the experiments. The resulting selection method is capable of separating different targets and capturing multiple scatterings from a human body. The selected range bins are stored in a set, named \emph{candidate bins}.

The second step is to select suitable range bins from the \emph{candidate bins} set. As shown in Figure~\ref{fig:examples1}, there are range bins containing high-quality vital sign signals and those containing mainly interference and noise. Instead of the maximum energy, we utilize frequency domain signal properties to select range bins corresponding to human chest-wall area that contain significant vita sign signals. The algorithm is summarized as Algorithm~\ref{alg:MSPbins}. In Algorithm~\ref{alg:MSPbins}, we use the peak frequency and the ratio of in-band energy and out-band energy. Specifically, in-band energy is calculated by accumulating frequency responses within a band of $[0.1, 0.8]Hz$ for respiration, and a band of $[0.8, 2]Hz$ for heartbeat. Out-band energy is calculated by accumulating frequency responses outside of these two frequency bands. The existence of respiration signal is determined by whether the peak fall inside the respiration band, and whether the thresholds $Th_{r}$ is surpassed by the ratio of in-band energy and out-band energy. If respiration signal is not detected, the existence of heartbeat signal is determined in a similar way. When either of respiration or heartbeat is detected, the range bin corresponding to the maximum energy will be selected as one scatter point in the MSP model. The respiration and heartbeat frequency bands above are chosen by considering most references and ground truth data from our experiments. The thresholds $Th_{r}$ and $Th_{h}$ are empirically set to be $5$. 

The selected range bins are labeled as \emph{MSP bins} and incorporated in the MSP model in Equation~\ref{eq:multiScatter}. The number of \emph{MSP bins} selected is not deterministic and is adaptive over different slow-time windows, individuals, sensing distances and angles.

\begin{algorithm}
	\caption{MSP range bin selection} \label{alg:MSPbins}
	\begin{algorithmic}[1]
		\For {$rangebin \in \{\text{range bins candidates}\}$}
		    \State extract unwrapped phase and remove mean
		    \State convert to frequency domain by FFT
		    \State Respiration detection:
		    \State $E_{in} \coloneqq$ signal energy within [0.1, 0.8]Hz
		    \State $E_{out} \coloneqq$ signal energy outside of [0.1, 2]Hz
		    \If{$f_{peak} \in [0.1, 0.8]Hz$  $\&$ $E_{in}/E_{out} >= Th_{r}$}
		        \State append this range bin to \{MSP bins\}
		        \Else
		        \State remove frequency band below 0.8Hz
		        \State Heartbeat detection:
		        \State $E_{in} \coloneqq$ signal energy within [0.8, 2]Hz
		        \If{$f_{peak} \in [0.8, 2]Hz$ $\&$ $E_{in}/E_{out} >= Th_{h}$}
		        \State append this range bin to \{MSP bins\}
		        \EndIf
		    \EndIf
		\EndFor
        \If{\{MSP bins\} = $\emptyset$}
            \State add range bin with the maximum energy to \{MSP bins\}
        \EndIf
	\end{algorithmic} 
\end{algorithm}

\subsection{Coherent Combining}
\subsubsection{Measurement study}
The MSP model approximate the received signal reflected from a human chest as the sum of signals from multiple range bins on the chest surface. Naturally, one  may ask how different areas of a chest surface move during respiration and cardiac activities. 

\begin{figure}[th]
    \centering
    \begin{subfigure}[b]{0.4\columnwidth}
        \centering
        \includegraphics[width=\textwidth]{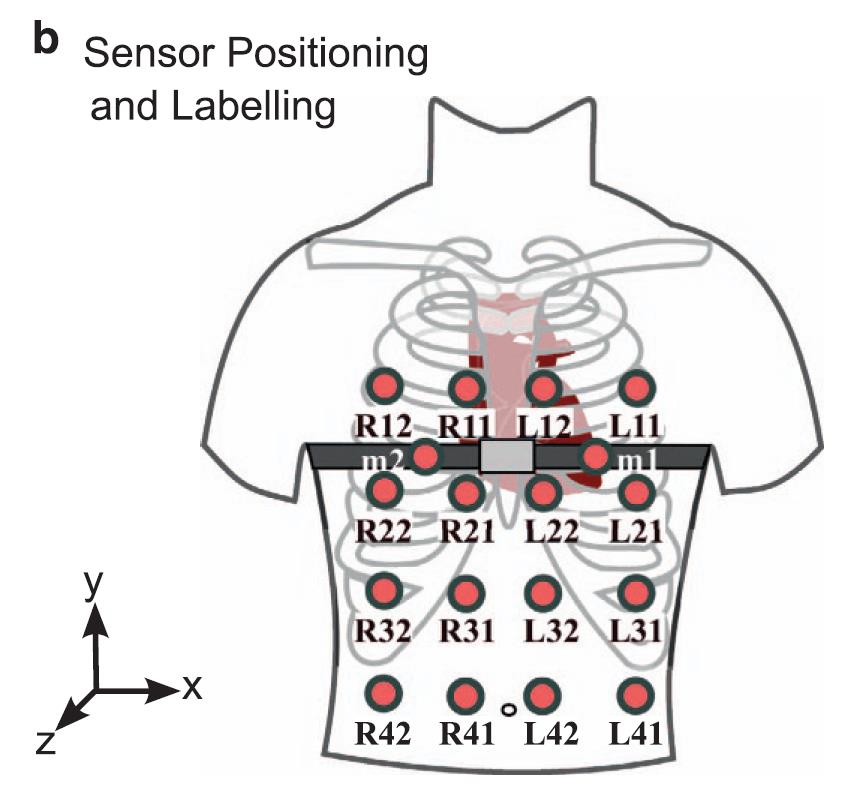}
        \caption{Marker placement.}
    \end{subfigure}
    \begin{subfigure}[b]{0.59\columnwidth}
        \centering
        \includegraphics[width=\textwidth]{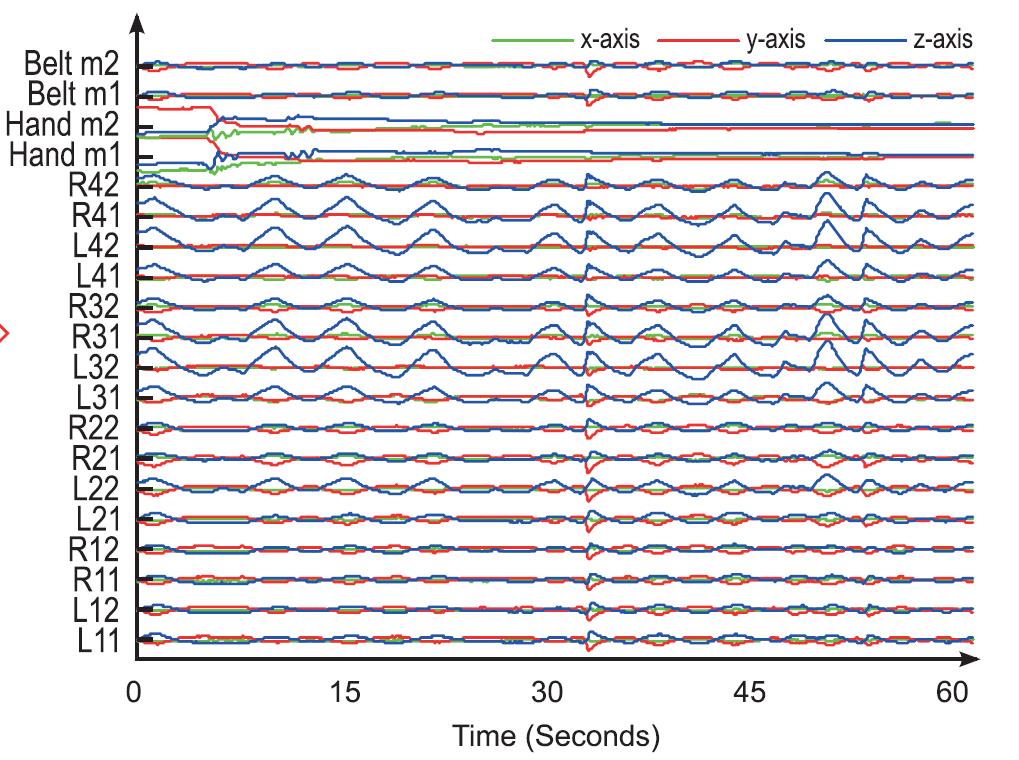}
        \caption{Displacements in a global frame}
    \end{subfigure}
\caption{Marker placement and 3D displacements~\cite{GS:2017}}
\label{fig:motioncapture}
\end{figure}

Fortunately, an open motion capture dataset for chest surface movements is available to answer this question~\cite{GSdata:2017}. In the dataset, where sixteen markers are placed on the surface of a human chest surface (Figure~\ref{fig:motioncapture}). A motion capture system is able to identify and track each marker with a displacement resolution of sub-millimeter level and time sampling frequency of 100Hz. 

The authors in~\cite{GS:2017} investigated the maximum phase coherence in a selected band (MPCB) by removing prior the drift and high-frequency noise. The MPCB between a respiration belt and all markers are highly correlated  with the coefficients higher than 0.8 in most cases. The results indicate that {\it the movements of different local areas on the chest-wall surface due to vital activities are highly in sync}. We further investigate the delays and the corresponding maximum correlations among the markers. Table~\ref{tab:table1} provides the correlations and delays among signals in the markers array.

\begin{table}[h]
\small
\begin{minipage}{\linewidth}
  \begin{center}
     \caption{Maximum correlation and delay of marker displacements}
     \label{tab:table1}
    \begin{subtable}[h]{0.49\linewidth}
    \centering
         \begin{tabular}{c c c c}
        \toprule
     	   \multicolumn{4}{c}{Correlation (Left to Right)} \\
     	\midrule
     		1.000   & 0.987  & 0.987 &  0.980\\
     		\hline
     		1.000   & 0.970  & 0.971   &0.986\\
     		\hline
     		1.000   & 0.850 & 0.835 &  0.982\\
     		\hline
     		1.000   & 0.970 & 0.964 &  0.995\\
        \toprule
 	   \multicolumn{4}{c}{Correlation (Top to Bottom)} \\
 	    \midrule
 		1.000   & 1.000  & 1.000 &  1.000\\
 		\hline
 		0.950   & 0.966  & 0.970   &0.957\\
 		\hline
 		0.901   & 0.856 & 0.879 &  0.933\\
 		\hline
 		0.873   & 0.806 & 0.829 &  0.896\\
 	    \bottomrule
 	    \end{tabular}
    \end{subtable}
    \begin{subtable}[h]{0.49\linewidth}
    \centering
        \begin{tabular}{c c c c}
        \toprule
     	   \multicolumn{4}{c}{Delay (ms) (Left to Right)} \\
     	\midrule
     		0   & -50  & -40 &  0 \\
     		\hline
     		0   & -70  & -70   & -10\\
     		\hline
     		0   & -50 & -70 &  -30\\
     		\hline
     		0   & -10 & -20 &  -10\\
        \toprule 
 	   \multicolumn{4}{c}{Delay (ms) (Top to Bottom)} \\
 	    \midrule
 		0   & 0  & 0 &  0 \\
 		\hline
 		10   & 0  & 0   & 0\\
 		\hline
 		110   & 100 & 80 &  70\\
 		\hline
 		160   & 200 & 170 &  150\\ 
        \bottomrule
 	    \end{tabular}
    \end{subtable}
  \end{center}
\end{minipage}
\end{table}

\begin{table}
\small
  \begin{center}
    \caption{Maximum correlation and delays of signals from seven  range bins w.r.t. to the signal from the reference range bin at sensing distance $0.3m$.}
    \label{tab:table2}
    \begin{tabular}{c |c c c c c c c}
    \toprule
 		Correlation & 0.661   & 0.978  & 0.860 &   0.851 & 0.737 & 0.707 & 0.704\\
 		\hline
 		Delay (ms) & 100   & 0  & 0   & 0 & -50 & 0 & 200\\
 	\bottomrule
    \end{tabular}
  \end{center}
\end{table}

We further examine an example, where the radar signals from seven neighbouring rang-bins on the chest surface around the sensing distance of $0.3m$.  Table~\ref{tab:table2} summarizes the respective correlations and delays. The results are consistent with the measurements from the motion capture system.

\subsubsection{Method}

After range FFT and range bin selection in Section~\ref{sec:MSPselection} , the received signal in $i^{th}$ bin from \emph{MSP bins} set is
\begin{align*}
y_i(t_m) = \alpha'_i\beta_i e^{j(4 \pi \frac{R_i(t_m)}{\lambda_{max}})},
\quad 1\leq i\leq I,  1\leq m\leq M,
\end{align*}
where $\beta_i$ is a complex weight introduced by range FFT. 

The measurement study inspires us to model the received signals in each range bin on the chest surface as the result of passing  the {\it same} source signal (a row vector $\mathbf{s}$) originated from one's respiration and cardiac activities through different channels (chest wall areas and over-the-air propagation).  Each channel has its own transfer function (due to thickness of the chest wall, reflectivity of the surface area and path length) and additive noise. 
Therefore, we can represent the received signal in $i^{th}$ range bin as:
\begin{align*}
\mathbf{y}_i = \mathbf{h_i}\mathbf{s} + \mathbf{n}_i,
\end{align*}  
and received signals from \emph{MSP bins} in a matrix form:
\begin{align}
\mathbf{Y} = \mathbf{h}\mathbf{s} + \mathbf{N},
\end{align}  
where $\mathbf{Y}$ is the signal matrix ($I\times M$) from the MSP model, with each row vector corresponding to the $i^{th}$ range bin, $\mathbf{s}$ is the signal vector with its phase containing vital activities, and $\mathbf{h} = [h_1,...,h_I]^T$ is the channel vector consisting of $I$ rows, and $h_i = \alpha'_i\beta_i = |h_i|e^{j\phi_i}$, where $|h_i|$ stands for the unknown fading amplitude of channel $i$, and $\phi_i$ stands for the phase delay of channel $i$. $\phi_i$ can be estimated using techniques such as cross-correlation. 

\begin{algorithm}
	\caption{Coherent Combining} \label{alg:cc}
	\begin{algorithmic}[1]
		\For {$i^{th}$ $rangebin \in \{\text{MSP bins}\}$}
		    \State calculate cross-correlations $corr_{i,j} = xcorr(y_i, y_j),$ for $j\neq i$ $\& j\in \{\text{MSP bins}\}$
		    \State sum up cross-correlations $corr_i = \sum_{j\neq i} corr_{i,j}$
		\EndFor
	\State choose the range bin with the max sum of cross-correlations $i^* = argmax(corr_i),$ for all $i$
    \State re-calculate cross-correlations and delays $corr_{i^*,j}, delay_j = xcorr(y_{i^*}, y_j),$ for $j\neq i^*$
	\State weighted sum according to $delay_j$ \Comment{Equation~(\ref{eq:cc})}
	\end{algorithmic} 
\end{algorithm}

Maximum ratio combining is a method of diversity combining in communication systems where the signals from different channel are added together coherently by taking the reciprocal of the channel matrix. It is optimal for additive white Gaussian noise channels when channel state information (CSI) is known perfectly, whether the channels are independent or not. In our scenario, CSI is only partially known. With estimated phase delays, we devise a coherent combining algorithm to take advantage of measurements in multiple range bins. The chest wall displacement is estimated as 
\begin{align}
\mathbf{\hat{s}} = \mathbf{e}^H\mathbf{Y}/I,
\label{eq:cc}
\end{align}
where $\mathbf{e} = [e^{j\phi_1},...,e^{j\phi_I}]^T$ is the phase delays vector. 
The SNR gain of coherent combining depends on the channel gain (power gain) of individual channels and their correlations (diversity gain). If channel responses follow independent, an identical distribution, equation~\eqref{eq:cc} gives the maximum diversity gain. 

For the resulting signal vector $\mathbf{\hat{s}}$, we unwrap the phase signal and remove mean along slow time. The obtained phase signals have a one-to-one mapping to the chest-wall displacement $R_i(t_m)$ in equation~\eqref{eq:multiScatter} at each time stamps. Finally, the estimated chest-wall displacement $\{ \mathbf{\hat{R}} | \hat{R}(t_m),  m=1,2,...,M\}$ is obtained.

\subsection{Discussion}
\label{sec:discussion}
\begin{figure}[th]
    \centering
    \begin{subfigure}[b]{0.49\columnwidth}
        \centering
        \includegraphics[width=\linewidth]{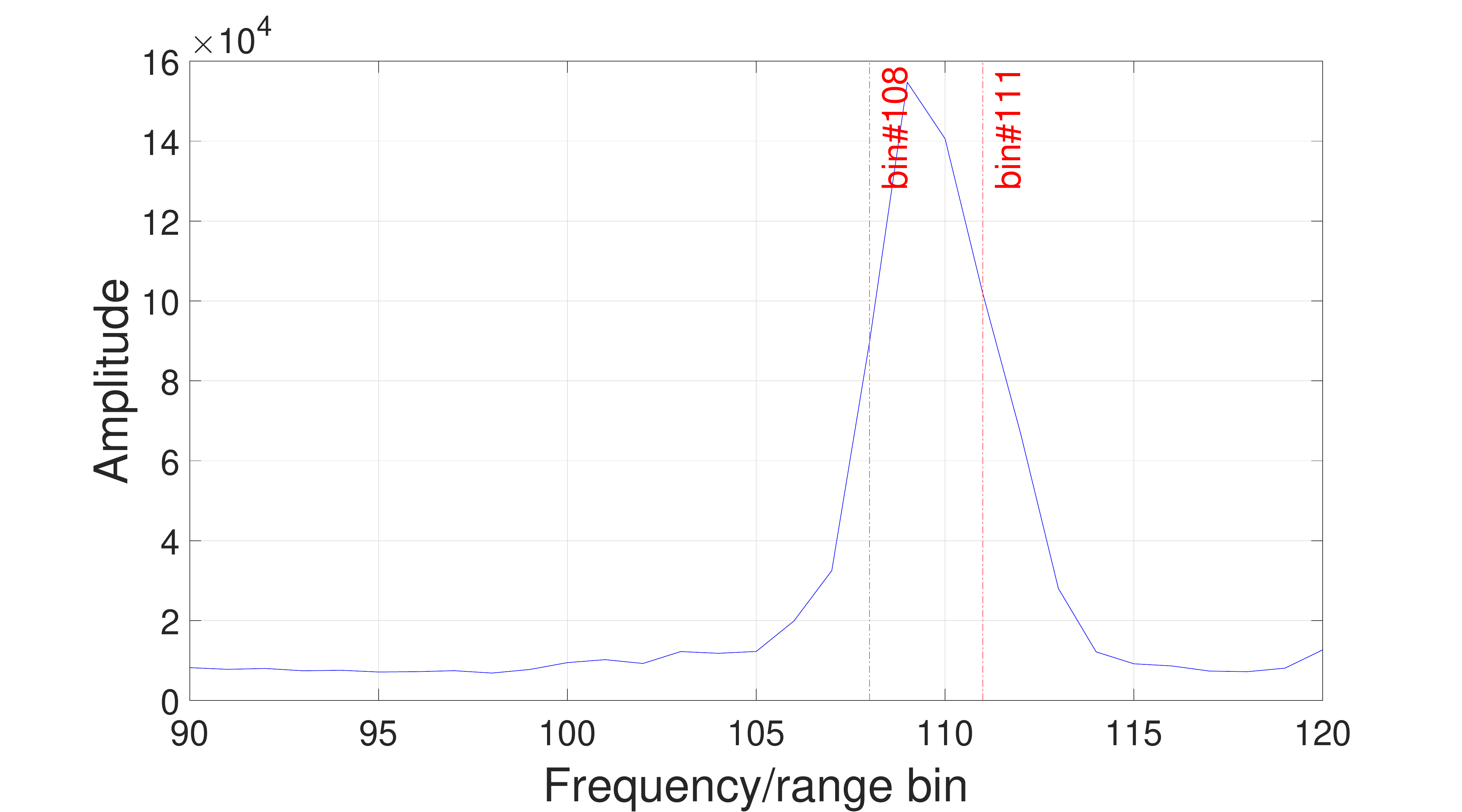}
        \caption{Frequency response over multiple range bins}
        \label{fig:example_5ma}
    \end{subfigure}
    \begin{subfigure}[b]{0.49\columnwidth}
        \centering
        \includegraphics[width=\linewidth]{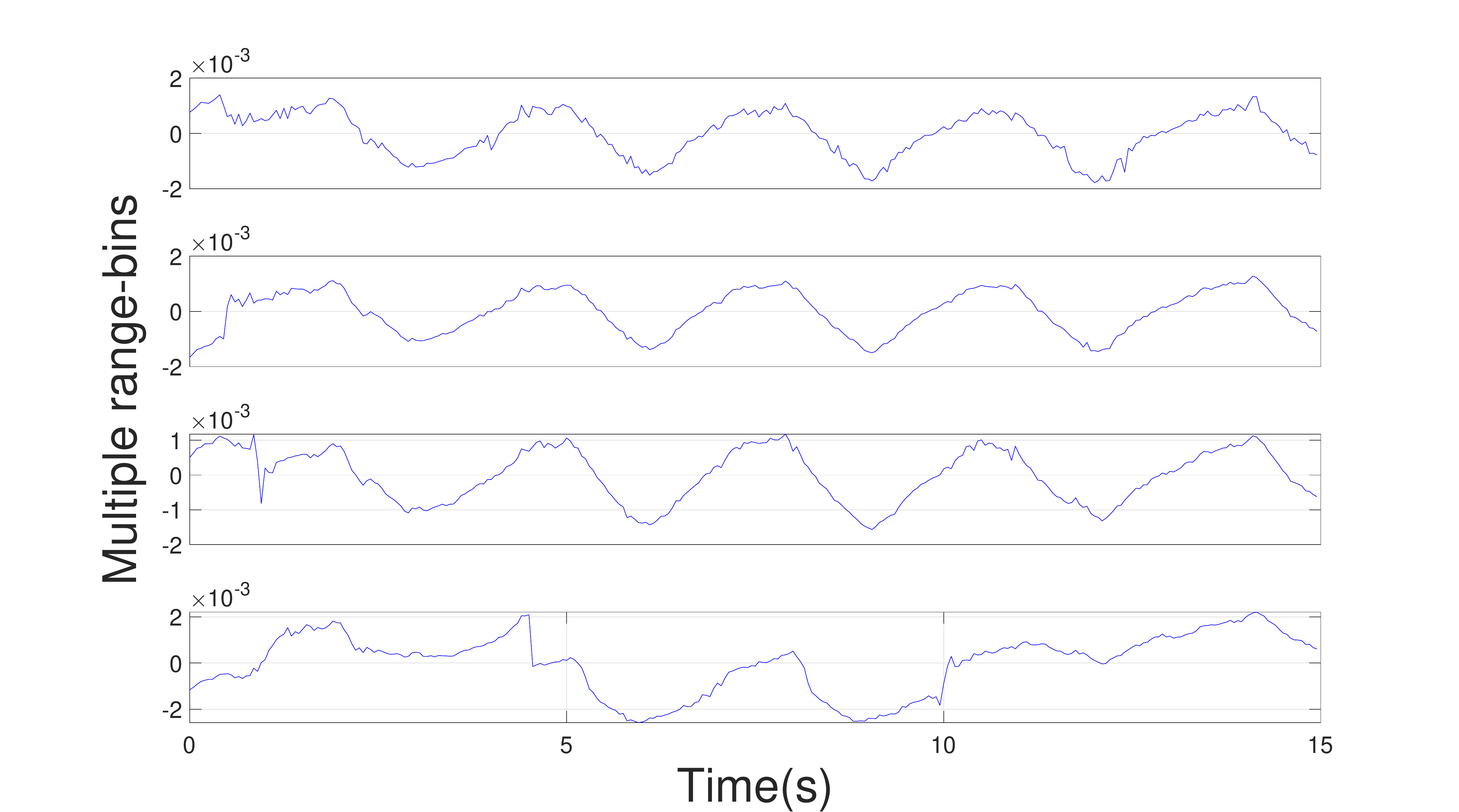}
        \caption{Chest wall movements (top 4 range bins)}
  \label{fig:example_5mb}
    \end{subfigure}
    \caption{Chest-wall displacement of multiple range bins at a long sensing distance.}
\label{fig:examples5m}
\end{figure}

The MSP model is motivated by the observation that at a close distance, one's chest area can occupy multiple range bins. However, it is still useful even when the subject is sufficiently far from the radar such that her chest area falls into a single range bin. This is because even for a single scatter point target, reflected energy can still be spread across multiple range bins. The reason is two-fold. First, due to motion, a subject may move from one range bin to another bin, a phenomenon called {\it range cell migration}. Second, finite number of sampling points windowing in discrete Fourier transformation (DFT) causes spectral leakage. As a result, it is beneficial to select and combine signals from multiple range bins. Figure~\ref{fig:examples5m} shows the frequency response when a subject sits still with her back against the back of chair at 5-meter distance from a radar. In Figure~\ref{fig:example_5ma}, The $3dB$-width of the main lobe roughly occupies four range bins, from range bin $\#108$ to range bin $\#111$. Figure~\ref{fig:example_5mb} plots the chest wall displacements from those range bins correspondingly. It is clear that the first three range bins contain vibration information from the same vital activity. Algorithm~\ref{alg:MSPbins} selects the first two range bins as multiple scattering points. The correlations with respect to the ground truth from NeuLog sensor are $[0.91, 0.96]$ for the selected two range bins, and the maximum delay span is $50ms$. In Section~\ref{sec:Experiments}, we provide further empirical evidence of the advantages of the MSP model when subjects are at different distances from a radar undergoing no or various RBMs.

\section{Physiological Models and Template Matching}
\label{sec:Modelsandtemplates}
Chest-wall displacement estimations from mmWave signals are inherently noisy and of low resolution (i.e., compared to those from optical motion capture systems). The presence of micro-RBMs further aggravates the problem. Our proposed solution to robust extraction of vital signals is motivated the key insight that though different from one subject to another, respiration and heartbeat waveforms are driven by bio-mechanical processes and thus share common characteristics across subjects. In this section, we first describe the mathematical models of respiratory and cardiac activities and derive templates for the two signal sources. Next, we present a novel template matching method for vital sign estimations.  

\subsection{Respiration Model}
\label{subsec:respirationmodel}
For respiration movements, an optimal chemical-mechanical respiratory control model is adopted in this work. The physical model is able to predict the ventilatory responses to chemical stimuli as well as muscular exercise, and simulate the instantaneous wave shapes the respiration behaviour. 
According to \cite{Shyan-LungL:2012} and the references therein, respiration behaviors can be modeled as a RC circuit governed by a differential equation as

\begin{align}
P(t) = \frac{\partial V(t)}{\partial t} R_{rs} + V(t)E_{rs},
\label{eq:Respiration Relaxation Oscillations}
\end{align}
where $R_{rs}$ represents the flow resistance and $E_{rs}$ represents volume elastic components. The inspiratory pressure is approximated by a quadratic function and the expiratory pressure is represented by an exponential discharging function of the form:
\begin{align}
P(t) =
\left\{
\begin{array}{ll}
      a_0 + a_1 t + a_2 t^2, & 0 \leq t \leq t_1, \\
      P(t_1) e^{\frac{t-t_1}{\tau}}, & t_1 \leq t \leq t_1+t_2. \\
\end{array} 
\right.
\label{eq:1}
\end{align}

The analytical solution for the lung volume can be obtained:
\begin{align}
\scriptstyle V(t) = 
\left\{
\begin{array}{ll}
\scriptstyle \frac{\tau_{RS}}{R_{rs}}[A_1t^2 + A_2t + A_3(1-e^{\frac{t}{\tau_{RS}}})] + V_0e^{\frac{t}{\tau_{RS}}}, & 
 0 \leq t \leq t_1, \\[1ex]
\scriptstyle \frac{P(t_1)}{R_{rs}(\frac{1}{\tau_{RS}}-\frac{1}{\tau})}[e^{\frac{t-t_1}{\tau}}-e^{\frac{t-t_1}{\tau_{RS}}}] + V(t_1)e^{\frac{t-t_1}{\tau_{RS}}}, & 
t_1 \leq t \leq t_1+t_2, \\
\end{array} 
\right.
\label{eq:2}
\end{align}
where $A_1 = a_2$, $A_2 = a_1 - 2a_2\tau_{RS}$, $A_3 = a_0 - a_1\tau_{RS} + 2a_2\tau^2_{RS}$, $\tau_{RS} = R_{rs}C_{rs}$, and $V_0 = V(t_0)=V(t_1+t_2)$.

\begin{figure}[th]
    \centering
    \begin{subfigure}[b]{0.49\columnwidth}
        \centering
        \includegraphics[width=\textwidth]{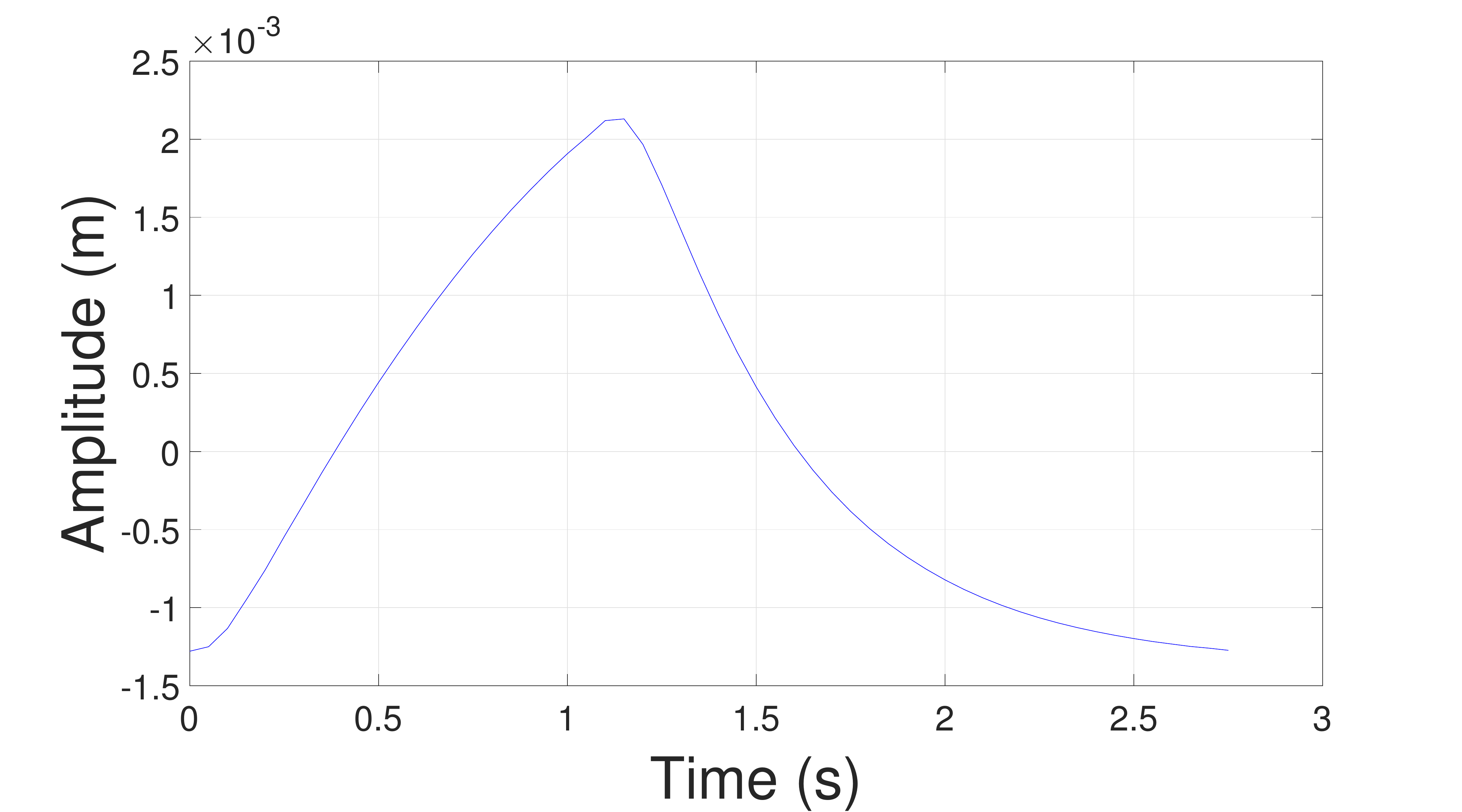}
        \caption{Respiration pulse.}
        \label{fig:respirationTemplate}
    \end{subfigure}
    \begin{subfigure}[b]{0.49\columnwidth}
        \centering
        \includegraphics[width=\textwidth]{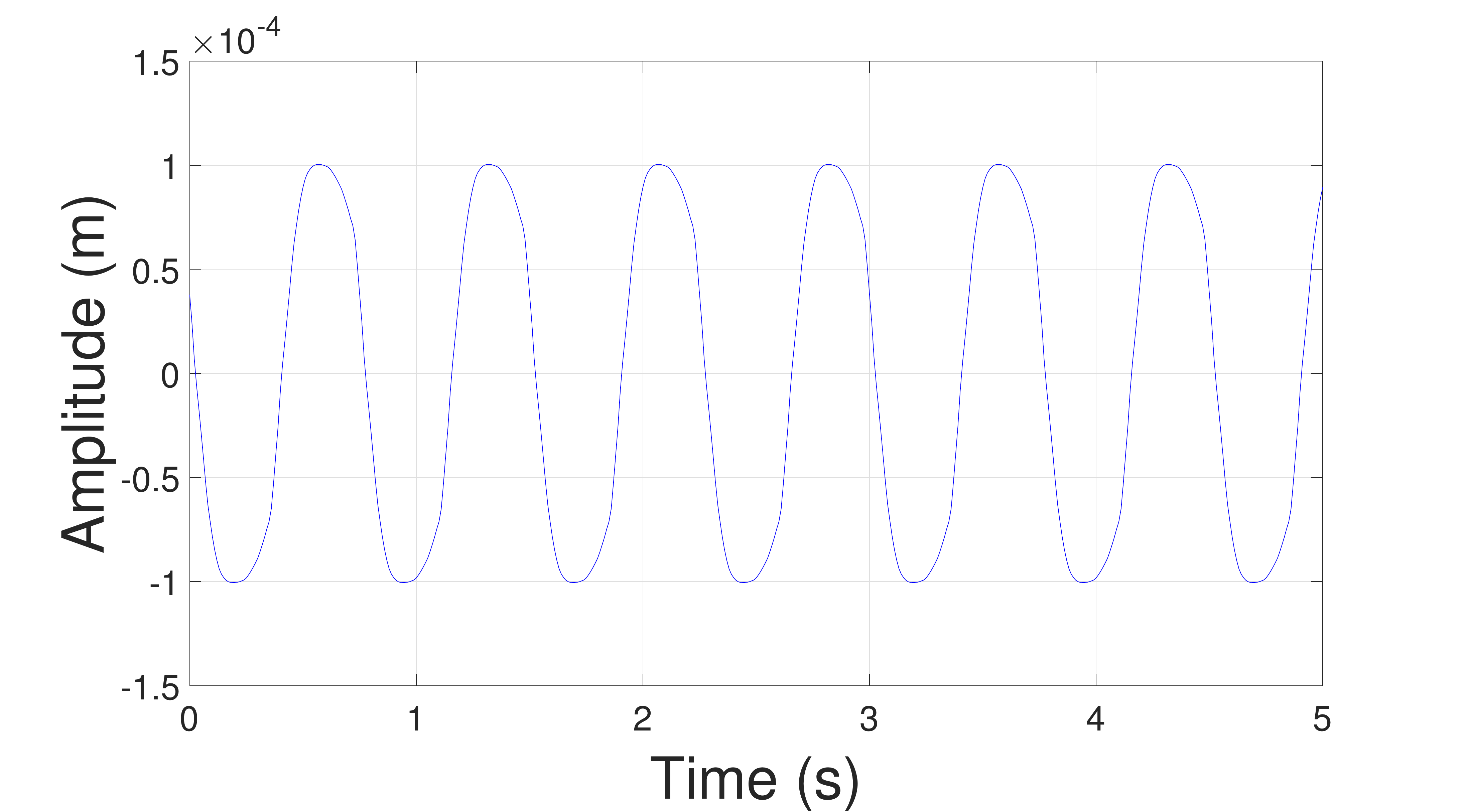}
        \caption{Heartbeat vibrations.}
        \label{fig:heartbeatTemplate}
    \end{subfigure}
\caption{Respiration and heartbeat templates: an example.}
\label{fig:Template1}
\end{figure}

Since human chest-wall movements is proportional to one's lung volume, we use the above equation to model the chest-wall movement caused by respiration. Setting the values of parameters $a_0, a_1, a_2$ as suggested in~\cite{AnuradhaS2:2020}, a respiration template has a fixed shape as illustrated in Figure~\ref{fig:respirationTemplate}. It is clear to see that the respiration process is modeled as a two-stage RC circuit behaviour. The charging/stimuli and discharging stages complete a whole respiration cycle.  

\subsection{Heartbeat Model}
\label{subsec:heartbeatmodel}
The mechanism of cardiac activity and mathematical models have been investigate intensely in the literature. The most popular model is a relaxed oscillatory system, and is described by a Van der Pol differential equation first proposed in~\cite{Balthv:1926,Balthv:1928}. This relaxed oscillatory system model is attractive due to the existence of efficient numerical solutions, and less control parameters. Moreover, it represents the sinoatrial (SA) node rhythmic behavior and the thoracic wall or chest wall movements~\cite{AnuradhaS2:2020,JF:2016,RC:2006,SandraG:2008}.

A relaxation Oscillation characteristic function~\cite{Balthv:1928} is
\begin{align}
\frac{\partial^2 V}{\partial t^2} - \alpha (1-V^2)\frac{\partial V}{\partial t} + \omega^2 V =0,
\label{eq:Relaxation Oscillations}
\end{align}
where $\alpha = \frac{R}{L}$, $\omega^2 = \frac{1}{LC}$, $T = 1.61\frac{\alpha}{\omega^2} = 1.61RC$. Change variables and re-write the equation in~\cite{Balthv:1926} as
\begin{align}
\frac{\partial^2 V}{\partial t^2} - \epsilon (1-V^2)\frac{\partial V}{\partial t} + V =0,
\label{eq:Relaxation Oscillations2}
\end{align}
where $\epsilon = \frac{\alpha}{\omega}$.

When $\epsilon >> 1$, there is no closed-form analytic solution. By fixing the parameters as suggested in ~\cite{AnuradhaS2:2020}, we can solve the differential equation numerically. The resulting heartbeat template has a fixed shape as illustrated in Figure~\ref{fig:heartbeatTemplate}. It is clear that the heartbeat process is modeled as generated by a simple oscillation mechanism, with one complete cycle corresponding to a RLC circuit vibration cycle. There exists more complicated models characterizing detailed heartbeat wave shapes such as ECG signals~\cite{SandraG:2008}. But they are out of the scope of this work and can be considered in the future research.

\begin{figure}[th]
    \centering
    \begin{subfigure}[b]{0.49\columnwidth}
        \centering
        \includegraphics[width=\textwidth]{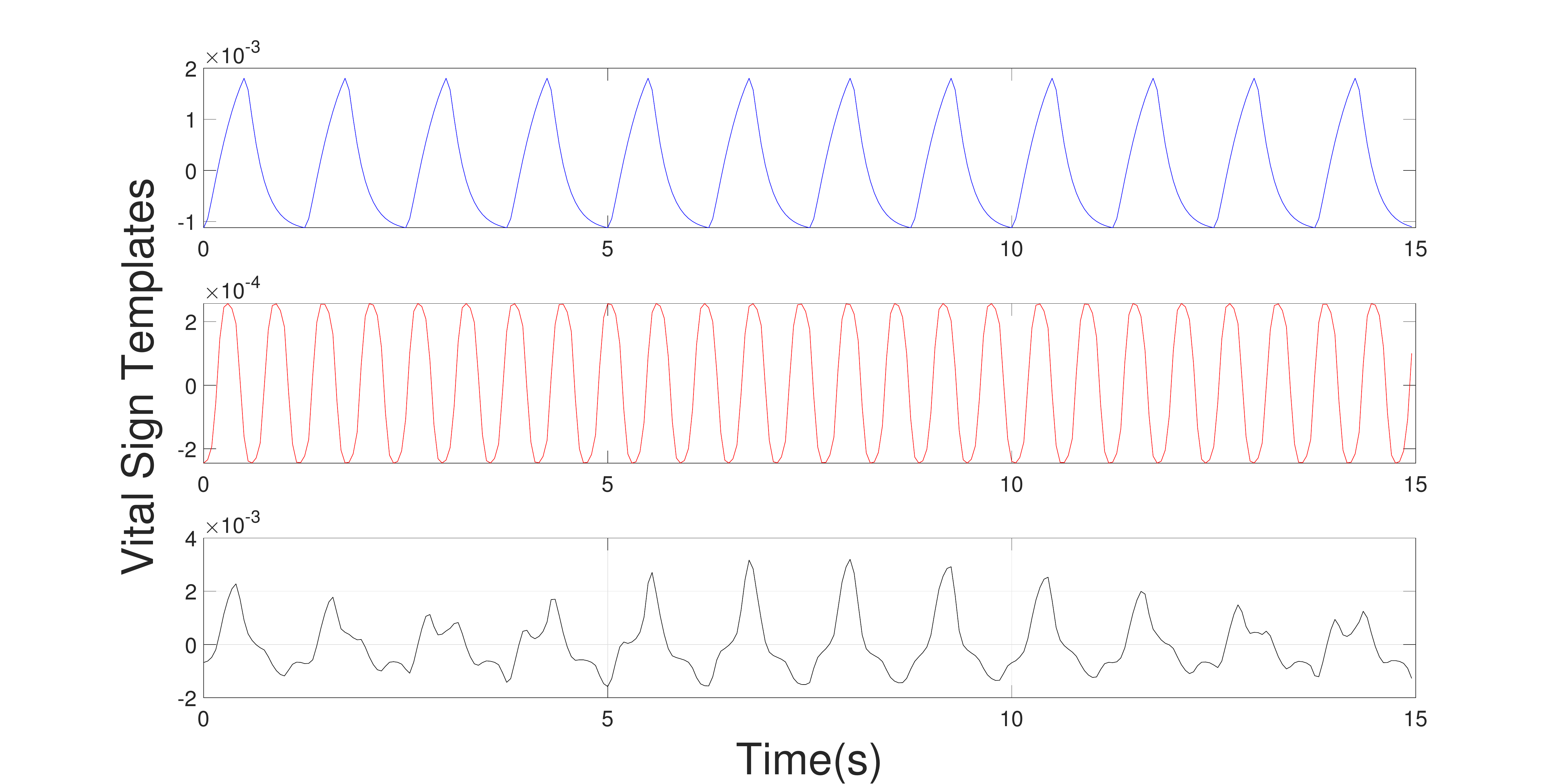}
        \caption{Time domain waveform.}
        \label{fig:Templatewave}
    \end{subfigure}
    \begin{subfigure}[b]{0.49\columnwidth}
        \centering
        \includegraphics[width=\textwidth]{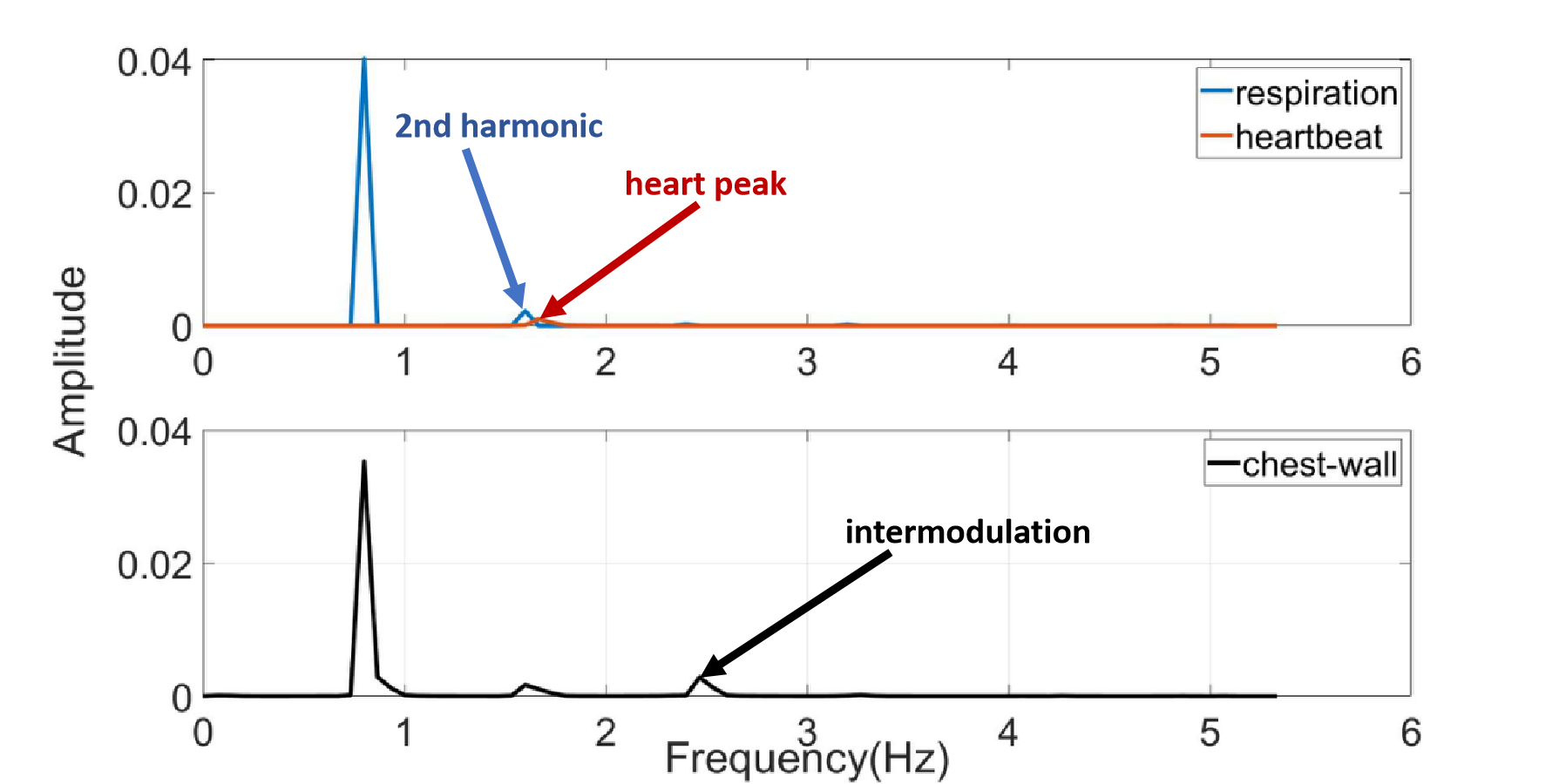}
        \caption{Frequency response.}
        \label{fig:Templatefrequency}
    \end{subfigure}
\caption{An example of the generated vital sign template with the parameters $(A_{h}=0.00025m, A_{res}=0.003m, T_{h}=0.59s, T_{res}=1.25s, t^h_{off}=0, t^r_{off}=0,
 y^h_{off}=0, y^r_{off}=0, c=2500)$. (a) shows the time domain waveform of the respiration signal, the heartbeat signal and the combined chest-wall vital sign template from top to bottom. (b) shows frequency responses of all three signals.}
\label{fig:Template2}
\end{figure}

\subsection{Template Matching Optimization}
Given the templates in Section~\ref{subsec:respirationmodel} and~\ref{subsec:heartbeatmodel}, we ``generate'' different respiration and heartbeat waves by scaling their amplitudes $A$, introducing time offsets $t_{off}$ and biases in measured values $y_{off}$, and varying the time duration of a single pulse $T$.  Specifically, the  piece-wise function within one pulse duration for respiration is given by:
\begin{align}
\scriptstyle y_{res}(t) = 
\left\{
\begin{array}{lll}
\scriptstyle A_{res}[a_1(t-t^r_{off})^2 + a_2(t-t^r_{off})
+ a_3e^{-\frac{t-t^r_{off}}{\tau_{RS}}}+a_4] + y^r_{off},  & \\[1ex]
\qquad 0 \leq t \leq \frac{t_1}{t_1+t_2}T_{res}, & \\[1ex]
\scriptstyle A_{res}[b_1e^{-\frac{t-t^r_{off}-\frac{t_1}{t_1+t_2}T_{res}}{\tau}}
+b_2e^{-\frac{t-t^r_{off}-\frac{t_1}{t_1+t_2}T_{res}}{\tau_{RS}}}]+ y^r_{off},  & \\[1ex]
\qquad \frac{t_1}{t_1+t_2}T_{res} \leq t \leq T_{res}, & \\
0, \quad t > T_{res} &
\end{array} 
\right.
\label{eq:3}
\end{align}
where $A_{res}, T_{res}, t^r_{off}, y^r_{off}$ are the amplitude, pulse duration, time offset, and bias, respectively. These variables are the parameters to be estimated and all other coefficients are known constants. For ease of presentation, we write $y_{res}(t)$  in one period as a function of these control parameters
\begin{align*}
y_{res}(t) = f(t, A_{res}, T_{res}, t^r_{off}, y^r_{off}).
\end{align*}

The respiration signal at time $t \in {\rm I\!R}^+$ is given by
\begin{align}
y_{res}(t) & = \sum_{i}f(t-i\cdot T_{res}, A_{res}, T_{res}, t^r_{off}, y^r_{off}). \nonumber
\end{align}

Applying similar steps, we can also ``generate'' heartbeat waveforms from a template with control parameters $A_{h}, T_{h}, t^h_{off}, y^h_{off}$. Finally, chest wall displacements are expressed as,
\begin{align} 
\scriptstyle y_{chest}(t) &= y_{res}(t) + y_{h}(t) + c\times y_{res}(t)y_{h}(t), \nonumber \\
 &= g(t, A_{h}, A_{res}, T_{h}, T_{res}, t^h_{off}, t^r_{off},
 y^h_{off}, y^r_{off}, c), 
\label{eq:chestTemplate}
\end{align}

where $y_{chest}(t)$ is the final template generated by physical respiration and heartbeat models, and $c$ allows for non-linear coupling. The final template is able to capture a wide variety of vital sign patterns, by adjusting the length of time windows, the periods, amplitudes, and biases of vital signs, etc. 
Figure~\ref{fig:Template2} shows one example template with a $15s$ time window with non-linear cross coupling modeled as a product of the respiration and the heartbeat as in Equation~(\ref{eq:chestTemplate}). Figure~\ref{fig:Templatewave} shows the time domain templates generated by the physical models. From top to bottom, the plots show the respiration wave, the heart wave, and the combined chest-wall template, respectively. Figure~\ref{fig:Templatefrequency} shows the magnitude spectra of the respiration wave (blue), the heartbeat wave (red) and the combined chest-wall wave (black). In this example, the peak frequency of the heart template is close to the second harmonic frequency of the respiration template but with a smaller amplitude. Therefore, the heartbeat is buried by the second harmonic frequency of the respiration in the combined frequency domain representation. Frequency domain methods on the chest-wall template are likely to confuse the second harmonic frequency of the respiration template with the correct base frequency of the heartbeat. The inter-modulation peak frequency at 2.5Hz can be problematic in frequency domain processing as well. On the other hand, a time domain method using 1D CNN in~\cite{UnsooH:2020} would extract the periodicity of the dominant second harmonic, unless the harmonic frequency is filtered out in the preprocessing steps. In contrast, the proposed template matching method operates in the time domain and is robust to harmonics and intermodulation in the frequency domain. 

The estimated chest-wall displacement $\mathbf{\hat{R}}$ can be written as $\mathbf{\hat{R}} = \mathbf{y_{chest}} + \mathbf{n}$ in a vector form, where $\mathbf{n}$ contains the interference from micro-RBMs and all noise sources, and is modelled as zero-mean Gaussian random variables. To determine $y_{res}(t)$ and $y_{h}(t)$, we can solve the following least square optimization problem,

\begin{align}
\begin{aligned}
\min_{para} \quad & ||\mathbf{\hat{R}} -  \mathbf{y_{chest}}||^2, \\
\textrm{s.t.} \quad &  1 \leq T_{res} \leq 10,
 0.5 \leq T_{h} \leq 1.25, \\
&  0 \leq A_{res} \leq 10^{-2}, 
 0 \leq A_{h} \leq 10^{-3}, \\
\end{aligned}
\label{eq:optimization}
\end{align}
where $para = \langle A_{h}, A_{res}, T_{h}, T_{res}, t^h_{off}, t^r_{off}, y^h_{off}, y^r_{off}, c\rangle$ are the control parameters. The constraints in \eqref{eq:optimization} are based on statistics from physiology literature. 

The objective function in \eqref{eq:optimization} is non-convex. With proper initial conditions, numerical methods such as gradient descent and Newton's methods are able to find good local minimum or the global optimal solutions. In our implementation, we calculate auto-correlations of the estimated chest-wall movements, to use the estimated respiration duration as the initial $T_{res}$. The initial $T_{res}$ allows fine-tuning zero crossing points of the estimated chest-wall movement . The initial offsite $t^r_{off}$ is estimated from $T_{res}$ and the zero-crossing points.
Then, we apply an exhaustive search with a coarse step-size (around $500$ points in the search space) to find $T_{h}$ and $t^h_{off}$ as initial parameters for heartbeat template. After the initialization, a standard numerical function fit is applied to jointly update the parameters in $para$ by the gradient descent method. The trust-region method~\cite{WenyuS:2006} is chosen to handle the constraints and the termination tolerance is set to be $10^{-8}$ for the sum of squared errors.

\section{Performance Evaluation}
\label{sec:Experiments}
In this section, we present the implementation of a prototype Pi-ViMo system using off-the-shelf mmWave radars and the experiment results from multi-subject testbed evaluations of the proposed system under various conditions, which are purposely chosen to closely mimic real-life situations.
\subsection{Experiment Setups}
\label{subsec:experimentsetups}
\paragraph*{Implementation.}
An IWR6843ISK board~\cite{iwr6843ISK} together with a DCA1000EVM board~\cite{DCA1000EVM} is used in the experiments. It operates at $60 \sim 64$ GHz (with 4-GHz bandwidth) and transmits FMCW signals. The radar front-end includes $3$ transmit antennas (Tx), $4$ receive antennas (Rx), with $120^{\circ}$ azimuth field of view (FoV) and $30^{\circ}$ elevation FoV at a range up to 10 meters. The acquired raw IF signal is sent to a host PC via Ethernet, where mmWave Studio~\cite{mmwavestudio} is used to initiate, configure, and control the radar boards.
Ground truth vital signs are collected from wearable NeuLog sensors~\cite{Log36,NUL208} in all scenarios, with respiration sensed by a chest-strapped pressure sensor, and heartbeats sensed by an optical sensor that can be clipped onto earlobes or finger tips. Both sensors are synchronized with the radar at the sampling rate (the frame rate in radar) of 20 Hz.

The processing chain of Pi-ViMo is implemented in MATLAB R2021a, which takes raw IF signals as input, and outputs vital sign signal waveforms and estimated respiration and heart rates. All  baseline methods are also implemented in MATLAB.

\paragraph*{Participants.}
To evaluate Pi-ViMo’s performance, we recruited 11 participants (4 females and 7 males), aging between 21 and 46, and with BMI in the range of $18.1 \sim 31.6$. Participants wore their daily attires such as T-shirts, blouses, and sweaters of different fabric materials. This research protocol has been approved by the research ethical board (REB) from our institution.

\begin{figure}[th]
    \centering
    \begin{subfigure}[b]{0.49\columnwidth}
        \centering
        \includegraphics[height=0.35\textheight]{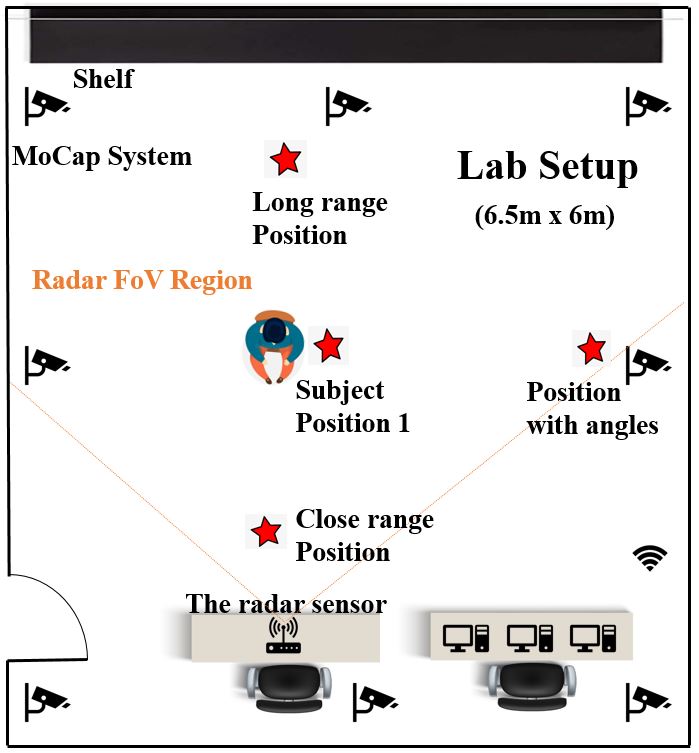}
        \caption{Lab}
        \label{fig:setup1}
    \end{subfigure}
    \begin{subfigure}[b]{0.49\columnwidth}
        \centering
        \includegraphics[width=\linewidth]{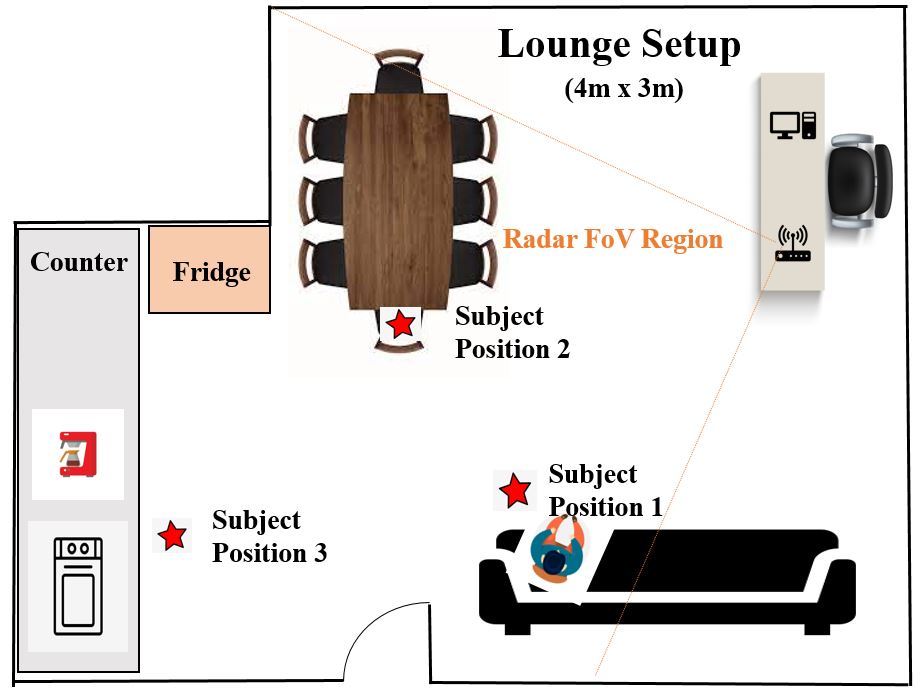}
        \caption{Lounge room}
        \label{fig:setup2}
    \end{subfigure}
\caption{Two environments for data collection.}
\label{fig:setup}
\end{figure}

\paragraph*{Experimental Environments.}
We conduct experiments in two environments, i.e., an open area ($6.5m \times 6m$) in a lab environment and a $4m \times 4m$ lounge room. The lab (Figure~\ref{fig:setup}a) has standard office furniture and many electronic equipment and wireless transceivers (WiFi, LTE, Bluetooth, etc.). 
The lounge room (Figure~\ref{fig:setup}b) is equipped with household appliances and furniture, similar to an apartment room. The radar sensor is placed on a desk in the corner of the room and its FoV region is showed in Figure~\ref{fig:setup2}. 
In the experiments, we use a single pair of TX and RX antennas to cover the whole space (FoV region) as shown in the figures. During the experiments, only one subject is present in the predefined positions. 

During data collection in the lab, subjects were asked to sit or stand at different locations in the test areas. At each location, subjects were asked to breathe normally and hold their breath for 15 seconds while siting against the back of a chair ({\it SS}) to reduce possible RBMs. To investigate scenarios with micro-RBMs, we consider 5 movements, namely, sitting with relaxed pose ({\it RP}), random head turning ({\it HT}), random limb stretching ({\it LBS}), relaxed standing ({\it RS}), and random leg shaking ({\it LS}). These cases are purposely chosen to be as realistic as possible to mimic real-life situations. Data were collected from all 11 participants involved in this set of experiments.
In the lounge, subjects were asked to sit in a sofa, sit in the chair beside a table and stand next to a counter. At the first two locations, the same trials as in the lab settings were conducted, except for the relaxed standing movements. At the third location, subjects stood next to a counter and only completed HT, LBM, and RS movements. Data were collected from 9 participants in this set of experiments.

In total, we conducted 679 experimental trials, with each lasting around 15 seconds. The total length of collected data is around 170 minutes. The ground truth data shows that subject respiration rates varied from 8 -- 28 bpm, and Subject heart rates varied from 56-110 bpm. 

\paragraph*{Evaluation metrics.}
The goal of Pi-ViMo  is to extract periodic information of vital signs from radar signals. We choose two most  metrics to quantify the accuracy of the proposed algorithms. The first one is frequency/rate ($bpm$) errors in percentage defined as:
\begin{align}
error = 100 \times \frac{|R_e-R_{gt}|}{R_{gt}},
\end{align}
where $R_e$ is the respiration or heart rate estimated by Pi-ViMo, and $R_{gt}$ is the corresponding rate from ground truth measurements.

The second metric is the Pearson Correlation Coefficient (PCC) that quantify the degree of  linear correlation between two time series~\cite{KarlP:1895}. It is defined as
\begin{align}
\rho = \frac{\sum^N_{i=1} (y^{e}_i - \bar{y^{e}})(y^{gt}_i - \bar{y^{gt}}) }{\sqrt{\sum^N_{i=1}(y^{e}_i - \bar{y^{e}})^2 \sum^N_{i=1} (y^{gt}_i - \bar{y^{gt}})^2}},
\end{align}
where $y_i^{e}$ stands for the $i^{th}$ sample estimated from Pi-ViMo, $y_i^{gt}$ is the corresponding sample from ground truth measurements, and $\bar{y}$ stands for mean value. We use PCC of respiration waves from Pi-ViMo and ground truth to measure their similarity. Unfortunately, there is no easy way to obtain the chest-wall expansions and contractions due to cardiac activities. The NeuLog sensor measure heart rates by measuring blood pressure, and the resulting waveform is close to ECG data as opposed to chest-wall movements. Consequently, we only measure heart rate errors and omit PCC for heart waveforms in subsequent performance evaluation.  

\paragraph*{Baseline methods.}
We have implemented the method in~\cite{MostafaA:2019}, which selects a single range bin with the maximum vibration energy and uses frequency-domain methods (BPF and FFT) for vital sign separation and estimation. Additionally, we implemented VMD in~\cite{ArindamR:2021}, a state-of-the-art signal decomposition method, for vital sign separation in conjunction with single range bin selection in~\cite{MostafaA:2019}. Since there is no open-source implementation and benchmark dataset available, we omit neural networks based approaches in~\cite{ZheC:2021,TianyueZ:2021} in the comparison.

\subsection{Experiment Results}
The main advantages of the proposed Pi-ViMo system are that it is capable of monitoring subjects at arbitrary locations in a radar’s FOV and it is robust against micro-RBMs. Therefore, we perform extensive experiment trials and evaluate the performance of Pi-ViMo at different sensing distances and under realistic micro-RBMs.

\subsubsection{Micro-benchmark for MSP Model Validation}
\label{sec:MSPVal}
In this section, we perform a micro-benchmark to validate the proposed MSP model and MSP-based range bin selection in Algorithm~\ref{alg:MSPbins}. In the experiments, different subjects sit still in front of a radar at different distances. The outputs from a NeuLog sensor serve as the ground truth for respiration waves. 

For comparison, we include the number of candidate range bins and MSP bins after the first and second steps of Algorithm~\ref{alg:MSPbins}. Among the candidate bins, we calculate the correlation of the signal in each bin and the ground truth waveform and list the values of the average of the top-2 correlation valued selected by an Oracle ($corr_{top2}$), the ones selected by Algorithm~\ref{alg:MSPbins} ($corr_{MSP}$), and the one according to the maximum vibration energy criterion ($corr_{MVE}$). Another metric of interests is the delay spread between two range bins. The maximum delay spread tells the phase differences of received vital sign signals. We expect that if Algorithm~\ref{alg:MSPbins} selects informative range bins, the results will be consistent with the measurement results in Table~\ref{tab:table2} from motion capture systems.

\paragraph*{Impact of distance.}
Table~\ref{tab:MSPvsdistanceStill} shows averaged metrics versus sensing distances. From the table we observe the number of selected MSP range bins drops as the sensing distance increases. At 5 meters, approximately 2 range bins are selected. This could be explained by the frequency leak phenomenon discussed in Section~\ref{sec:discussion}. Interestingly, the correlations do not decrease with sensing distance. This is because the vital sign information is embedded in the phase instead of magnitude of received signals, and a sensitivity in the scale of millimeters is enough to detect chest-wall movement caused by vital signs. When comparing $corr_{MSP}$, $corr_{top2}$ and $corr_{MVE}$, we can infer that almost always the MSP bins would include the top 2 bins with the highest correlation with the ground truth. In contrast,  the signal in the bin with the maximum variation energy generally has lower correlations, and thus fails to capture vital sign waveform well. 

\paragraph*{Impact of RBMs.}
Micro random body movements has slight impact on the number of MSP range bins due to the selection Algorithm~\ref{alg:MSPbins}. The RBMs introduce interference or noise with the chest-wall movement caused by purely vital signs. As shown in Table~\ref{tab:MSPvsdistanceRBM}, signals from all the range bins are affected and have decreased correlations with ground truth measurements. However, the range bins selected by Algorithm~\ref{alg:MSPbins} have higher correlations than the one by the maximum vibration energy criterion. 

\paragraph*{Impact of subjects.}
Table~\ref{tab:MSPvssubject} summarizes the results from 11 subjects in case of sitting still. The chest-wall sizes of different subjects are the main factor that affect the number of selected MSP range bins. Another source of variability comes from subject behaviors. Some subjects may not remain still through the entire data collection duration. As a result, the correlation values even in the case of top-2 bins vary a lot. However, we observe the correlation values attained by Algorithm~\ref{alg:MSPbins} are mostly consistent with the Oracle, while $corr_{MVE}$ tends to much smaller. 

Due to our experiment design aiming to relax Lab control conditions, the subjects are instructed to sit comfortable instead of strictly still to introduce more diversity. Individual behaviours of different subjects during the experiment shows the effect on the correlations.

\paragraph*{Summary.}
To this end, we conclude that selecting informative range bins that contain vital sign signals is important to ensure the quality of data for subsequent processing. It is a non-trival problem. Simple criterion such as maximum vibration energy does not yield good results. The proposed MSP-based range bin selection algorithm is effective at different distances, across subjects with and without RBM. 

\begin{table*}[ht]
\begin{minipage}{\textwidth}
  \begin{center}
  \caption{Metrics of MSP model over sensing distances in case of staying still}
  \label{tab:MSPvsdistanceStill}
  \begin{tabular}{cccccccl}
    \toprule
    distances(m) & \texttt{$\#$}candidate bins & \texttt{$\#$}MSP bins & $corr_{MSP}$ & $corr_{MVE}$ & $corr_{top2}$ &
    $delay_{spread}$(s) \\ 
    \midrule
    0.3 & 7.83 & 2.58 & 0.69 & 0.61 & 0.74 & 0.5  \\
    1 & 8.64 & 2.4 & 0.73 & 0.63 & 0.75 & 0.75  \\
    2 & 8.09 & 1.91 & 0.69 & 0.55 & 0.69 & 0.3  \\
    5 & 7.64 & 1.9 & 0.73 & 0.58 & 0.73 & 0.61  \\
    \bottomrule
  \end{tabular}
      \end{center}
  \bigskip
  {\footnotesize Results are averaged on multiple trials and only include a single subject. Correlations and delays are calculated between radar data and ground truth waveform. $corr_{MSP}$: averaged correlations of range bins in \{MSP bins\}. $corr_{MVE}$: correlation of a single range bin selected by maximum vibration energy. $corr_{top2}$: averaged correlations of range bins with top 2 highest correlations.}
\end{minipage}
\end{table*}

\begin{table*}[ht]
\begin{minipage}{\textwidth}
  \begin{center}
  \caption{Metrics of MSP model over sensing distances in presence of RBMs}
  \label{tab:MSPvsdistanceRBM}
  \begin{tabular}{cccccccl}
    \toprule
    distances(m) & \texttt{$\#$}candidate bins & \texttt{$\#$}MSP bins & $corr_{MSP}$ & $corr_{MVE}$ & $corr_{top2}$ &
    $delay_{spread}$(s) \\ 
    \midrule
    0.3 & 7.27 & 2 & 0.54 & 0.41 & 0.58 & 0.74  \\
    1 & 7.89 & 2.09 & 0.56 & 0.48 & 0.61 & 0.91  \\
    2 & 8.33 & 2.18 & 0.52 & 0.45 & 0.57 & 0.73  \\
    5 & 7.78 & 1.88 & 0.53 & 0.41 & 0.56 & 0.58  \\
    \bottomrule
  \end{tabular}
      \end{center}
  \bigskip
  {\footnotesize Results are averaged on multiple trials and only include a single subject. Correlations and delays are calculated between radar data and ground truth waveform. $corr_{MSP}$: averaged correlations of range bins in \{MSP bins\}. $corr_{MVE}$: correlation of a single range bin selected by maximum vibration energy. $corr_{top2}$: averaged correlations of range bins with top 2 highest correlations.}
\end{minipage}
\end{table*}

\begin{table*}[ht]
\begin{minipage}{\textwidth}
  \begin{center}
  \caption{Metrics of MSP model over different subjects in case of staying still}
  \label{tab:MSPvssubject}
  \begin{tabular}{cccccccl}
    \toprule
    subject(\texttt{\#}) & \texttt{$\#$}candidate bins & \texttt{$\#$}MSP bins & $corr_{MSP}$ & $corr_{MVE}$ & $corr_{top2}$ &
    $delay_{spread}$(s) \\ 
    \midrule
    S1 & 11.5 & 3.25 & 0.69 & 0.64 & 0.77 & 0.37  \\
    S2 & 7 & 1.5 & 0.85 & 0.58 & 0.85 & 0.7  \\
    S3 & 9.75 & 2.5 & 0.69 & 0.66 & 0.75 & 0.63  \\
    S4 & 8.5 & 3 & 0.56 & 0.63 & 0.64 & 0.4  \\
    S5 & 8 & 1.25 & 0.77 & 0.54 & 0.77 & 0.6  \\
    S6 & 7.5 & 1.25 & 0.66 & 0.47 & 0.66 & 0.25  \\
    S7 & 7 & 1.75 & 0.59 & 0.56 & 0.68 & 0.21  \\
    S8 & 7.5 & 2 & 0.79 & 0.71 & 0.80 & 0.26  \\
    S9 & 6.75 & 4.25 & 0.76 & 0.73 & 0.76 & 1.1  \\
    S10 & 8.4 & 1.5 & 0.56 & 0.43 & 0.63 & 0.4  \\
    S11 & 6.5 & 2.25 & 0.84 & 0.58 & 0.84 & 0.4  \\
    \bottomrule
  \end{tabular}
      \end{center}
  \bigskip
  {\footnotesize Results are averaged on multiple trials and multiple sensing distances. Correlations and delays are calculated between radar data and ground truth waveform. $corr_{MSP}$: averaged correlations of range bins in \{MSP bins\}. $corr_{MVE}$: correlation of a single range bin selected by maximum vibration energy. $corr_{top2}$: averaged correlations of range bins with top 2 highest correlations.}
\end{minipage}
\end{table*}

\subsubsection{A Case Study}
\begin{figure*}[!ht]
    \centering
    \begin{subfigure}[b]{0.325\textwidth}
        \centering
        \includegraphics[width=\textwidth]{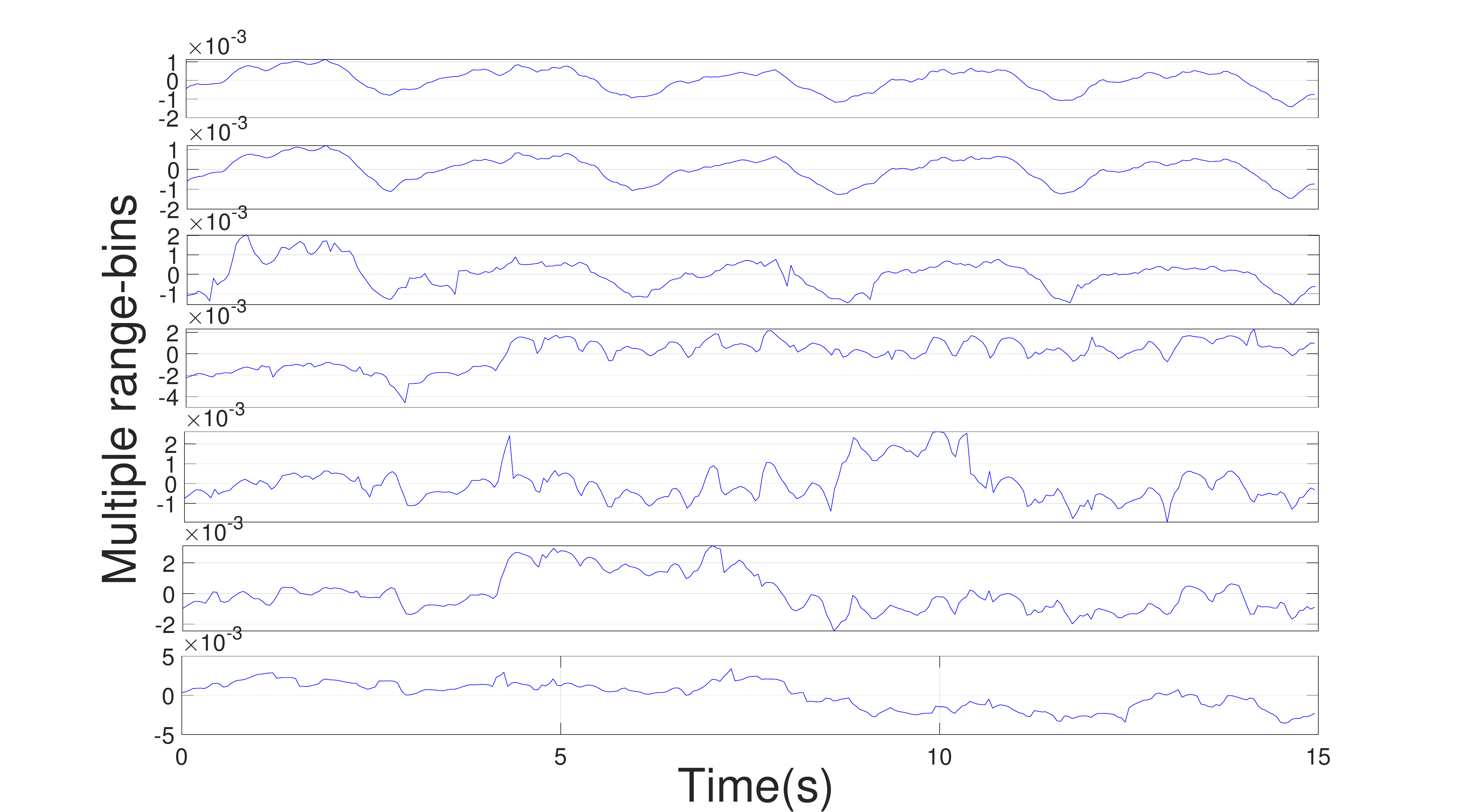}
        \caption{Multiple range bins.}
        \label{fig:exampleA}
    \end{subfigure}
    \hspace{0.1em}
    \begin{subfigure}[b]{0.325\textwidth}
        \centering
        \includegraphics[width=\textwidth]{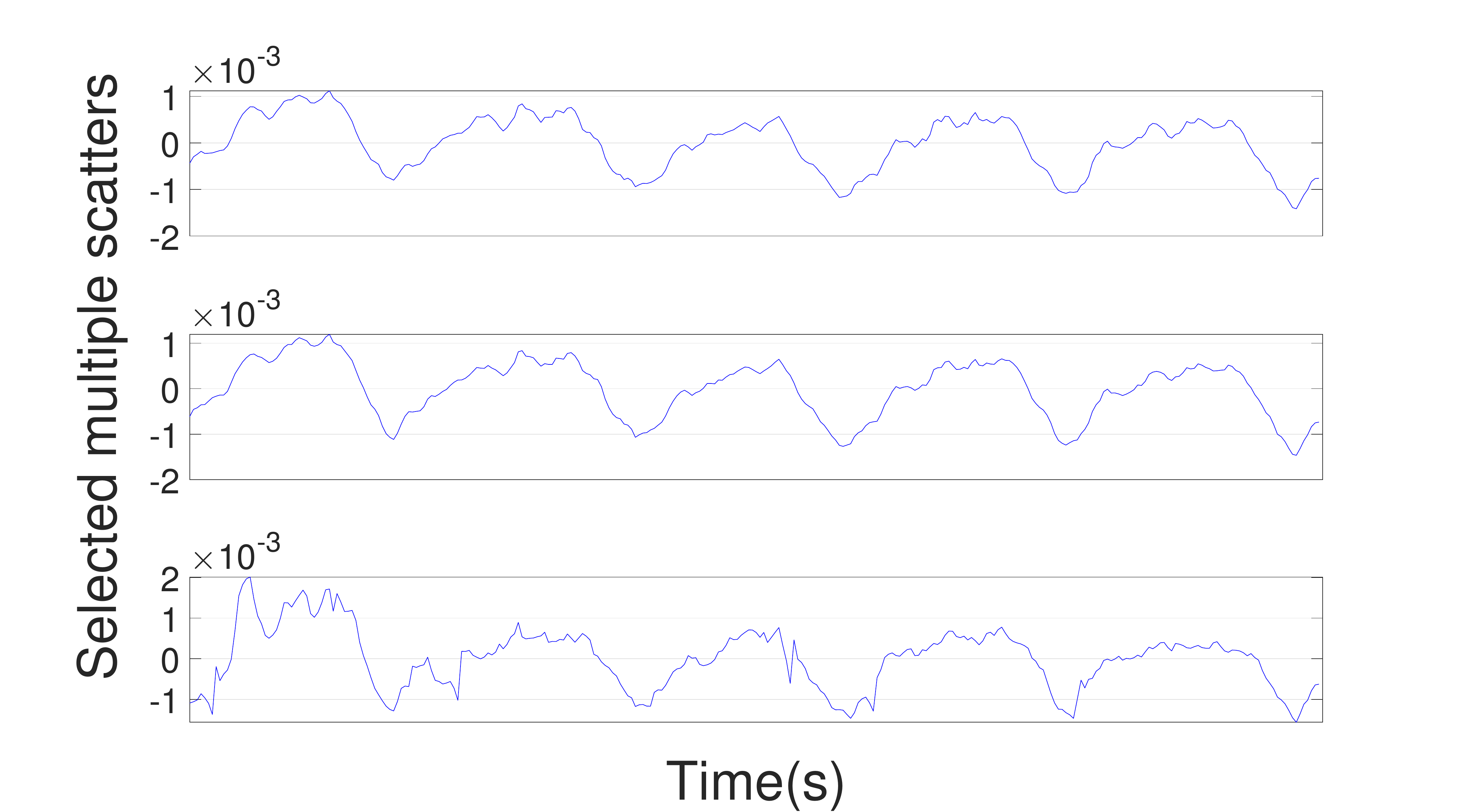}
        \caption{Selected range bins.}
        \label{fig:exampleB}
    \end{subfigure}
    \hspace{0.1em}
    \begin{subfigure}[b]{0.325\textwidth}
        \centering
        \includegraphics[width=\textwidth]{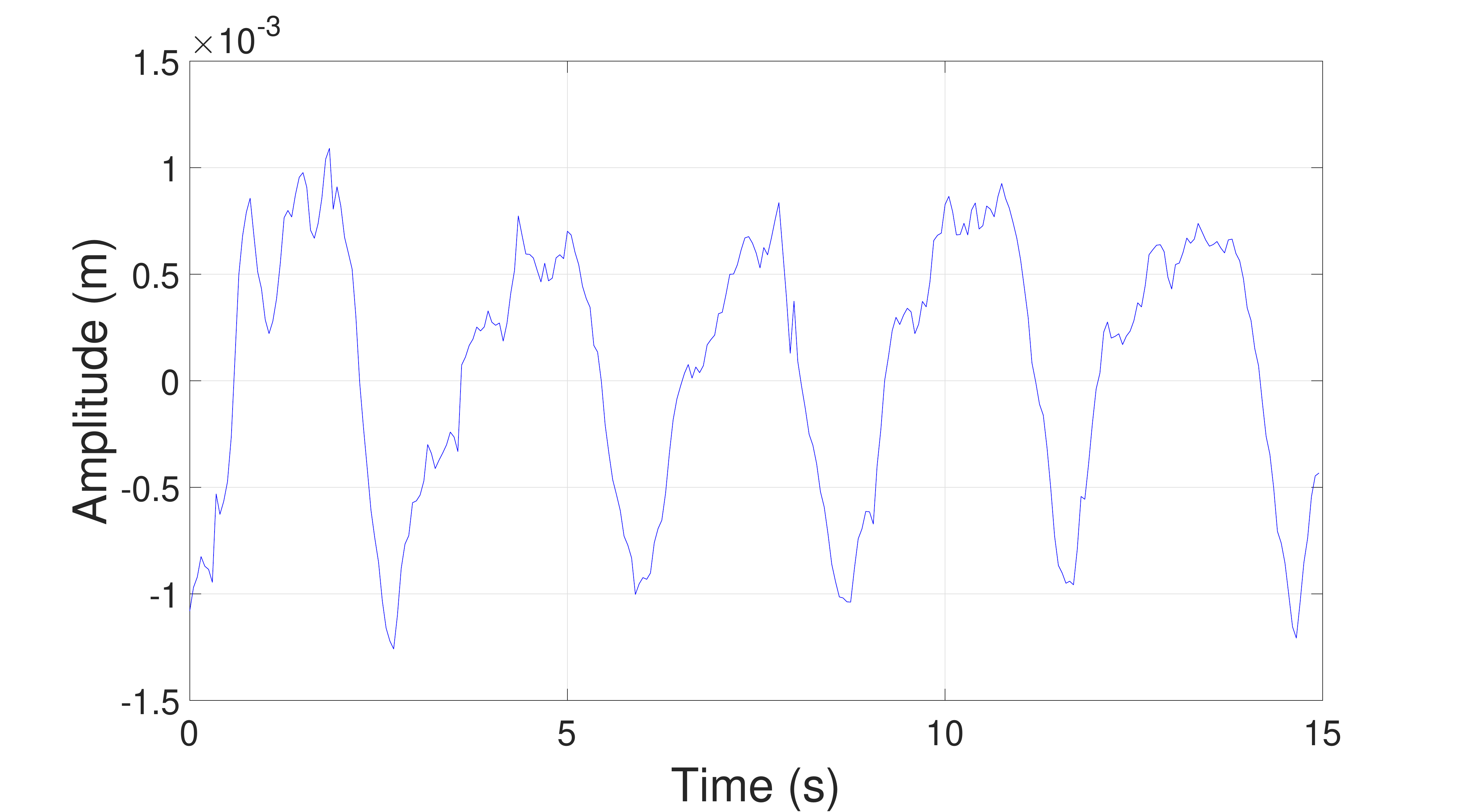}
        \caption{Coherent combining.}
        \label{fig:exampleC}
    \end{subfigure}
    \begin{subfigure}[b]{0.325\linewidth}
        \centering
        \includegraphics[width=\textwidth]{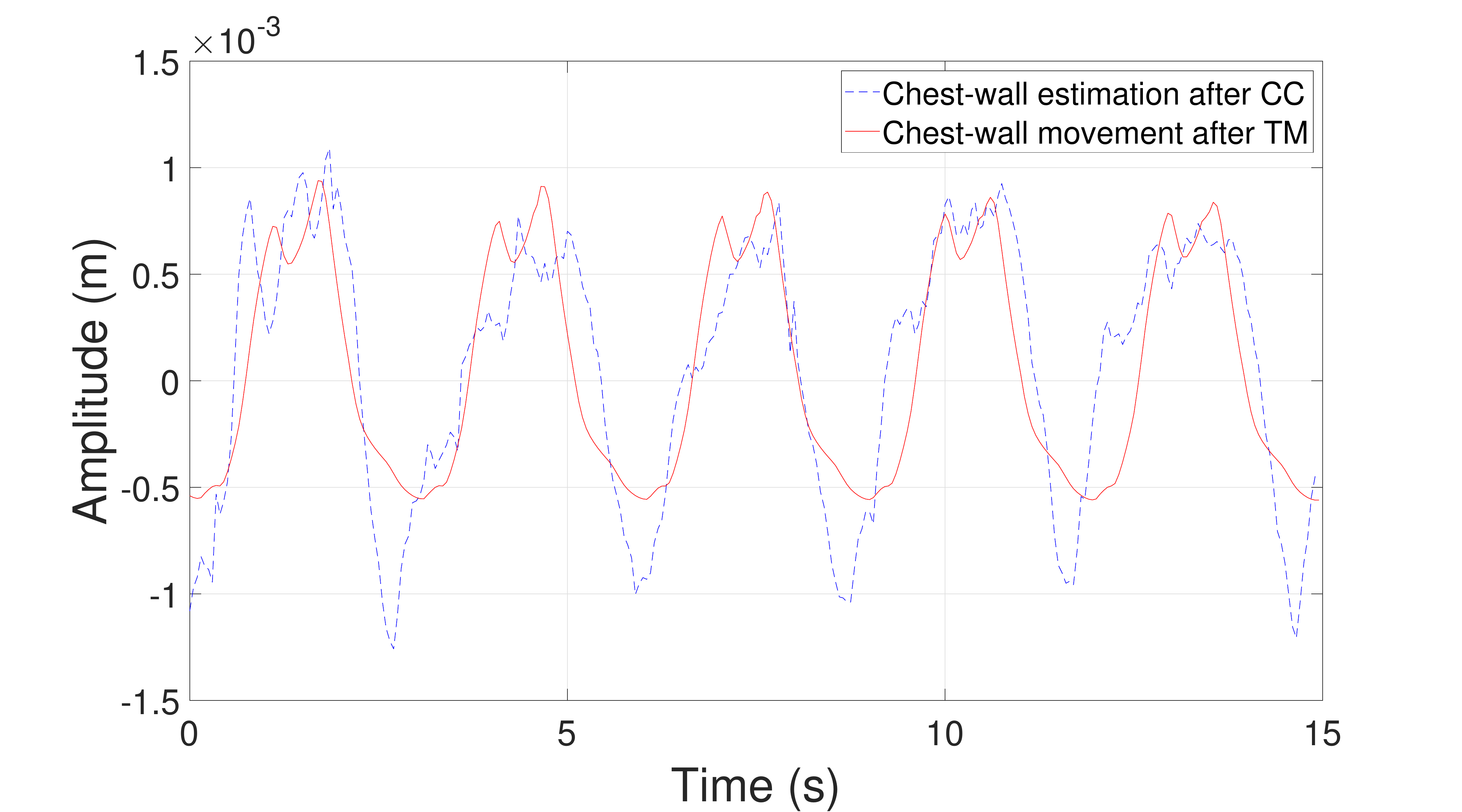}
        \caption{Template matching.}
        \label{fig:exampleF}
    \end{subfigure}
    \begin{subfigure}[b]{0.325\linewidth}
        \centering
        \includegraphics[width=\textwidth]{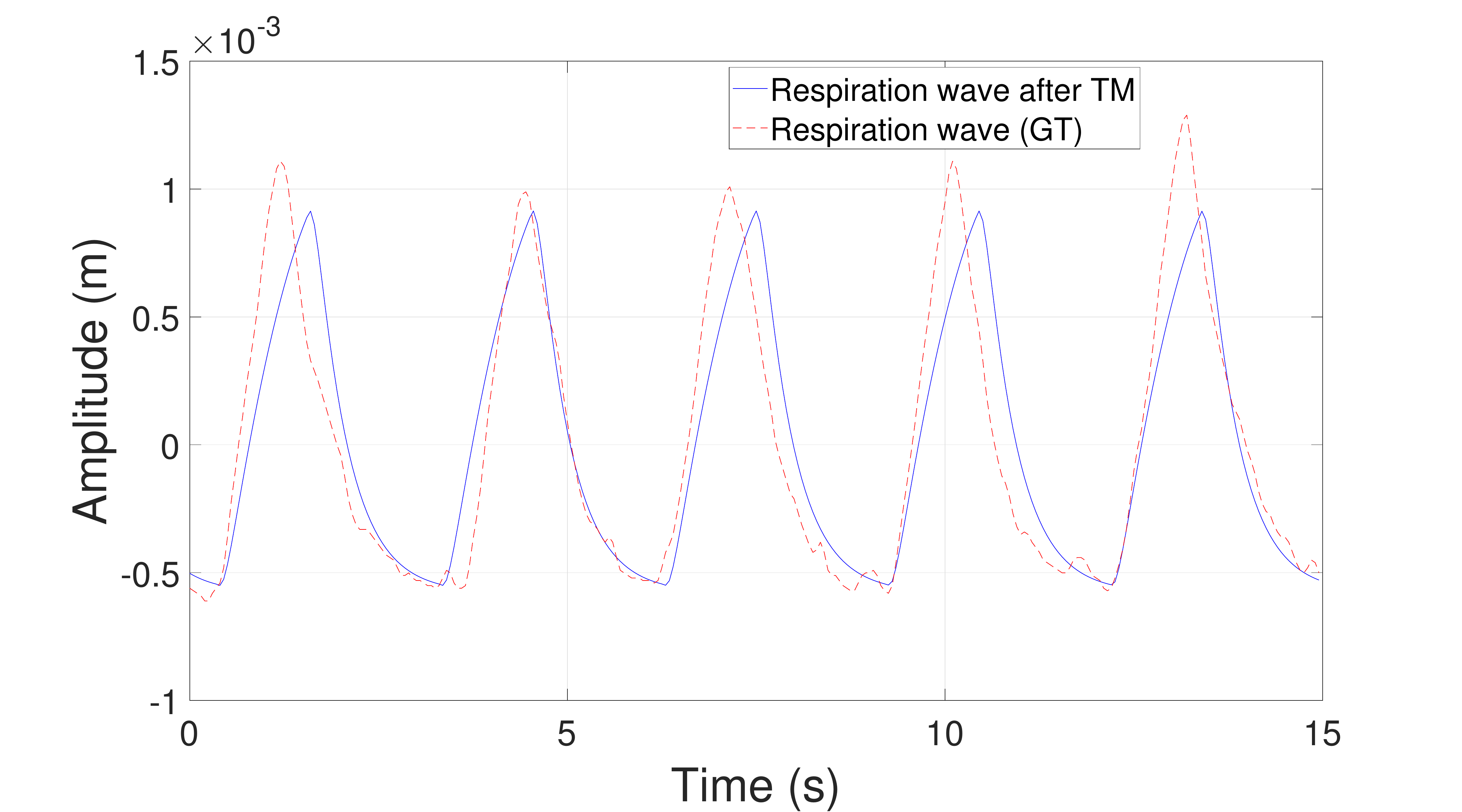}
        \caption{Respiration waves}
        \label{fig:exampleG}
    \end{subfigure}
    \begin{subfigure}[b]{0.325\linewidth}
        \centering
        \includegraphics[width=\textwidth]{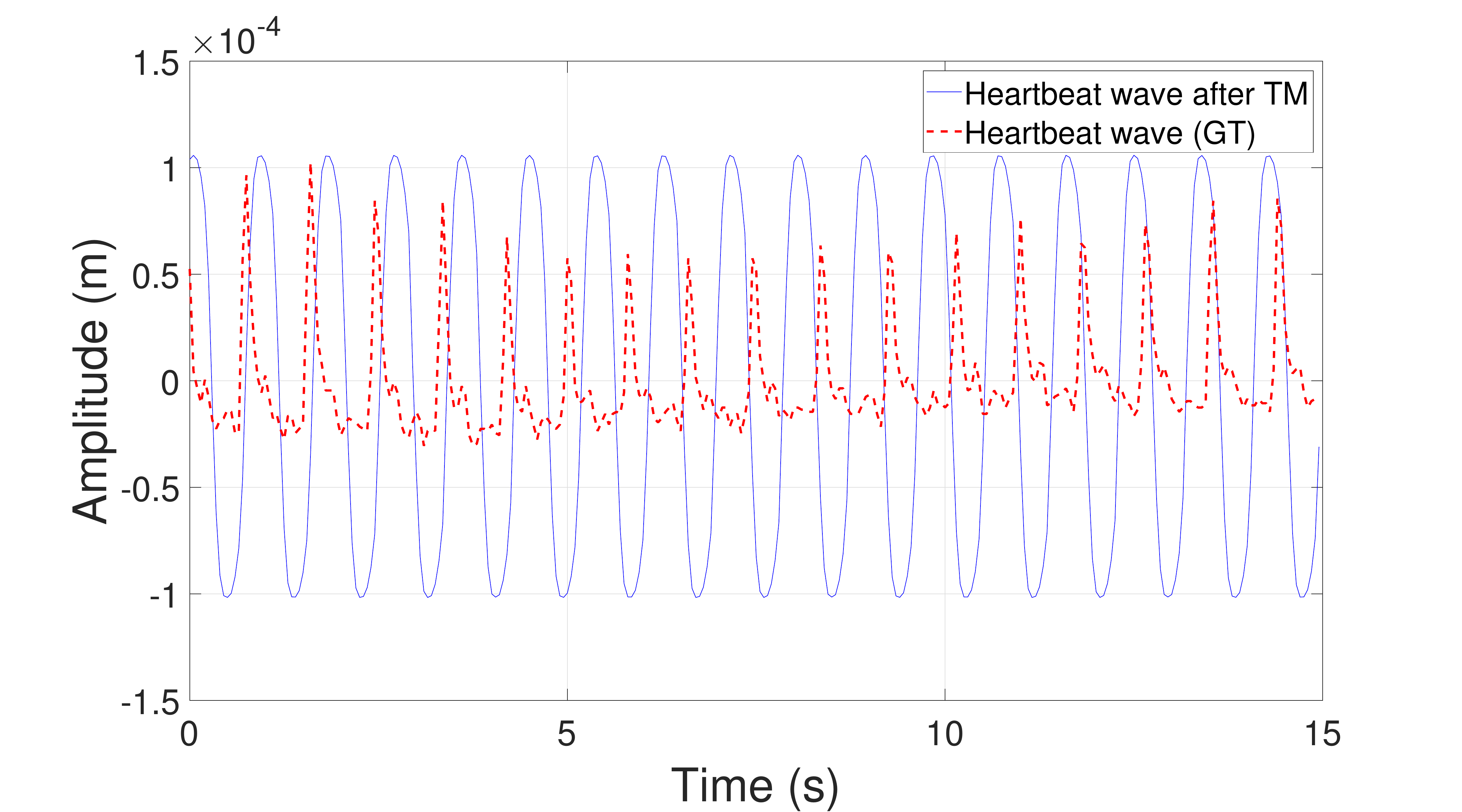}
        \caption{Heartbeat waves}
        \label{fig:exampleH}
    \end{subfigure}
\caption[Outputs from each processing step in Pi-ViMo.]{Outputs from each processing step in Pi-ViMo in one experimental trial. A subject is sitting in a chair with persistent legs shaking for $15$ seconds at a sensing distance of $0.3$ meters.}
\label{fig:example}
\end{figure*}

We hereby illustrate the operations of Pi-ViMo using a case study. In the trial, a subject sits in a chair facing the radar sensor and the sensing distance is around $0.3$ meters. The subject purposely shakes her legs constantly during the $15$-second data collection period.  The output of each step of the Pi-ViMo pipeline in Figure~\ref{fig:sys1} is shown in Figure~\ref{fig:example}. 

Radar signals from 7 neighboring range bins and preprocessed. The raw chest-wall displacement estimations in these bins are shown in Figure~\ref{fig:exampleA}. Waveforms in the last four range bins are severely distorted, whereas waveforms in the first three range bins well preserve chest-wall movements. MSP selection is able to choose the first three range bins and the corresponding waveforms are showed in Figure~\ref{fig:exampleB}. High frequency fluctuations around respiration waves can be seen clearly in all three waveforms. These fluctuations are the combined effect of movements from leg shaking and cardiac activities. We apply the proposed coherent combining algorithm on the selected waveforms and the estimated chest-wall movements are showed in Figure~\ref{fig:exampleC}.
Figure~\ref{fig:exampleF} shows the fitted chest-wall movement (red line) after template matching. The estimated respiration waveform and the heart waveform can be found in Figure~\ref{fig:exampleG} and~\ref{fig:exampleH}, respectively. In both figures, the blue line is the estimated waveform and the red line is the ground truth waveform recorded by NeuLog sensors. For respiration, both radar signals and the NeuLog respiration belt measure the chest-wall expansions and contractions. The two match very well. As mentioned in Section~\ref{subsec:experimentsetups}, the NeuLog sensor measures changes in blood pressure. Therefore, the shape of the estimated heart wave differs from NeuLog measurements. However, the periods (or equivalently rates) are almost identical.

\subsubsection{Impact of locations}
\begin{figure*}[!ht]
    \centering
    \begin{subfigure}[b]{0.325\textwidth}
        \centering
        \includegraphics[width=\textwidth]{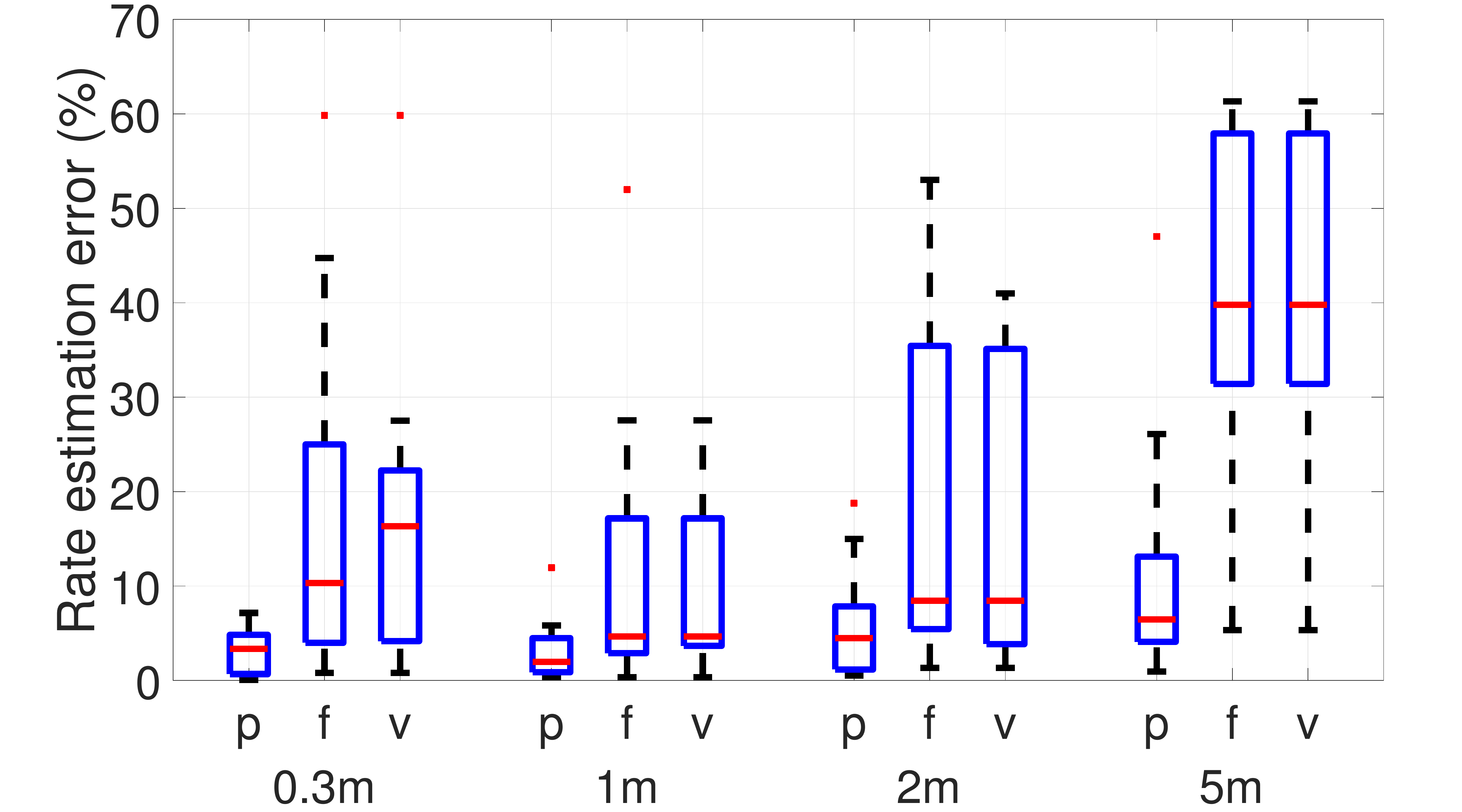}
        \caption{Respiration rate.}
        \label{fig:performanceRangesStillA}
    \end{subfigure}
    \begin{subfigure}[b]{0.325\textwidth}
        \centering
        \includegraphics[width=\textwidth]{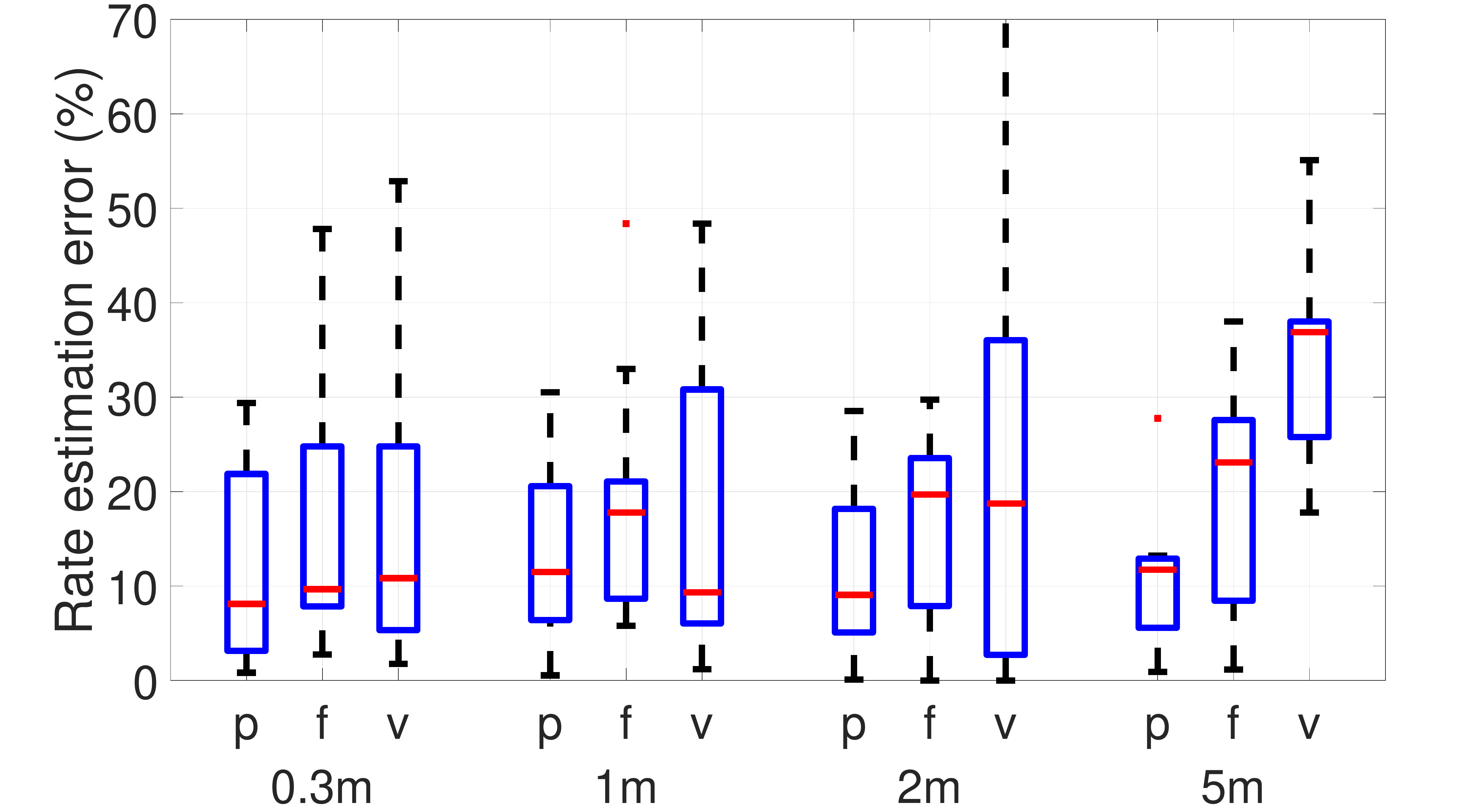}
        \caption{Heartbeat rate.}
        \label{fig:performanceRangesStillB}
    \end{subfigure}
    \begin{subfigure}[b]{0.325\textwidth}
        \centering
        \includegraphics[width=\textwidth]{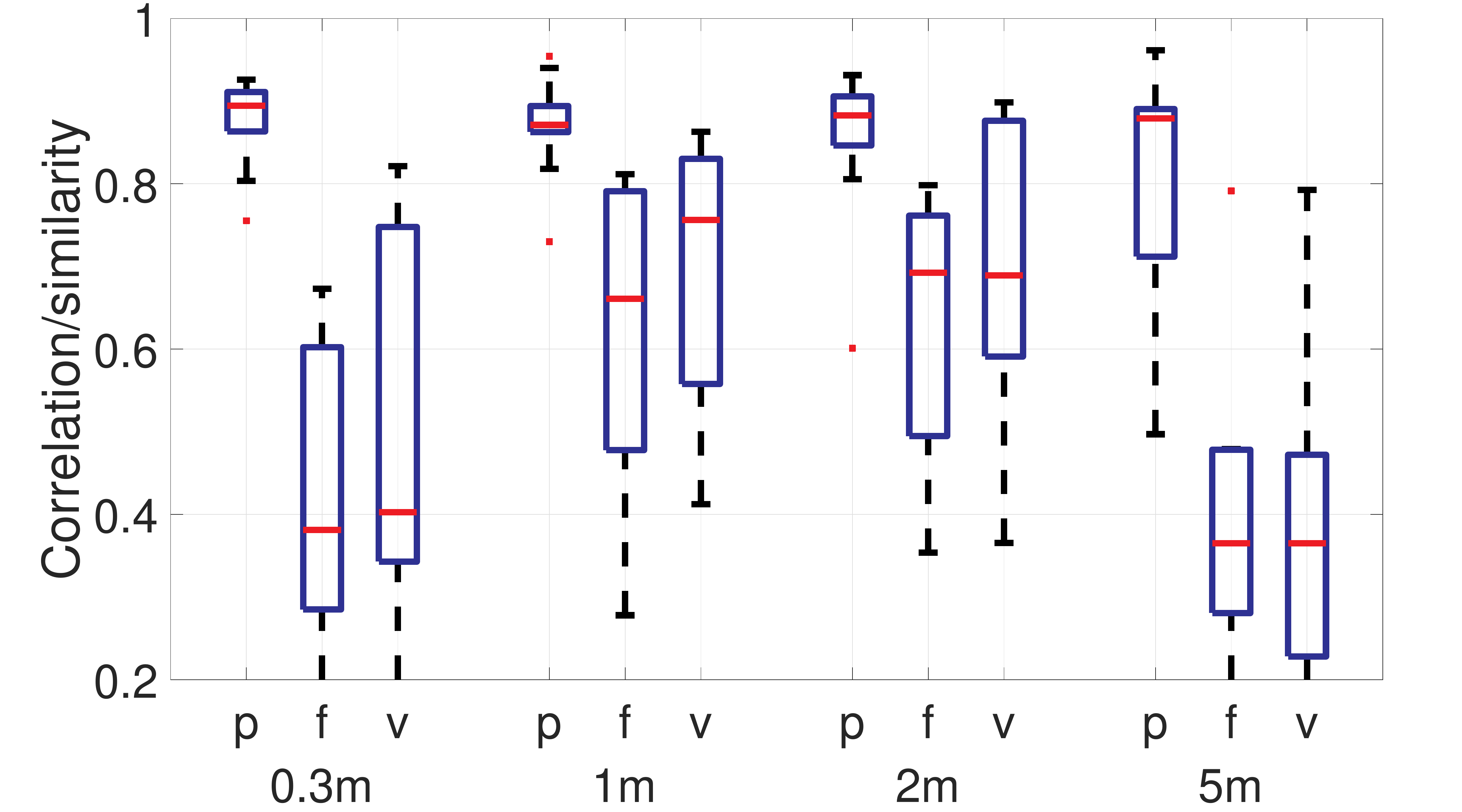}
        \caption{Correlation/similarity.}
        \label{fig:performanceRangesStillC}
    \end{subfigure}
\caption{Performance at different sensing distances. Subjects sit still. Prefixes: p -- Pi-ViMo; f -- FFT; v -- VMD.}
\label{fig:performanceRangesStill}
\end{figure*}

Sensing distance is a major limitation in prior works and the range is typically from 1 to 2 meters in most related work. In this set of experiments, we evaluate the performance of Pi-ViMo in stationary settings. In the experiments, measurements are taken when subjects sit still in four locations, which are 0.3m, 1m, 2m, and 5m away from the radar sensor. Figure~\ref{fig:performanceRangesStill} shows results for all subjects. In Figure~\ref{fig:performanceRangesStillA} -- ~\ref{fig:performanceRangesStillC}, it is evident that Pi-ViMo performs consistently better than baseline methods. Both respiration rate errors and correlations of Pi-ViMo have similar median values and variances at all distances. Even at distance of $5m$, the median respiration and heart rate errors are around $5\%$ and $12\%$. In contrast, the performance of FFT and VMD degrades at distance $0.3m$ and $5m$. 

\subsubsection{Effects of micro-RBMs}
\begin{figure*}[!ht]
    \centering
    \begin{subfigure}[b]{0.325\textwidth}
        \centering
        \includegraphics[width=\textwidth]{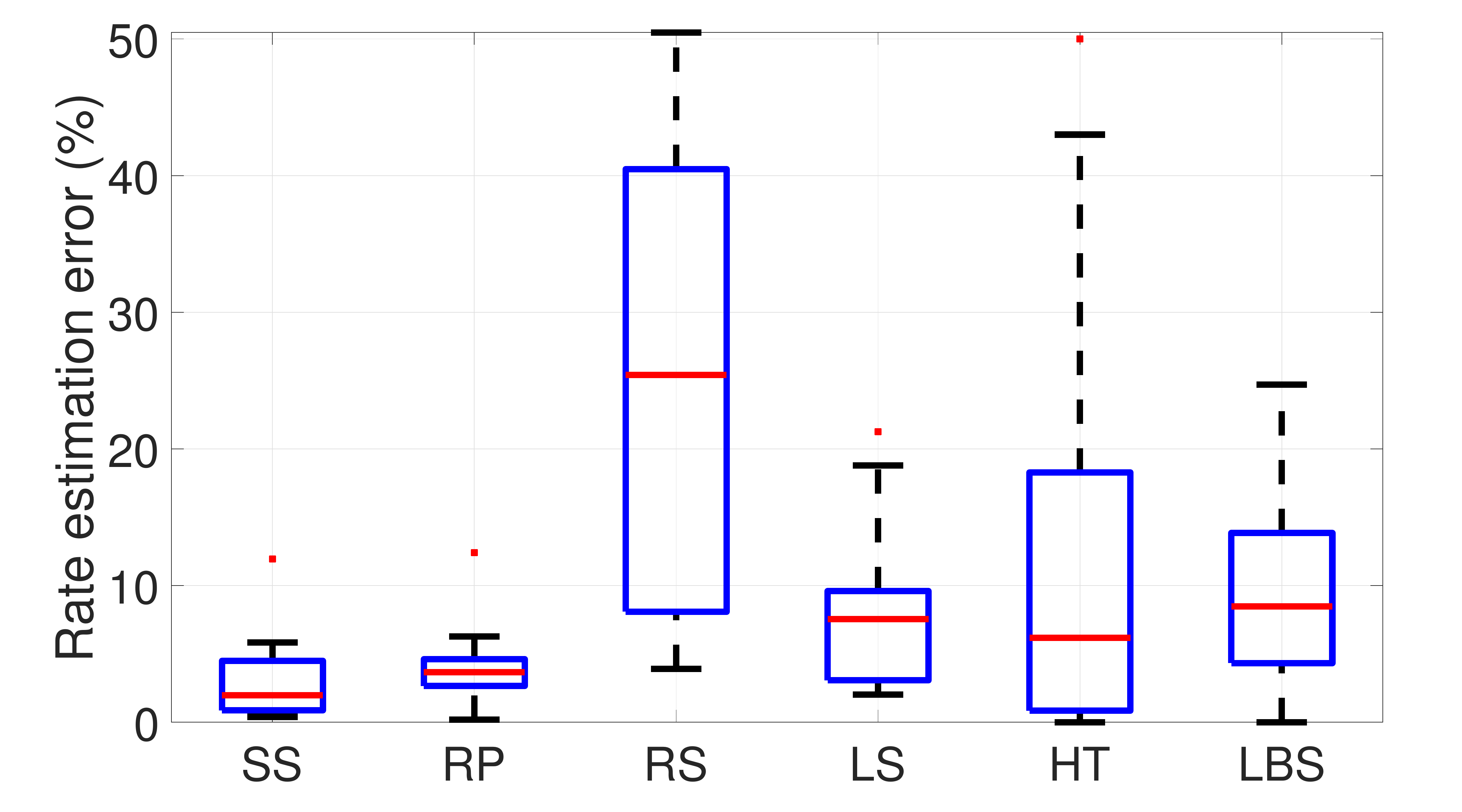}
        \caption{Respiration rate.}
        \label{fig:performanceRBMA}
    \end{subfigure}
    \begin{subfigure}[b]{0.325\textwidth}
        \centering
        \includegraphics[width=\textwidth]{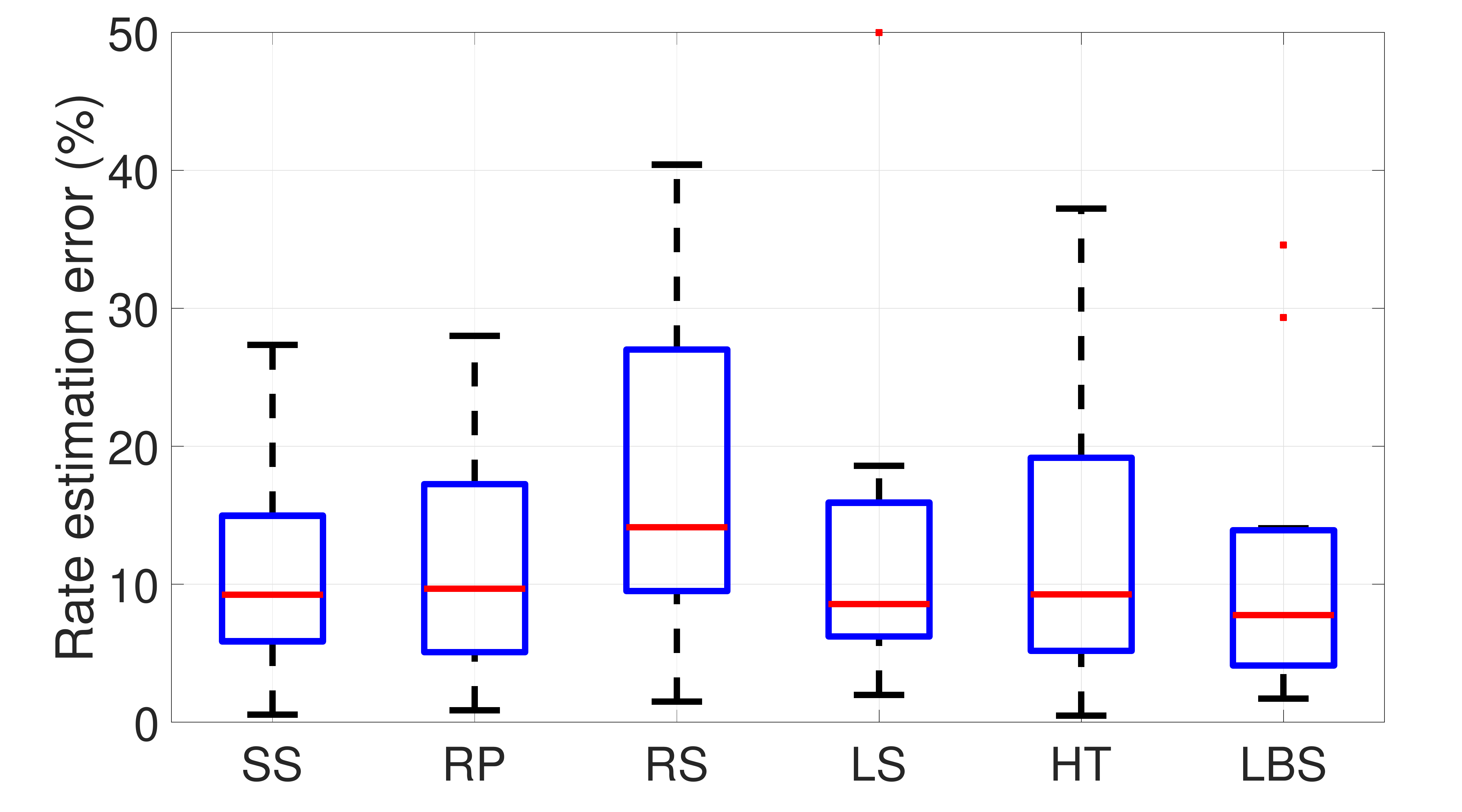}
        \caption{Heartbeat rate.}
        \label{fig:performanceRBMB}
    \end{subfigure}
    \begin{subfigure}[b]{0.325\textwidth}
        \centering
        \includegraphics[width=\textwidth]{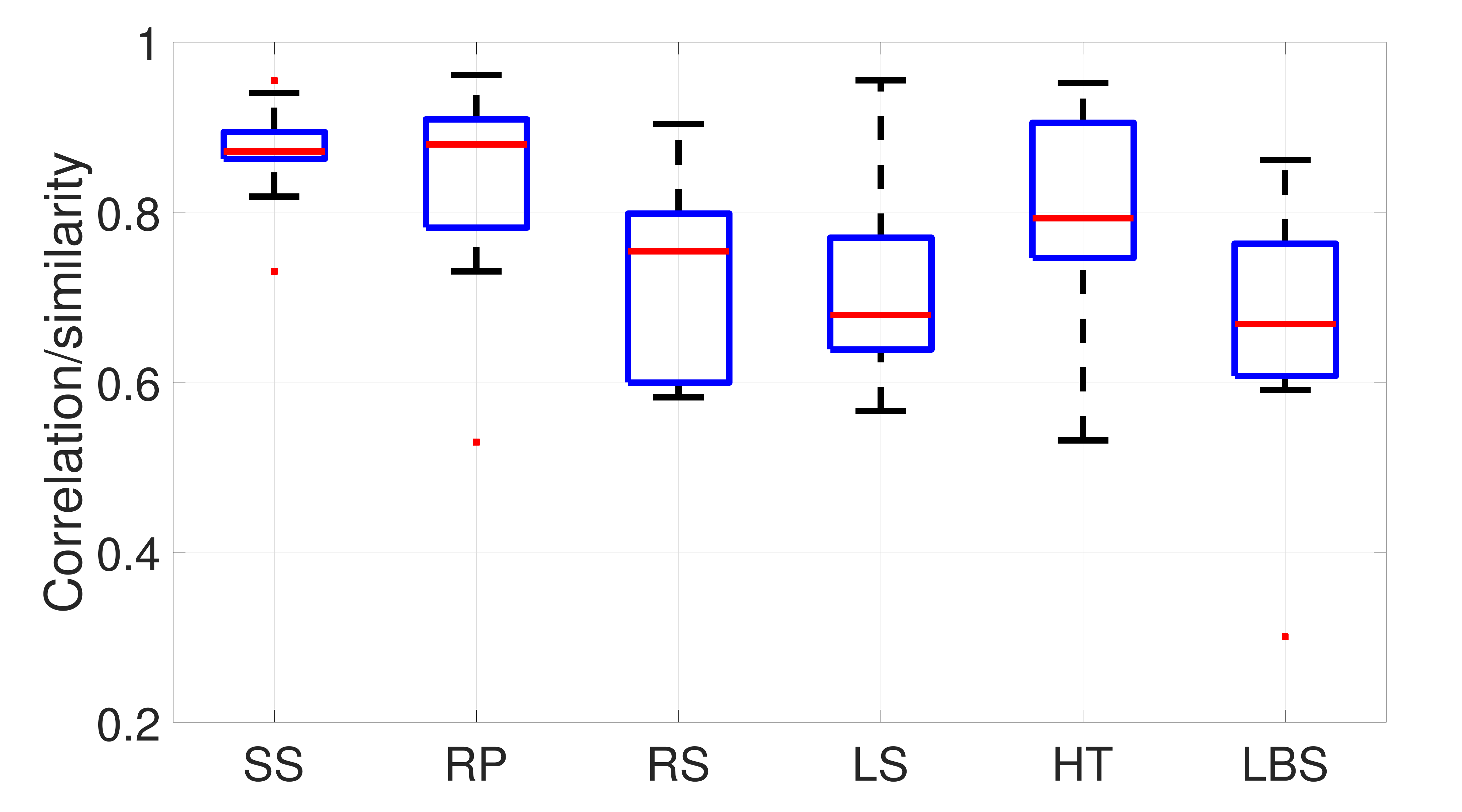}
        \caption{Correlation/similarity.}
        \label{fig:performanceRBMC}
    \end{subfigure}
\caption{Performance of Pi-ViMo under different micro-RBMs. SS: sitting still, RP: relaxed pose, RS: relaxed standing, LS: leg shaking, HT: head turning, LBS: limb stretching.}
\label{fig:performanceRBM}
\end{figure*}

\begin{figure*}[!ht]
    \centering
    \begin{subfigure}[b]{0.325\textwidth}
        \centering
        \includegraphics[width=\textwidth]{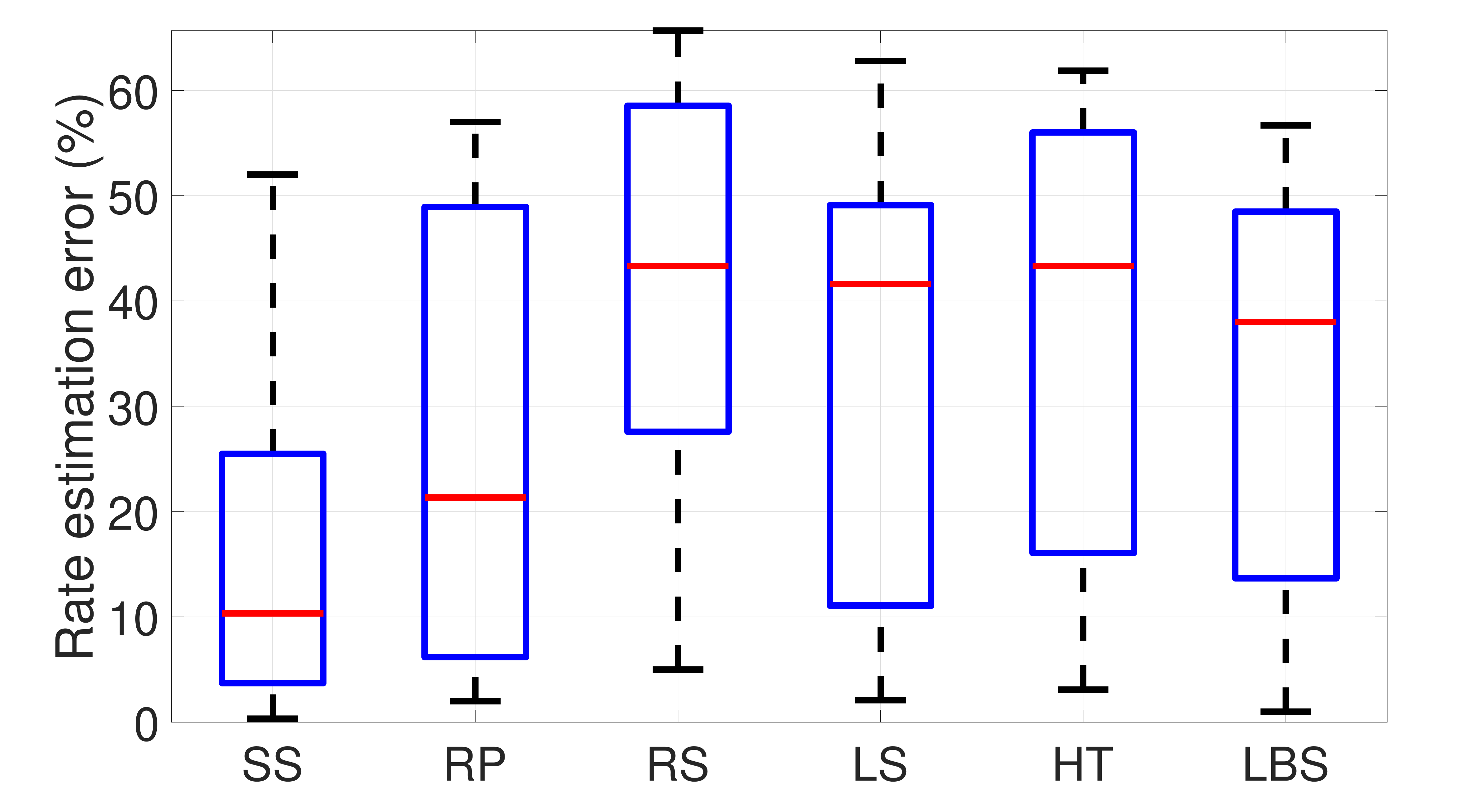}
        \caption{Respiration rate.}
        \label{fig:performanceRBMfftA}
    \end{subfigure}
    \begin{subfigure}[b]{0.325\textwidth}
        \centering
        \includegraphics[width=\textwidth]{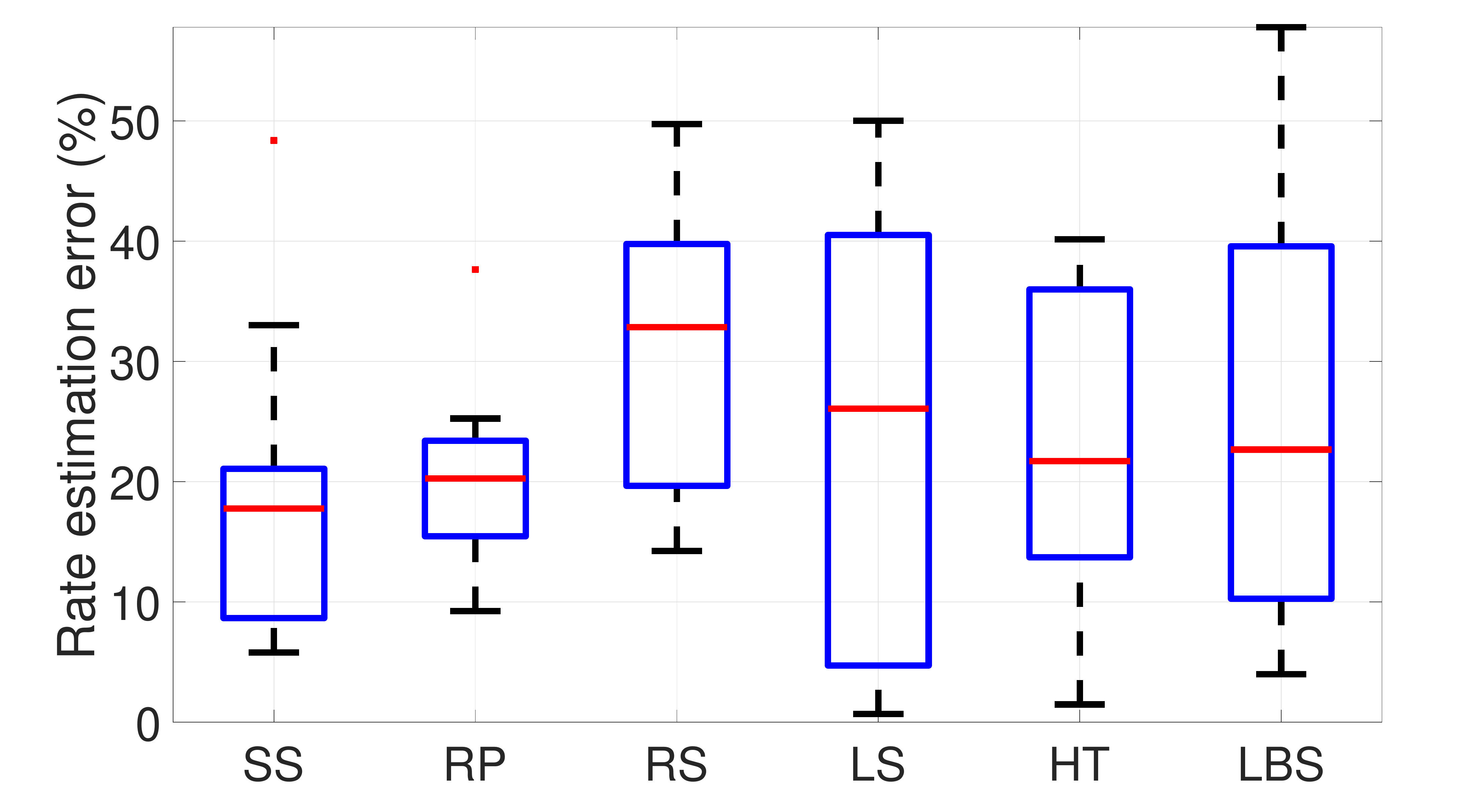}
        \caption{Heartbeat rate.}
        \label{fig:performanceRBMfftB}
    \end{subfigure}
    \begin{subfigure}[b]{0.325\textwidth}
        \centering
        \includegraphics[width=\textwidth]{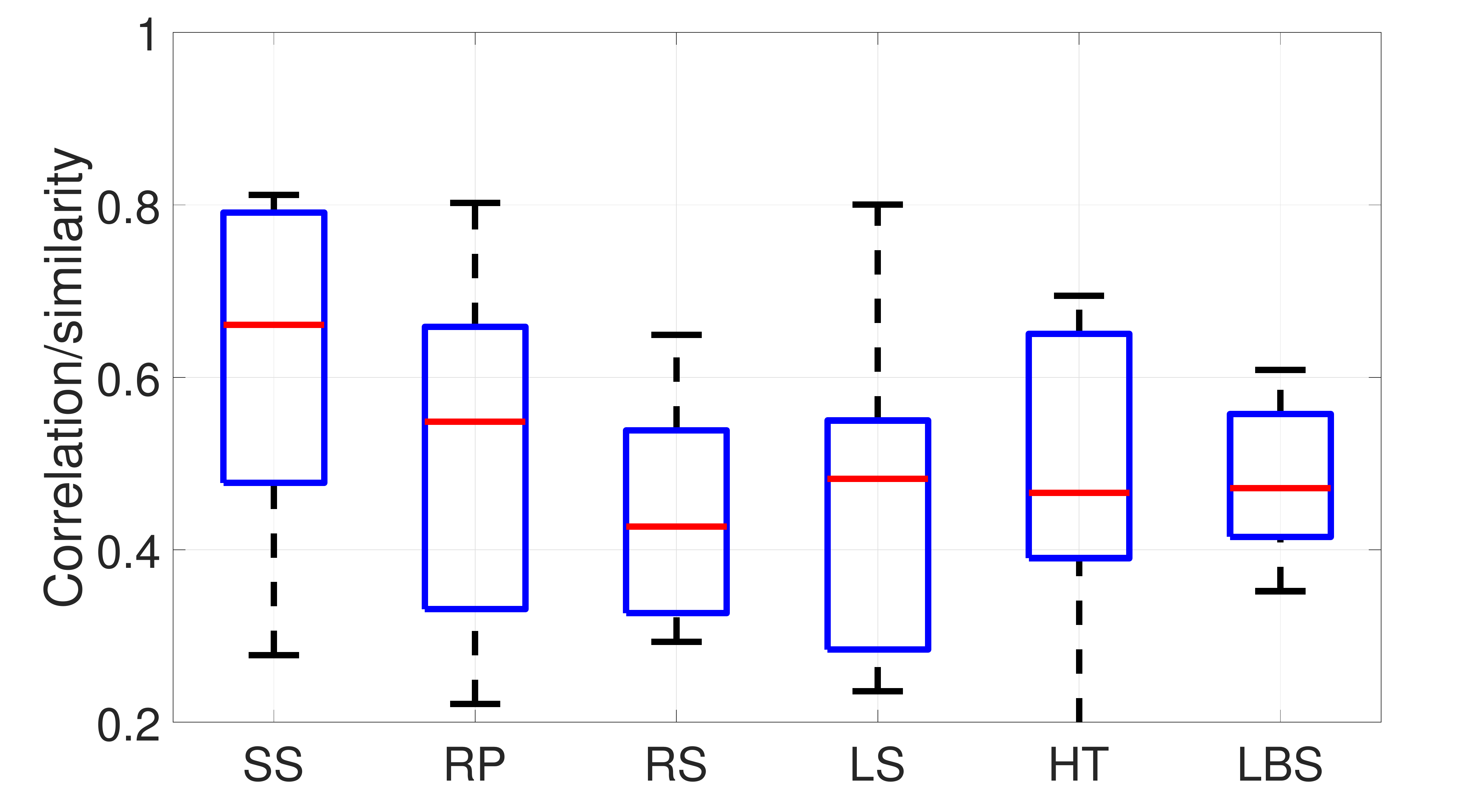}
        \caption{Correlation/similarity.}
        \label{fig:performanceRBMfftC}
    \end{subfigure}
\caption{Performance of baseline FFT under different micro-RBMs. SS: sitting still, RP: relaxed pose, RS: relaxed standing, LS: leg shaking, HT: head turning, LBS: limb stretching.}
\label{fig:performanceRBMfft}
\end{figure*}

\begin{figure*}[!ht]
    \centering
    \begin{subfigure}[b]{0.325\textwidth}
        \centering
        \includegraphics[width=\textwidth]{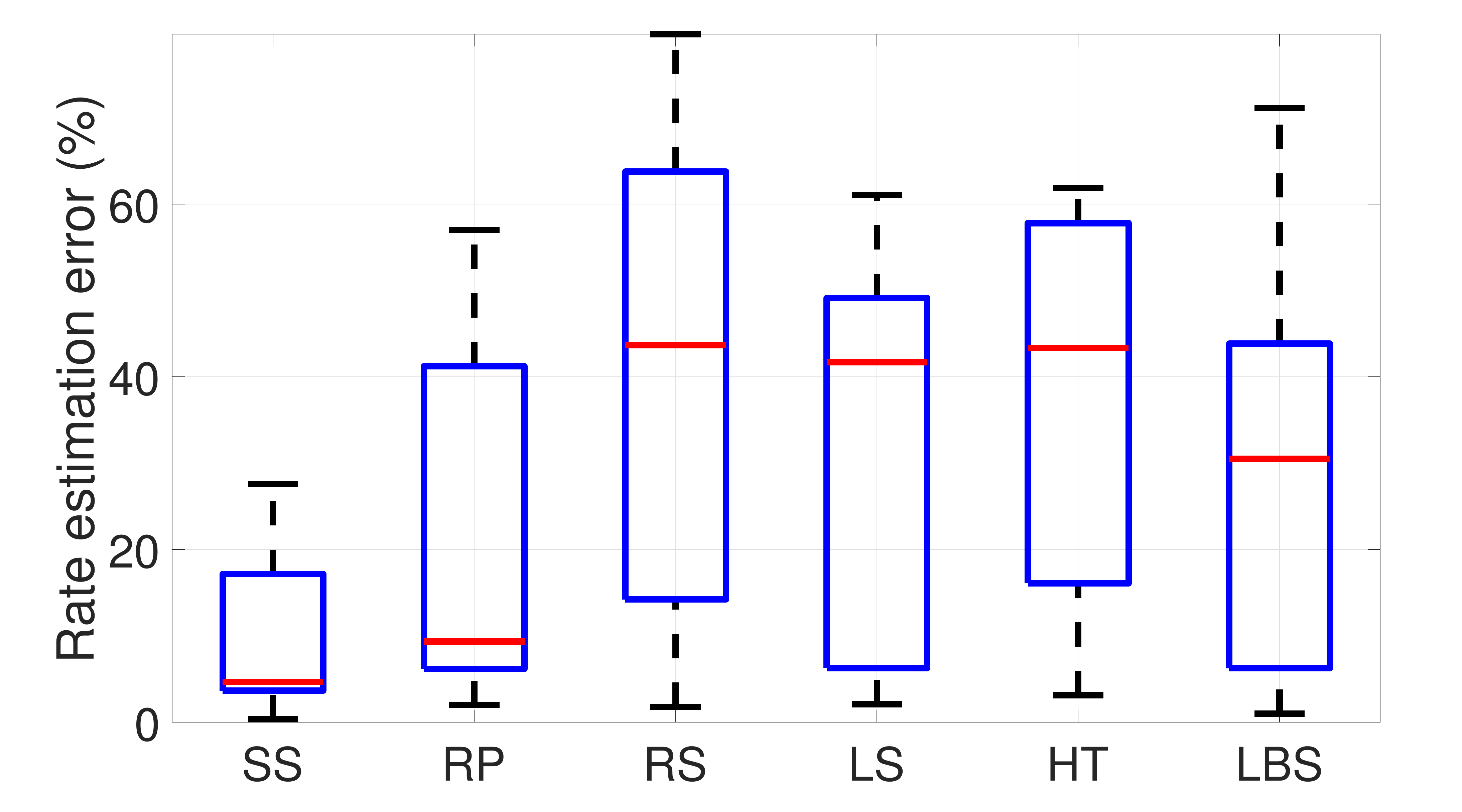}
        \caption{Respiration rate.}
        \label{fig:performanceRBMvmdA}
    \end{subfigure}
    \begin{subfigure}[b]{0.325\textwidth}
        \centering
        \includegraphics[width=\textwidth]{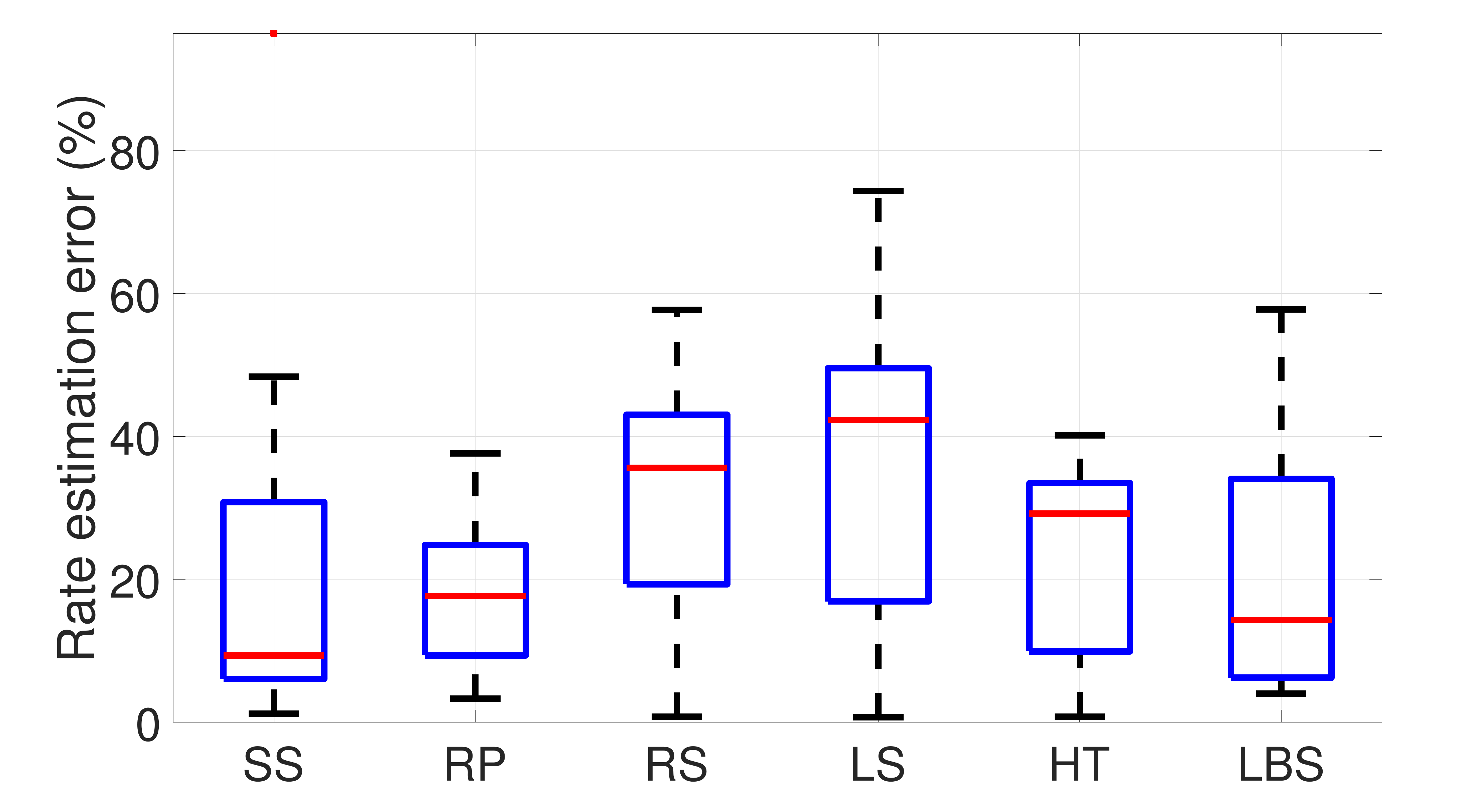}
        \caption{Heartbeat rate.}
        \label{fig:performanceRBMvmdB}
    \end{subfigure}
    \begin{subfigure}[b]{0.325\textwidth}
        \centering
        \includegraphics[width=\textwidth]{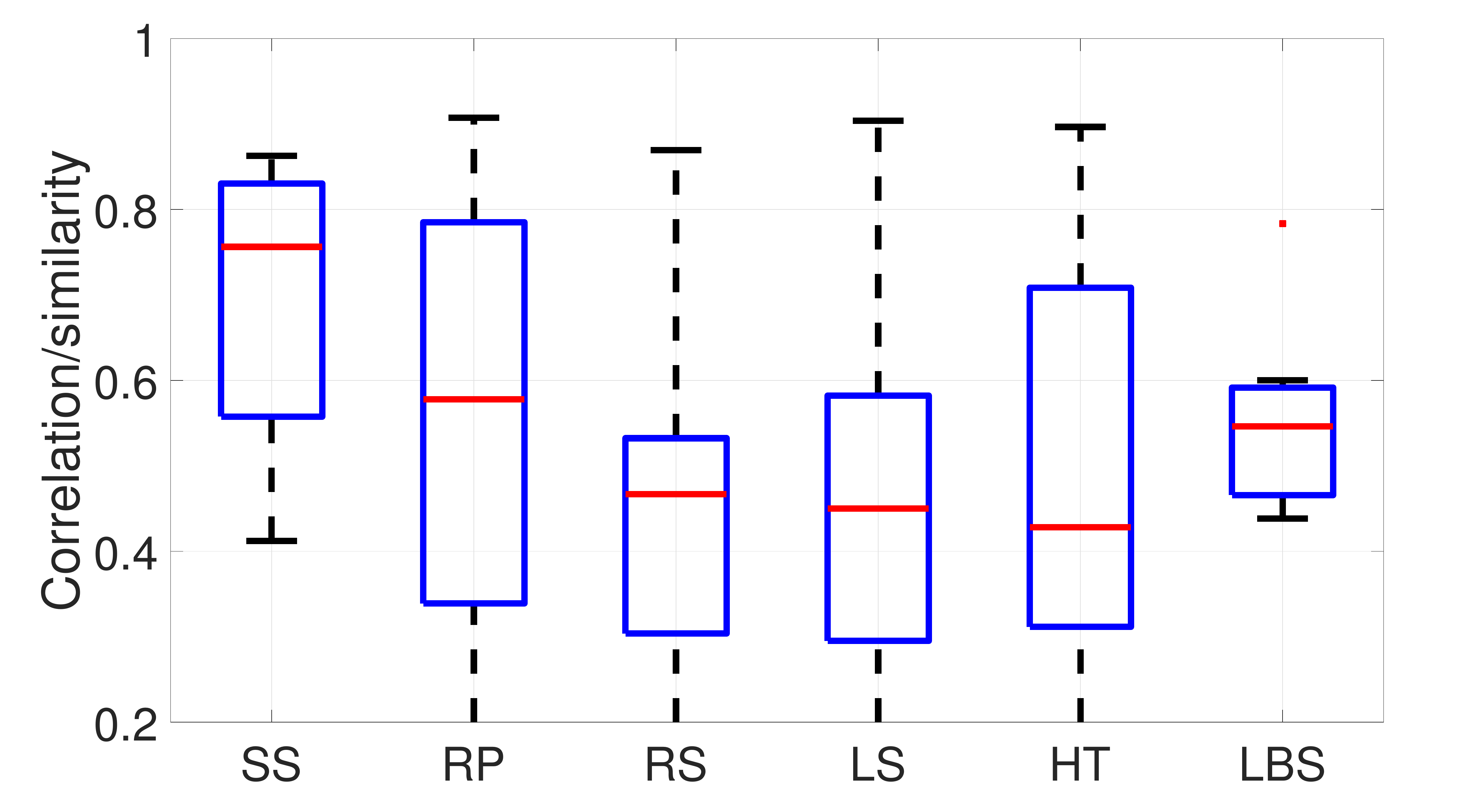}
        \caption{Correlation/similarity.}
        \label{fig:performanceRBMvmdC}
    \end{subfigure}
\caption{Performance of baseline VMD under different micro-RBMs. SS: sitting still, RP: relaxed pose, RS: relaxed standing, LS: leg shaking, HT: head turning, LBS: limb stretching.}
\label{fig:performanceRBMvmd}
\end{figure*}

To investigate the effect of micro-RBM, we conduct experiments with 5 different movements (i.e., RP, RS, LS, HT and LBS). Figure~\ref{fig:performanceRBM} show the results of Pi-ViMo when a subject is at 1 meters away from the radar in the lab environment. For comparison, results from sitting still is also included. From the figure, we observe the accuracy of respiration rate estimations drop with micro-RBMs. Among the five movements, relaxed standing has the most impact on respiration rate estimation followed by leg shaking, head turning, limb stretching and relaxing pose. This is because relaxed standing has the most torso movements due to lack of supports or constraints among all 5 micro-RBMs. The degradation of heart rate estimations due to micro-RBMs is less pronounced. With the exception of leg shaking, the other movements are relatively slow. Leg shaking, though at a high frequency, generates chest displacement patterns different from cardiac activities, and can be suppressed by template matching. 

Figure~\ref{fig:performanceRBMfft} and ~\ref{fig:performanceRBMvmd} show the results from the two baseline methods under the stationary pose and 5 different micro-RBMs. The subjects are at 1 meters away from the radar in the same lab environment. We find micro-RBMs impose similar effects on both the base line methods. The accuracy of respiration rate estimations decrease with micro-RBMs. Compared to the stationary pose, the median respiration errors doubled in relaxing pose which represent the slightest RBM. The median respiration errors in relaxed standing, leg shaking, and head turning are beyond $40\%$ for both baseline methods. Unlike Pi-ViMo, the heart rate estimations also degrade with micro-RBMs. Relaxed standing shows the most impact on heart rate estimations for FFT method and leg shaking has the most impact on heart rate estimations for VMD method. The impact of micro-RBM is significant on both baseline methods.

\subsubsection{Overall Performance}
\begin{table*}
\begin{center}
  \caption{Performance comparison of Pi-ViMo and baselines with average rate errors and average PCCs in a lab environments}
  \label{tab:overallLab}
  \begin{tabular}{c c c c c c c c c c} 
    \toprule
    & \multicolumn{3}{c}{\textbf{Respiration error $\%$}} & \multicolumn{3}{c}{\textbf{Heartbeat error $\%$}} & \multicolumn{3}{c}{\textbf{PCC of respiration}}\\
    & Pi-ViMo & FFT & VMD & Pi-ViMo & FFT & VMD & Pi-ViMo & FFT & VMD \\
    \midrule
    \textbf{Sitting still} & {\bf 6.03} & 22.0 & 20.8 & {\bf 11.9} & 18.4 & 23.7 & {\bf 0.85} & 0.59 & 0.61 \\
    \textbf{RBM} & {\bf 13.5} & 27.9 & 26.7 & {\bf 13.6} & 23.4 & 22.4 & {\bf 0.76} & 0.54 & 0.56 \\
  \bottomrule
\end{tabular}
\end{center}
\bigskip
\end{table*}

Table~\ref{tab:overallLab} summarizes the performance of all methods when subjects sit still and carry out micro-RBMs. The results are the averages of all locations, multiple trials and 5 micro-RBMs. In the stationary case, the average respiration error of Pi-ViMo is $6.3\%$ and the average heart error is $11.9\%$. The average PCC of respiration is $0.85$, which indicates high similarity between the recovered waveform and the ground truth. In comparison, both baseline methods show average errors around $20\%$ and average PCCs around $0.6$. In presence of micro-level RMBs, the average respiration error of Pi-ViMo increases to $13.5\%$, the average heart error increases $2\%$, and the average PCC drops to $0.76$. Similar observations apply to both baseline methods, except the average heart error for VMD decreases with micro-RBMs. Overall, Pi-ViMo significantly outperforms both baseline methods for all three metrics in all scenarios. 

\subsubsection{Ablation study.}
There are two main components in Pi-ViMo: the MSP model and the template matching algorithm. The MSP model makes use of the vibration information in neighboring range bins and is able to eliminate the drawbacks of selecting single range bin. The template matching algorithm can mitigate the effects of interfering motions and weak thermal noise. In order to investigate their respective contributions in improving vital sign estimation compared to baseline algorithms, we conduct an ablation study and present the experiment results in this section.

\begin{figure*}[!ht]
    \centering
    \begin{subfigure}[b]{0.325\textwidth}
        \centering
        \includegraphics[width=\textwidth]{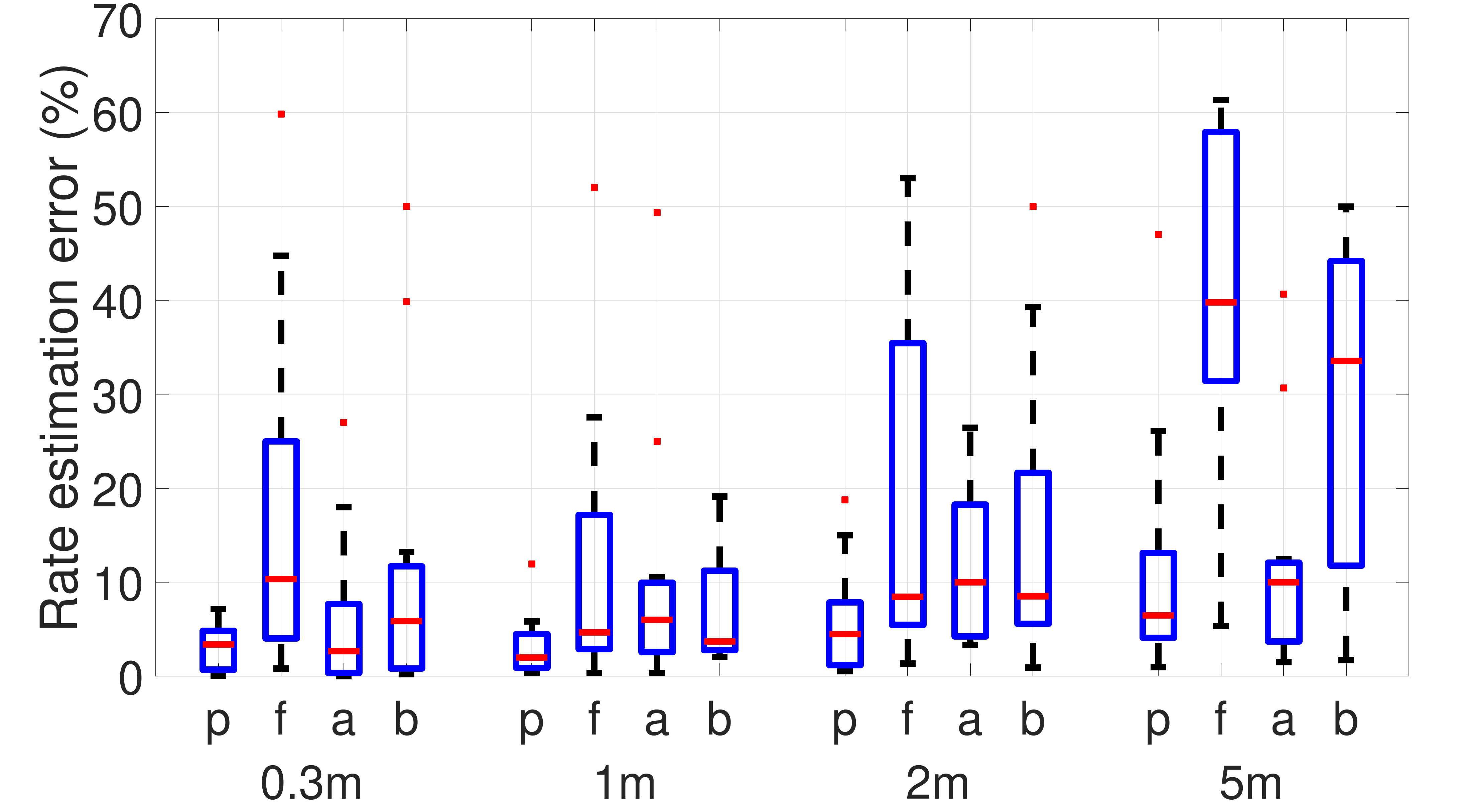}
        \caption{Respiration rate.}
        \label{fig:performanceRangesStillabA}
    \end{subfigure}
    \begin{subfigure}[b]{0.325\textwidth}
        \centering
        \includegraphics[width=\textwidth]{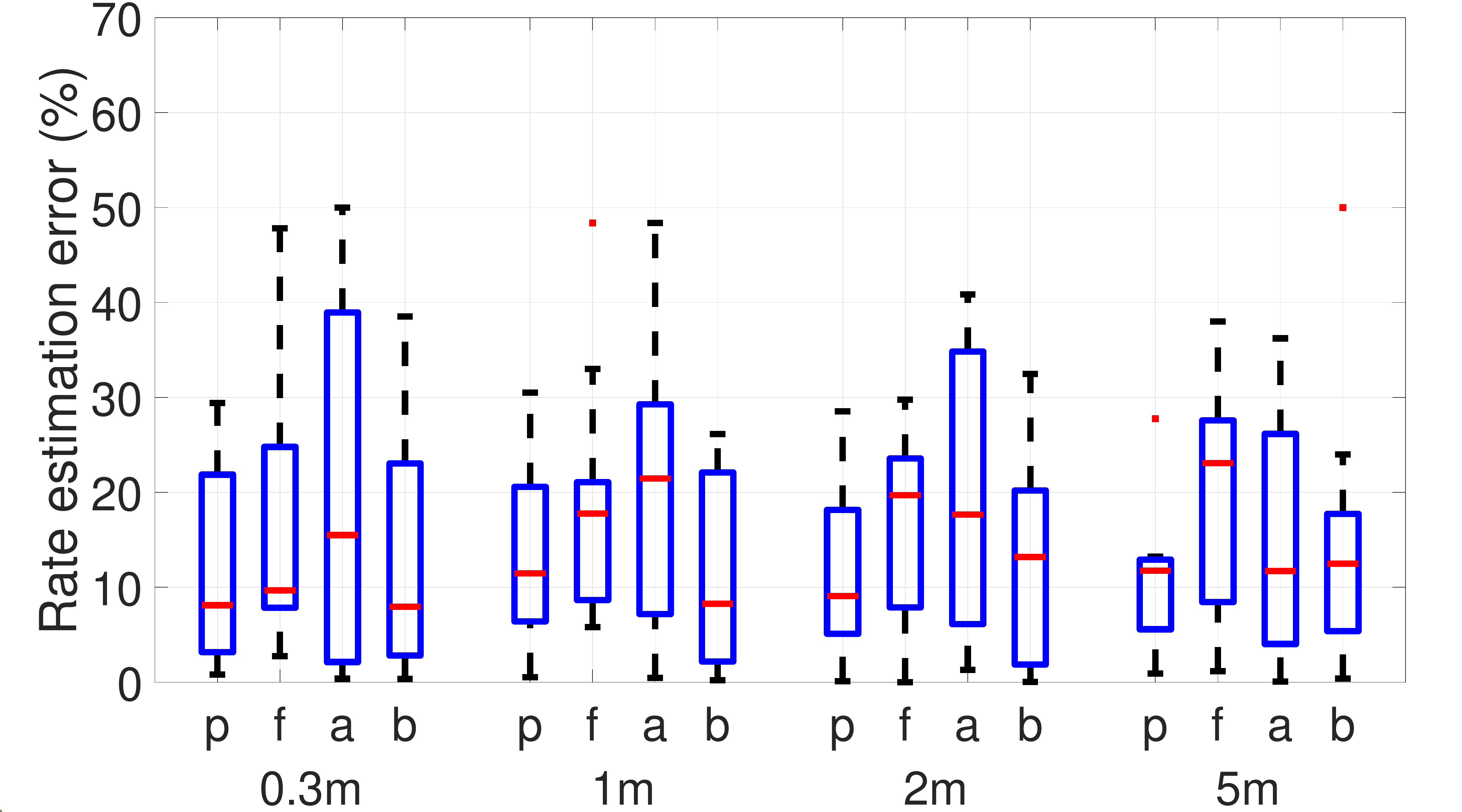}
        \caption{Heartbeat rate.}
        \label{fig:performanceRangesStillabB}
    \end{subfigure}
    \begin{subfigure}[b]{0.325\textwidth}
        \centering
        \includegraphics[width=\textwidth]{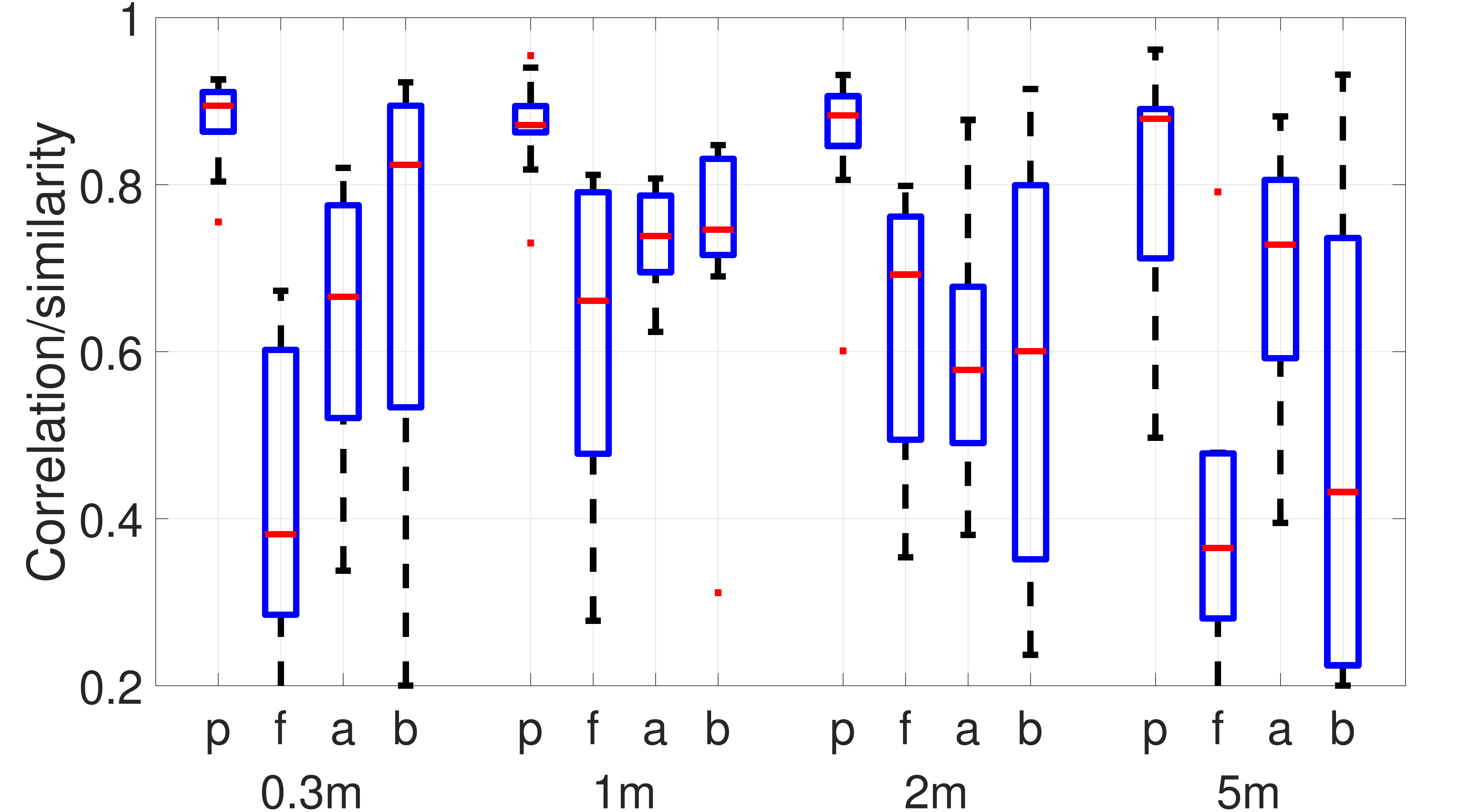}
        \caption{Correlation/similarity.}
        \label{fig:performanceRangesStillabC}
    \end{subfigure}
\caption{Performance at different sensing distances. Subjects sit still. Prefixes: p -- Pi-ViMo; f -- FFT; a -- MSP+FFT; b -- single-bin+TM.}
\label{fig:performanceRangesStillab}
\end{figure*}

We compare four approaches: Pi-ViMo with both MSP and template matching, MSP and FFT, single range bin and template matching, and single range bin and FFT. First, we evaluate the performances of different approaches when the radar is at different sensing distances in stationary settings. In Figure~\ref{fig:performanceRangesStillabA} -- ~\ref{fig:performanceRangesStillabC}, it is clearly {\it Pi-ViMo} performs the best and {\it single-bin+FFT} performs the worst. {\it MSP+FFT} shows similar performance gains as {\it single-bin+TM} at $1m$ and $2m$ distances over {\it single-bin+FFT}. At the close sensing distance of $0.3m$ and the far sensing distance of $5m$, the improvements with {\it MSP+FFT} are significant, which indicates the advantages of the MSP model at these sensing ranges. It is evident that with the adoption of MSP, the resulting estimations are robust across different sensing distances.     

Table~\ref{tab:overallLabab} summarizes the performance of the four approaches averaged over all locations, multiple trials and 5 micro-RBMs. When subjects sit still, the introduction of MSP model outperforms schemes with template matching alone by $6\%$ in both respiration and heartbeat errors. When subjects carry out micro-RBMs, the incorporation of MSP model over the baseline {\it single-bin+FFT} is less pronounced especially for heart rate estimation. This may be attributed to inclusion of unwanted signals due to body movements. Overall, we observe that both MSP and template matching contribute to more accurate estimations in both sitting still and RBM scenarios. The combination of the two in Pi-ViMo provides the most gain.  

\begin{table*}
\begin{center}
  \caption{Ablation study of Pi-ViMo and baselines with average rate errors and average PCCs in a lab environments.}
  \label{tab:overallLabab}
  \begin{tabular}{c c c c c c c} 
    \toprule
    & \multicolumn{2}{c}{\textbf{Respiration error $\%$}} & \multicolumn{2}{c}{\textbf{Heartbeat error $\%$}} & \multicolumn{2}{c}{\textbf{PCC of respiration}}\\
    & Sitting still & RBM & Sitting still & RBM & Sitting still & RBM   \\
    \midrule
    \textbf{Pi-ViMo} & {\bf 6.03} & {\bf 13.5} & {\bf 11.9} & {\bf 13.6} & {\bf 0.85} & {\bf 0.76} \\
    \textbf{MSP+FFT} & 10.1 & 20.6 & 12.6 & 20.6 & 0.66 & 0.61 \\
    \textbf{single-bin+TM} & 15.8 & 21.3 & 19.1 & 18.3 & 0.62 & 0.58 \\
    \textbf{single-bin+FFT} & 22.0 & 27.9 & 18.4 & 23.4 & 0.59 & 0.54 \\
  \bottomrule
\end{tabular}
\end{center}
\bigskip
\end{table*}

\subsubsection{Evaluations in the lounge room.}
To further study the robustness of Pi-ViMo, we conduct experiments in the lounge room.
Table~\ref{tab:overallLounge} summarizes the performance of three methods at three different locations, averaged over all the movements and the stationary case. In Pi-ViMo, the average respiration error is around $7\%$ and the average heart error is $13\%$ for subjects sitting in a chair beside the table. The corresponding average errors are $12.5\%$ and $12.9\%$ when subjects sit in the sofa. The average errors are $13.9\%$ for respiration and $16\%$ for heartbeat when subjects stand near the counter. The  average correlations in the three cases are comparable. The standing case gives the worst performance due to longer distance to the radar and large torso movements during standing; whereas sitting beside the table achieves the best performance. The results are consistent with those in the lab settings. It is interesting to observe that the performance for subjects sitting in the sofa is worse than that from sitting in a chair though both positions are at the same distance to the radar. This may be attributed to partial blockage of abdomen areas by one's legs in the radar's FOV. 

The performance of the two baseline methods are similar to each other. Both perform best when subjects sit at the table and the worst when subjects sit in the sofa. Overall, Pi-ViMo significantly outperforms both  methods at all  locations with or without micro-RBMs.
\begin{table*}
\begin{center}
  \caption{ Performance comparison of Pi-ViMo and baseline methods with average rate errors and average PCCs in the lounge room. All metrics are averages over multiple trials under different movements.}
  \label{tab:overallLounge}
  \begin{tabular}{c c c c c c c c c c} 
    \toprule
    & \multicolumn{3}{c}{\textbf{Respiration error $\%$}} & \multicolumn{3}{c}{\textbf{Heartbeat error $\%$}} & \multicolumn{3}{c}{\textbf{PCC of respiration}}\\
    & sofa & counter & table & sofa & counter & table & sofa & counter & table \\
    \midrule
    \textbf{Pi-ViMo} & {\bf 12.5} & {\bf 13.9} & {\bf 7.3} & {\bf 12.9} & {\bf 16.1} & {\bf 13.2} & {\bf 0.82} & {\bf 0.76} & {\bf 0.85} \\
    \textbf{FFT} & 26.2 & 26.4 & 20.9 & 22.9 & 21.3 & 24.3 & 0.56 & 0.60 & 0.64 \\
    \textbf{VMD} & 28.3 & 26.4 & 19.5 & 23.8 & 21.5 & 21.0 & 0.55 & 0.60 & 0.65 \\
  \bottomrule
\end{tabular}
\end{center}
\bigskip
\end{table*}

\section{Related Work}
\label{sec:Relatedwork}
Continuous monitoring of human vital signs can provide information in tracking health conditions, emotion recognition, sleep monitoring and predicting or providing early warnings on emergencies such as critical dizziness, arrhythmias, apnea, and so on~\cite{AdamVB:2019,DavidB:2015,GeorgeAM:2019}. 

Radar-based systems for vital sign detection are developed based on measuring movements of chest-wall area due to respiration and cardiac activities. Specifically, the chest-wall displacement modulates the received RF signals in both the amplitude and the phase. The range map is discretized by the range resolution and each range sample point is called a ``range bin''. 
Most existing work selects a single range bin to represent chest-wall movements from radar signals by certain criteria. For example, in~\cite{JM:2019}, Muñoz-Ferreras {\it et al.} chose a single rang-bin with the maximum energy within an observation window; In~\cite{MostafaA:2019}, the range bin associated with the maximum vibration energy from the unwrapped phase information is selected; Choi {\it et al.} incorporated amplitude information by choosing one range bin according to the magnitude-phase coherency (MPC) index in~\cite{Ho-IkC:2020}. However, the underlying assumption of single point target is inappropriate when sensing distance is small, and thus the resulting estimation accuracy decreases dramatically in case of mis-selection.

In general, the amplitude of RF signals suffer more from severe fading and distortion than phase. Therefore, most radar-based vital sign detection solutions rely on phase information~\cite{AdeelA:2018,MostafaA:2019,Ho-IkC:2020}. The phases of reflection signals are extracted, unwrapped and interpreted as changes in range, proportional to the mechanical motions induced by respiration and heartbeats. In \cite{XingzheS:2020,ZheC:2021}, both phase and amplitude features are utilized in neural network models for vital signal monitoring. However, in absence of proper procedures to handle fading and interference, generalizability of such models requires further investigation. 

To separate vital signs from noise or interference, and to decompose respiration and heartbeat from  estimated chest-wall movements, various signal processing algorithms have been applied.In ~\cite{AdeelA:2018}, Ahmad {\it et al.} used band-pass filter and FFT to extract vital signs, which are advantageous in fast computational speed. Other prior works have applied classical signal decomposition methods, such as EMD~\cite{NordenEH:1998}, VMD~\cite{KonstantinD:2013}, and their variances~\cite{MariaET:2011,ZhenyuL:2020,DmytroI:2015,ArindamR:2021,FYWang:2022,TianyueZ:2020}. The decomposed IMFs are fundamentally of very narrow bandwidth and can not directly represent vital signs. Therefore, it is necessary to select and combine several IMFs to synthesize respiration and heartbeat signals, which remains an open problem. In another direction, dynamic models, widely used for stochastic random process and time series analysis, are adopted and proposed in~\cite{HyunjaeL:2019} with an AR model, in~\cite{ZengyangM:2020} with a Markov chain model, and in~\cite{GuigengS:2021} with a extended Kalman filter. Those models are useful to predict vital signs in a short time period, but they need to be adapted to handle long time period and differences among individuals. Recently, Unsoo Ha {\it et al.} in~\cite{UnsooH:2020} proposed a complex processing chain with template matching, beamforming, and neural-nets for seismocardiogram signals. They targeted detailed heartbeat waveforms only and reduced radar signal observation windows to one second. In~\cite{ZheC:2021,TianyueZ:2021}, motion robust vital sign extractions were achieved by applying a deep contrastive neural-net and a novel encoder-decoder model. They focus on a challenging scenario with presence of major body movements. Their system can handle eight pre-defined types of movements but is restricted to a short sensing distance from $0.6$ to $2$ meters. Another work in~\cite{FYWang:2022} applies vital sign monitoring in a vehicle setting, and propose to use range cell migration techniques to compensate motions during driving. Their work is an extension of~\cite{TianyueZ:2020}.

Template matching has been used in radar-based vital sign monitoring. The templates in use can be categorized into data driven templates and bio-mechanical templates. Data driven templates are learned from measurement data and thus account for subject differences and time variances. However, training stages are required to generate the templates. In~\cite{Khamis:2020} a $10s$ calibration period is required before each test, to capture a subject’s respiration pattern. Unsoo Ha {\it et al.} in~\cite{UnsooH:2020} utilized a 1D CNN to learn a template "T" as its convolution filter to estimate heartbeat intervals and perform data segmentation accordingly. The 1D CNN based template method extracts the periodicity of the dominant periodic signal as defined by Equation (1) in~\cite{UnsooH:2020}. When the heart wave is dominated by a stronger periodic interference, such as the harmonic components of respiration and RBMs, the extracted periodicity can be incorrect. Therefore, multi-antenna beamforming and other preprocessing techniques are required to remove such strong interferences. Bio-mechanical templates, in contrast, utilize knowledge regarding the bio-mechanical process underlying behind vital activities. The main limitation is that such templates do not capture subtle individual differences and can result in large estimation errors in the case of abnormal vital sign patterns. Will {\it et al.} in~\cite{ChrisW:2016} used the template of a single heartbeat, and applied the cross-correlation for heart rate estimation. To account for subjects variability and measurement data from different point locations on the body, the same authors extended their work to utilize five types of single heartbeat templates with 10 templates for each type in ~\cite{ChristophW:2017}. As a consequence, multiple subjects and sufficient training time are required to ensure that all the templates are acquired.

In this work, we choose the bio-mechanical templates for both respiration and heartbeat as discussed in Section~\ref{sec:Modelsandtemplates}. Chest-wall movements are modeled as the sum of respiration and heartbeat waveforms from the respective templates and a second order non-linear term representing the coupling between the two. By finding the control parameters for stretches and biases in both amplitude and time domain, our method adapts simple templates to model individual differences without the need for training.

Key features of state-of-the-art vital sign monitoring methods and Pi-ViMo are summarized in Table~\ref{tab:summary}.

\begin{table*}[t]
\tiny
\begin{minipage}{\textwidth}
  \begin{center}
    \caption{A Comparison of Existing mmWave-based Solutions to Vital Signs.}
    \label{tab:summary}
    \begin{tabular}{|c|c|c|c|c|c|c|}
    \hline
 	   \textbf{Reference} & \textbf{Key Technologies} & \textbf{Range Map} & \textbf{\# of Subjects} & \textbf{Phase/amplitude} & \textbf{Range} & \textbf{Motion} \\
 	\hline
 		\cite{MostafaA:2019}   & BPF, FFT, model-based  & single range bin & single person & phase & $1\sim 2m$ & static \\
 		\hline
 		\cite{AdeelA:2018,JM:2019}   & BPF, FFT, model-based  & single range bin   & two people & phase & $1\sim 2m$ & static \\
 	   \hline
 	   \cite{Ho-IkC:2020}   & MPC, model-based  & single range bin   & single person & phase,amplitude & $1\sim 2m$ & static \\
 	   \hline
 	   \cite{HyunjaeL:2019} & MUSIC, AR, model-based  & range integration 30cm   & two people & phase & $1\sim 3m$ & static \\
 	   \hline
 	   \cite{ZengyangM:2020} & HMM, model-based  & single range bin   & two people & phase & $1\sim 3m$ & RBMs \\
 	   \hline
 	   \cite{UnsooH:2020} & BF, TM, DNN, data-driven  & single range bin   & single person & phase & $0.3m$ & static \\
 	   \hline
 	   \cite{TianyueZ:2020,FYWang:2022} & MC, MS-VMD, data-driven  & multiple range bins   & four people & phase & $0.5\sim 1m$ & driving in a vehicle \\
 	   \hline
 	   \cite{ZheC:2021,TianyueZ:2021} & BF, DNN, data-driven  & range $\pm 30cm$   & single person & phase, amplitude & $0.5\sim 2m$ & eight motions \\
 	   \hline\hline
 	   {\it Pi-ViMo} & {\it MSP model, CC, TM}  & {\it multiple range bins}   & {\it single person} & {\it phase} & {\it any location} & {\it micro-RBMs} \\
 	    & {\it model-based}  &    &  &  & {\it within radar FOV} &  \\
 	\hline
    \end{tabular}
  \end{center}
  \bigskip
  {\footnotesize BPF: bandpass filter; MPC: magnitude-phase coherency; BF: beamforming; TM: template matching; DNN: deep neural network; HMM: hidden Markov model; RBM: random body movement; MC: motion compensation.}
\end{minipage}
\end{table*}

\section{Discussion}
\label{sec:Discussion}
In this section, we discuss limitations in this work. This work targets scenarios where the subject sits at different positions within a radar's FoV. For a single radar board in an indoor environment, the typical FoV covers 120 degrees and the farthest distance is limited to 6 meters. Moreover, though subjects can sit with arbitrary postures, their chest should face the radar sensor. Pi-ViMo cannot handle large RBMs, locomotions such as walking or jogging or other activities that causes significant body movements. 

Current set-ups target vital signs monitoring of a single subject using a single transceiver pair. Multi-antenna beamforming and steering can improve signal quality and allow vital sign monitoring of multiple subjects. Pi-ViMo can be easily extended to handle multiple subjects as long as the subjects are sufficiently separated in range and angle.

Template matching optimization in Section~\ref{sec:Modelsandtemplates} requires that the radar signals contain at least one complete cycle of respiration and heartbeat waveforms. A complete cycle of respiration is typically around $4$ to $6$ seconds, and spans over multiple heartbeat cycles. The time window of the radar signals in this work is set to be $15$ seconds, which covers $2$ to $3$ respiration cycles. The current implementation of Pi-ViMo is more suitable for off-line processing since the computation time of the template matching optimization on average takes $4.3$ seconds on a host PC with an i7-8700(3.20GHz) processor, 16GB RAM.
 
\section{Conclusion}
\label{sec:Conclusions}
In this paper, we proposed Pi-ViMo, a physiology-inspired robust vital sign monitoring solutions. Pi-ViMo is fundamentally a white-box approach through the modeling of multi-point scattering from human chests and the bio-mechanical processes of respiration and cardiac activities. Multi-subject experimental evaluations in two different environments demonstrated the superior performance of Pi-ViMo over baseline methods.

For future work, we will investigate the incorporation of range-angle maps from multi-antenna systems for multi-subject vital signal monitoring. Another direction is to further improve the performance of Pi-ViMo under micro-RBMs and RBMs. Significant torso movements can cause large errors in respiration monitoring, and impose a more challenging scenario in vital sign monitoring. One possible solution is to combine data-driven approaches such as neural networks and model-driven approaches such as template matching. 

\bibliographystyle{unsrt}  
\bibliography{references}  

\begin{thebibliography}{10}

\bibitem{JM:2019}
José-María Muñoz-Ferreras, Jing Wang, Zhengyu Peng, Changzhi Li, and Roberto
  Gómez-García.
\newblock Fmcw-radar-based vital-sign monitoring of multiple patients.
\newblock In {\em MTT-S International Microwave Biomedical Conference
  (IMBioC)}, pages 1--3. IEEE, May 2019.

\bibitem{MostafaA:2019}
Mostafa Alizadeh, George Shaker, João Carlos Martins~De Almeida,
  Plinio~Pelegrini Morita, and Safeddin Safavi-Naeini.
\newblock Remote monitoring of human vital signs using mm-wave fmcw radar.
\newblock {\em IEEE Access}, 7:54958--54968, April 2019.

\bibitem{Ho-IkC:2020}
Ho-Ik Choi, Heemang Song, and Hyun-Chool Shin.
\newblock Target range selection of fmcw radar for accurate vital information
  extraction.
\newblock {\em IEEE Access}, 9:1261--1270, December 2020.

\bibitem{Liang:2018}
Xiaolin Liang, Hao Zhang, Shengbo Ye, Guangyou Fang, and T.~Aaron Gulliver.
\newblock Improved denoising method for through-wall vital sign detection using
  uwb impulse radar.
\newblock {\em Digital Signal Processing}, 74:72--93, March 2018.

\bibitem{He:2017}
Mi~He, Yongjian Nian, and Yushun Gong.
\newblock Novel signal processing method for vital sign monitoring using fmcw
  radar.
\newblock {\em Biomedical Signal Processing and Control}, 33:335--345, March
  2017.

\bibitem{NordenEH:1998}
Norden~E. Huang, Zheng Shen, Steven~R. Long, Manli~C. Wu, Hsing~H. Shih, Quanan
  Zheng, Nai-Chyuan Yen, Chi~Chao Tung, and Henry~H. Liu.
\newblock The empirical mode decomposition and the hilbert spectrum for
  nonlinear and non-stationary time series analysis.
\newblock {\em Mathematical, Physical and Engineering Sciences},
  454(1971):903--995, March 1998.

\bibitem{KonstantinD:2013}
Konstantin Dragomiretskiy and Dominique Zosso.
\newblock Variational mode decomposition.
\newblock {\em IEEE Transactions on Signal Processing}, 62(3):531--544,
  November 2013.

\bibitem{MariaET:2011}
María~E. Torres, Marcelo~A. Colominas, Gastón Schlotthauer, and Patrick
  Flandrin.
\newblock A complete ensemble empirical mode decomposition with adaptive noise.
\newblock In {\em 2011 IEEE International Conference on Acoustics, Speech and
  Signal Processing (ICASSP)}, pages 4144--4147. IEEE, May 2011.

\bibitem{DmytroI:2015}
Dmytro Iatsenko, Peter V.~E. McClintock, and Aneta Stefanovska.
\newblock Nonlinear mode decomposition: A noise-robust, adaptive decomposition
  method.
\newblock {\em Phys. Rev. E}, 92(3):032916, September 2015.

\bibitem{ArindamR:2021}
Arindam Ray, Anwesha Khasnobish, Smriti Rani, Arijit Chowdhury, and Tapas
  Chakravarty.
\newblock Exploration of mode decomposition for concurrent cardiopulmonary
  monitoring using dual radar.
\newblock In {\em 2020 28th European Signal Processing Conference (EUSIPCO)},
  pages 1140--1144. IEEE, January 2021.

\bibitem{ZhenyuL:2020}
Zhenyu Liu, Yongan Kong, Xin Zhang, Jiayan Wu, and Wei Lu.
\newblock Vital sign extraction in the presence of radar mutual interference.
\newblock {\em IEEE Signal Processing Letters}, 27:1745--1749, September 2020.

\bibitem{FYWang:2022}
Fengyu Wang, Xiaolu Zeng, Chenshu Wu, Beibei Wang, and K.~J.~Ray Liu.
\newblock Driver vital signs monitoring using millimeter wave radio.
\newblock {\em IEEE Internet of Things Journal}, 9(13):11283--11298, 2022.

\bibitem{TianyueZ:2020}
Tianyue Zheng, Z.~Chen, C.~Cai, J.~Luo, and X.~Zhang.
\newblock V2ifi: in-vehicle vital sign monitoring via compact rf sensing.
\newblock {\em Proc. ACM Interact. Mob. Wearable Ubiquitous Technologies},
  4(2):1--27, June 2020.

\bibitem{ZheC:2021}
Zhe Chen, Tianyue Zheng, Cai Chao, and Jun Luo.
\newblock Movi-fi: Motion-robust vital signs waveform recovery via deep
  interpreted rf sensing.
\newblock In {\em Proceedings of the 27st Annual International Conference on
  Mobile Computing and Networking}. MobiCom '21, December 2021.

\bibitem{TianyueZ:2021}
Tianyue Zheng, Zhe Chen, S.~Zhang, C.~Cai, and J.~Luo.
\newblock More-fi: Motion-robust and fine-grained respiration monitoring via
  deep-learning uwb radar.
\newblock In {\em ACM Conference on Embedded Networked Sensor Systems
  (SenSys)}. ACM, November 2021.

\bibitem{ZhichengY:2016}
Zhicheng Yang, Parth~H. Pathak, Yunze Zeng, Xixi Liran, and Prasant Mohapatra.
\newblock Monitoring vital signs using millimeter wave.
\newblock In {\em Proceedings of the 17th ACM International Symposium on Mobile
  Ad Hoc Networking and Computing}, page 211–220. MobiHoc '16, July 2016.

\bibitem{AmaniYousefO:2020}
Amani~Yousef Owda, Neil Salmon, Alexander~J Casson, and Majdi Owda.
\newblock The reflectance of human skin in the millimeter-wave band.
\newblock {\em Sensors (Basel, Switzerland)}, 20(5):1480, March 2020.

\bibitem{GSdata:2017}
G.~Shafiq and K.~C. Veluvolu.
\newblock {Figshare}.
\newblock \url{https://dx.doi.org/10.6084/m9.figshare.c.3258022}, 2017.

\bibitem{GS:2017}
G.~Shafiq and K.~C. Veluvolu.
\newblock Multimodal chest surface motion data for respiratory and
  cardiovascular monitoring applications.
\newblock {\em Sci Data}, 4:170052, April 2017.

\bibitem{Shyan-LungL:2012}
Shyan-Lung Lin, Nai-Ren Guo, and C.~Chiu.
\newblock Modeling and simulation of respiratory control with labview.
\newblock {\em Journal of Medical and Biological Engineering}, 32:51--60, 2012.

\bibitem{AnuradhaS2:2020}
Anuradha Singh, Saeed~Ur Rehman, Sira Yongchareon, and Peter Han~Joo Chong.
\newblock Modelling of chest wall motion for cardiorespiratory activity for
  radar-based ncvs systems.
\newblock {\em Sensors}, 20(18):5094, September 2020.

\bibitem{Balthv:1926}
Balth. van~der Pol and Jun. D.Sc.
\newblock Lxxxviii. on “relaxation-oscillations”.
\newblock {\em The London, Edinburgh, and Dublin Philosophical Magazine and
  Journal of Science}, 2(11):978--992, 1928.

\bibitem{Balthv:1928}
Balth. van~der Pol and J.~van~der Mark.
\newblock Lxxii. the heartbeat considered as a relaxation oscillation, and an
  electrical model of the heart.
\newblock {\em The London, Edinburgh, and Dublin Philosophical Magazine and
  Journal of Science}, 6(38):763--775, 1928.

\bibitem{JF:2016}
J.~Feher.
\newblock {\em Quantitative Human Physiology}, page 446–454.
\newblock Elsevier Science Publishing Co Inc., San Diego, CA, USA, 2nd edition,
  2016.

\bibitem{RC:2006}
R.~Carroll.
\newblock {\em Elsevier’s Integrated Physiology}, page 65–75.
\newblock Mosby Elsevier, Philadelphia, PA, USA, 1st edition, 2006.

\bibitem{SandraG:2008}
Sandra~R.F.S.M. Gois and Marcelo~A. Savi.
\newblock An analysis of heart rhythm dynamics using a three-coupled oscillator
  model.
\newblock {\em Chaos, Solitons and Fractals}, 41(5):2553--2565, September 2008.

\bibitem{UnsooH:2020}
Unsoo Ha, Salah Assana, and Fadel Adib.
\newblock Contactless seismocardiography via deep learning radars.
\newblock In {\em Proceedings of the 26st Annual International Conference on
  Mobile Computing and Networking}, pages 1--14. MobiCom '20, April 2020.

\bibitem{WenyuS:2006}
Wenyu Sun and Ya-Xiang Yuan.
\newblock {\em Optimization Theory and Methods: Nonlinear Programming}.
\newblock Springer, Boston, MA, 2006.

\bibitem{iwr6843ISK}
Texas Instruments.
\newblock Iwr6843isk, 2020.

\bibitem{DCA1000EVM}
Texas Instruments.
\newblock Dca1000evm, 2020.

\bibitem{mmwavestudio}
Texas Instruments.
\newblock mmwave studio, 2020.

\bibitem{Log36}
NeuLog.
\newblock Respiration monitor belt logger sensor nul-236, 2020.

\bibitem{NUL208}
NeuLog.
\newblock Heart rate and pulse logger sensor nul-208, 2020.

\bibitem{KarlP:1895}
Karl Pearson.
\newblock Vii. note on regression and inheritance in the case of two parents.
\newblock {\em Proc. R. Soc. Lond.}, 58(347-352):240--242, January 1895.

\bibitem{AdamVB:2019}
Adam~V Benjafield, Najib~T Ayas, Peter~R Eastwood, Raphael Heinzer, Mary S~M
  Ip, Mary~J Morrell, Carlos~M Nunez, Sanjay~R Patel, Thomas Penzel,
  Jean-Louis~D Pépin, Paul~E Peppard, Sanjeev Sinha, Sergio Tufik, Kate
  Valentine, and Atul Malhotra.
\newblock Estimation of the global prevalence and burden of obstructive sleep
  apnoea: a literature-based analysis.
\newblock {\em The Lancet. Respiratory medicine}, 7(8):687–698, August 2019.

\bibitem{DavidB:2015}
David Blumenthal, Elizabeth Malphrus, and J.~Michael McGinnis.
\newblock {\em Vital Signs: Core Metrics for Health and Health Care Progress}.
\newblock National Academies Press (US), Washington (DC), 2015.

\bibitem{GeorgeAM:2019}
George~A Mensah, Gregory~A Roth, and Valentin Fuster.
\newblock The global burden of cardiovascular diseases and risk factors: 2020
  and beyond.
\newblock {\em J Am Coll Cardiol}, 74(20):2529--2532, November 2019.

\bibitem{AdeelA:2018}
Adeel Ahmad, June~Chul Roh, Dan Wang, and Aish Dubey.
\newblock Vital signs monitoring of multiple people using a fmcw
  millimeter-wave sensor.
\newblock In {\em 2018 IEEE Radar Conference (RadarConf18)}, pages 1450--1455.
  IEEE, April 2018.

\bibitem{XingzheS:2020}
Xingzhe Song, Boyuan Yang, Ge~Yang, Ruirong Chen, Erick Forno, Wei Chen, and
  Wei Gao.
\newblock Spirosonic: monitoring human lung function via acoustic sensing on
  commodity smartphones.
\newblock In {\em Proceedings of the 26st Annual International Conference on
  Mobile Computing and Networking}, pages 1--14. MobiCom '20, September 2020.

\bibitem{HyunjaeL:2019}
Hyunjae Lee, Byung-Hyun Kim, Jin-Kwan Park, and Jong-Gwan Yook.
\newblock A novel vital-sign sensing algorithm for multiple subjects based on
  24-ghz fmcw doppler radar.
\newblock {\em Remote Sensing}, 11(10):1237, May 2019.

\bibitem{ZengyangM:2020}
Zengyang Mei, Qisong Wu, Zhengyu Hu, and Jun Tao.
\newblock A fast non-contact vital signs detection method based on regional
  hidden markov model in a 77ghz lfmcw radar system.
\newblock In {\em ICASSP 2020 - 2020 IEEE International Conference on
  Acoustics, Speech and Signal Processing (ICASSP)}, pages 1145--1149. IEEE,
  May 2020.

\bibitem{GuigengS:2021}
Guigeng Su, Nikita Petrov, and Alexander Yarovoy.
\newblock Dynamic estimation of vital signs with mm-wave fmcw radar.
\newblock In {\em 2020 17th European Radar Conference (EuRAD)}, pages 206--209.
  IEEE, January 2021.

\bibitem{Khamis:2020}
Abdelwahed Khamis, Branislav Kusy, Chun~Tung Chou, and Wen Hu.
\newblock Wirelax: Towards real-time respiratory biofeedback during meditation
  using wifi.
\newblock {\em Ad Hoc Networks}, 107:102226, 2020.

\bibitem{ChrisW:2016}
Christoph Will, Kilin Shi, Fabian Lurz, Robert Weigel, and Alexander Koelpin.
\newblock Instantaneous heartbeat detection using a cross-correlation based
  template matching for continuous wave radar systems.
\newblock In {\em 2016 IEEE Topical Conference on Wireless Sensors and Sensor
  Networks (WiSNet)}, pages 31--34. IEEE, 2016.

\bibitem{ChristophW:2017}
Christoph Will, Kilin Shi, Robert Weigel, and Alexander Koelpin.
\newblock Advanced template matching algorithm for instantaneous heartbeat
  detection using continuous wave radar systems.
\newblock In {\em 2017 First IEEE MTT-S International Microwave Bio Conference
  (IMBIOC)}, pages 1--4. IEEE, May 2017.

\end{thebibliography}

\end{document}